\documentclass{article}
\usepackage[utf8]{inputenc}
\usepackage{sidecap}
\usepackage{xcolor}
\usepackage{graphicx}
\usepackage{hyperref}
\usepackage{enumitem}
\graphicspath{ {./figs/} }
\usepackage[a4paper, total={6in, 10.6in}]{geometry}
\usepackage{longtable, array}
\usepackage{geometry}
\usepackage{multirow}
\usepackage{lipsum}        
\usepackage{authblk}
\usepackage{caption}
\usepackage{subcaption}
\usepackage{etoolbox}
\usepackage{lmodern}
\usepackage[T1]{fontenc}
\usepackage{babel}
\usepackage[ruled,linesnumbered,commentsnumbered,vlined]{algorithm2e}
\usepackage{mathtools}
\usepackage{verbatim}
\usepackage{lineno}
\usepackage{epigraph}
\usepackage{comment}
\usepackage[normalem]{ulem} 
\usepackage{soul}           
\usepackage{textcomp}       
\usepackage{amsmath}

\makeatletter
\patchcmd{\@maketitle}{\LARGE \@title}{\fontsize{20}{29}\selectfont\@title}{}{}

\title{Exploration of Dark Chemical Genomics Space via Portal Learning: Applied to Targeting the Undruggable Genome and COVID-19 Anti-Infective Polypharmacology}
\author[1]{Tian Cai}
\author[2]{Li Xie}
\author[3]{Muge Chen}
\author[2]{Yang Liu}
\author [1] {Di He}
\author [1]{Shuo Zhang}
\author [4]{Cameron Mura}
\author [4]{Philip E. Bourne}
\author[1,2,5,*]{Lei Xie}

\affil[1]{Ph.D. Program in Computer Science, The Graduate Center, The City University of New York, New York, 10016, USA}
\affil[2]{Department of Computer Science, Hunter College, The City University of New York, New York, 10065, USA}
\affil[3]{Master Program in Computer Science, Courant Institute of Mathematical Sciences, New York University }
\affil[4]{School of Data Science \& Department of Biomedical Engineering, University of Virginia, Virginia, 22903, USA}
\affil[5]{Helen and Robert Appel Alzheimer’s Disease Research Institute, Feil Family Brain \& Mind Research Institute, Weill Cornell Medicine, Cornell University, New York, 10021, USA} 
\affil[*]{lei.xie@hunter.cuny.edu}
\begin{document}

\maketitle
\begin{abstract}
Advances in biomedicine are largely fueled by exploring uncharted territories of human biology. Machine learning can both enable and accelerate discovery, but faces a fundamental hurdle when applied to unseen data with  distributions that differ from previously observed ones---a common dilemma in scientific inquiry. We have developed a new deep learning framework, called {\textit{Portal Learning}}, to explore dark chemical and biological space. Three key, novel components of our approach include: (i) end-to-end, step-wise transfer learning, in recognition of biology's sequence-structure-function paradigm,  (ii) out-of-cluster meta-learning, and (iii) stress model selection. Portal Learning provides a practical solution to the out-of-distribution (OOD) problem in statistical machine learning. Here, we have implemented Portal Learning to predict chemical-protein interactions on a genome-wide scale. Systematic studies demonstrate that Portal Learning can effectively  assign ligands to unexplored gene families (unknown functions), versus existing state-of-the-art methods, thereby allowing us to target previously ``undruggable'' proteins and design novel polypharmacological agents for disrupting interactions between SARS-CoV-2 and human proteins. Portal Learning is general-purpose and can be further applied to other areas of scientific inquiry.
\end{abstract}

\clearpage

\section{Introduction}

The central aim of scientific inquiry has been to deduce new concepts from existing knowledge or generalized observations. The biological sciences offer  numerous such challenges. The rise of deep learning has spurred major  interest in using machine learning to explore uncharted molecular and functional spaces in biology and medicine, ranging from `deorphanizing' G-protein coupled receptors\cite{DISAE} and translating cell-line screens to patient drug responses\cite{few-shot-repurposing}\cite{he2021robust}, to predicting novel protein structures\cite{alphafold1}\cite{alphafold2}\cite{rosetta-fold}, to identifying new cell types from single-cell omics data\cite{cellline}. Illuminating the dark space of human knowledge is a fundamental problem that one can attempt to address via deep learning---that is, to generalize a ``well-trained'' model to unseen data that lies out-of-the-distribution (OOD) of the training data, in order to successfully predict outcomes from conditions that the model has never before encountered. While deep learning is capable, in theory, of simulating any functional mapping, its generalization power is notoriously limited in the case of distribution shifts\cite{causal-repr}.

The training of a deep learning model starts with a domain-specific model architecture. The final model instance that is selected, and its performance, are determined by a series of data-dependent design choices, including model initialization, data used for training, validation, and testing, optimization of loss function, and evaluation metrics. Each of these design choices impacts the generalization power of a trained model. The development of several recent deep learning-based approaches---notably transfer learning\cite{transfer-OOD-1}, self-supervised representation learning\cite{albert},{\linebreak[0]} and meta-learning{\linebreak[0]}\cite{maml}\cite{DBLP:meta-survey}{\linebreak[0]}---has been motivated by the OOD challenge. However, each of these methods focuses on only one aspect in the training pipeline of a deep model. Causal learning and mechanism-based modeling could be a more effective solution to the OOD problem \cite{causal-repr}, but at present these approaches can be applied only on modest scales because of data scarcity and limited domain knowledge.  Solving large-scale OOD problems in biomedicine, via machine learning, would benefit from a systematic framework for integrative, beginning-to-end model development, training, and testing.

Here, we propose a new deep learning framework, called {\textit{Portal Learning}}, that systematically addresses  the three OOD vulnerabilities in a training pipeline: specifically, we employ biology-inspired model initialization, optimization on an OOD loss, and model selection methods. We define `{\textit{portal}}' as a model with an initialized instance that is (preferably) close to the global optimum in some learning `{\textit{universe}}'. The {\textit{universe}} includes a specific input data-set, specific tasks, and a model architecture that provides a functional mapping from the data-set (and associated distributions) to the tasks. Note that, even with the same model architecture, changes in a pipeline's associated data-set correspond to changes in the universe. Portal Learning takes a global view to design training schemes that are task-specific and use domain knowledge as constraints to guide the exploration of the learning space. 

To assess the utility of Portal Learning, we implemented this concept as a concrete framework, termed {\textit{PortalCG}}, for predicting small-molecule binding to dark gene families (i.e., those with no annotated ligands). Despite tremendous progress in high-throughput screening, the majority of chemical genomics space remains unexplored or `dark' \cite{dark2019} (more details in results).  Elucidating dark gene families can illuminate many fundamental but only poorly characterized biological pathways, such as microbiome-host interactions mediated by metabolite-protein interactions. Such efforts could also provide novel approaches for identifying new druggable targets and discovering effective therapeutic strategies for currently incurable diseases; for instance, in Alzheimer's disease (AD) many disease-associated genes have been identified from multiple omics studies, but are currently considered un-druggable \cite{disgenet}. Accurately predicting chemical-protein interactions (CPIs) on a genome-wide scale is a challenging OOD problem\cite{DISAE}. If one considers only the reported area under the receiver operating characteristic curve (AUROC), which has achieved 0.9 in many state-of-the-art methods\cite{deepaffinity}\cite{DeepDTA}, it may seem the problem has been solved. However, the performance has been primarily measured in scenarios where the data distribution in the test set does not differ significantly from that in the training set, in terms of identities of proteins or types of chemicals. Few sequence-based methods have been developed and evaluated for an out-of-gene family scenario, where proteins in the test set belong to different (non-homologous) gene families than in the training set; this sampling bias is even more severe in considering cases where the new gene family does not have any reliable three-dimensional (3D) structural information. Therefore, one can fairly claim that all existing work has been confined to just narrow regions of chemical genomics space, without validated generalizability into the dark genome.

Rigorous benchmarking studies, reported herein, show that PortalCG significantly outperforms the leading  methods that are available for predicting ligand binding to (dark) proteins. We applied PortalCG to predict candidate drug compounds for undrugged disease genes in the dark human genome, and we prioritized hundreds of undrugged genes that can be efficaciously targeted by existing drugs (notably, many of which involve alternative splicing and transcription factor). These novel genes and their lead compounds provide new opportunities for drug discovery. Furthermore, using PortalCG, we identified polypharmacological agents that might leverage novel drug targets in order to disrupt interactions between SARS-CoV-2 and human proteins. The rapid emergence of SARS-CoV-2 variants has posed a significant challenge to existing vaccine and anti-viral development paradigms. Gordon et al. experimentally identified 332 human proteins that interact with the SARS-CoV-2 virus\cite{sars2-interactor}. This PPI map provides unique opportunities for anti-SARS-CoV-2 drug discovery: targeting the host proteins involved in PPIs can disrupt human SARS-CoV-2 interactions, thereby thwarting the onset of COVID-19. By not aiming to directly kill virions, this indirect strategy should lessen the selection pressure on viral genome evolution. A polypharmacological agent that interacts moderately strongly with multiple human proteins could be a potentially quite effective and safe anti-COVID-19 therapeutic: on the one hand, the normal functions of human proteins should not be significantly perturbed while, on the other hand, the interactions required for successful SARS-CoV-2 infection would be inhibited. Here, we virtually screened compounds in the Drug Repurposing Hub\cite{drug-hub} against the 332 human SARS-CoV-2 interactors. Two drugs, Fenebrutinib and NMS-P715, ranked highly; interestingly, both of these anti-tumorigenic compounds inhibit kinases. Their interactions with putative human targets were supported by further (structure-based) analyses of protein-ligand binding poses. 

In summary, the contributions of this work are three-fold: 
\begin{enumerate}

    \item A novel, generalized training scheme, {\textit{Portal Learning}}, is proposed as a way to guide biology-inspired systematic design in order to improve the generalization power of machine learning on OOD problems, such as is found in the dark  regions of molecular/functional space.
    
    \item To concretely illustrate the Portal Learning approach, a specific algorithm, PortalCG, is proposed and implemented. Comprehensive benchmark studies demonstrate the promise of PortalCG when applied to OOD problems, specifically for exploring the dark regions of chemical genomics space. 
    
    \item Using PortalCG, we shed new light on unknown protein functions in dark genomes (viz. small molecule-binding properties), and open new avenues in polypharmacology and drug repurposing; as demonstrated by identifying novel drug targets and lead compounds for AD and anti-SARS-CoV-2 polypharmacology.  
\end{enumerate}

\section{Conceptual basis of Portal Learning}

\begin{figure}[ht]
    \centering
    \includegraphics[width=0.90\linewidth]{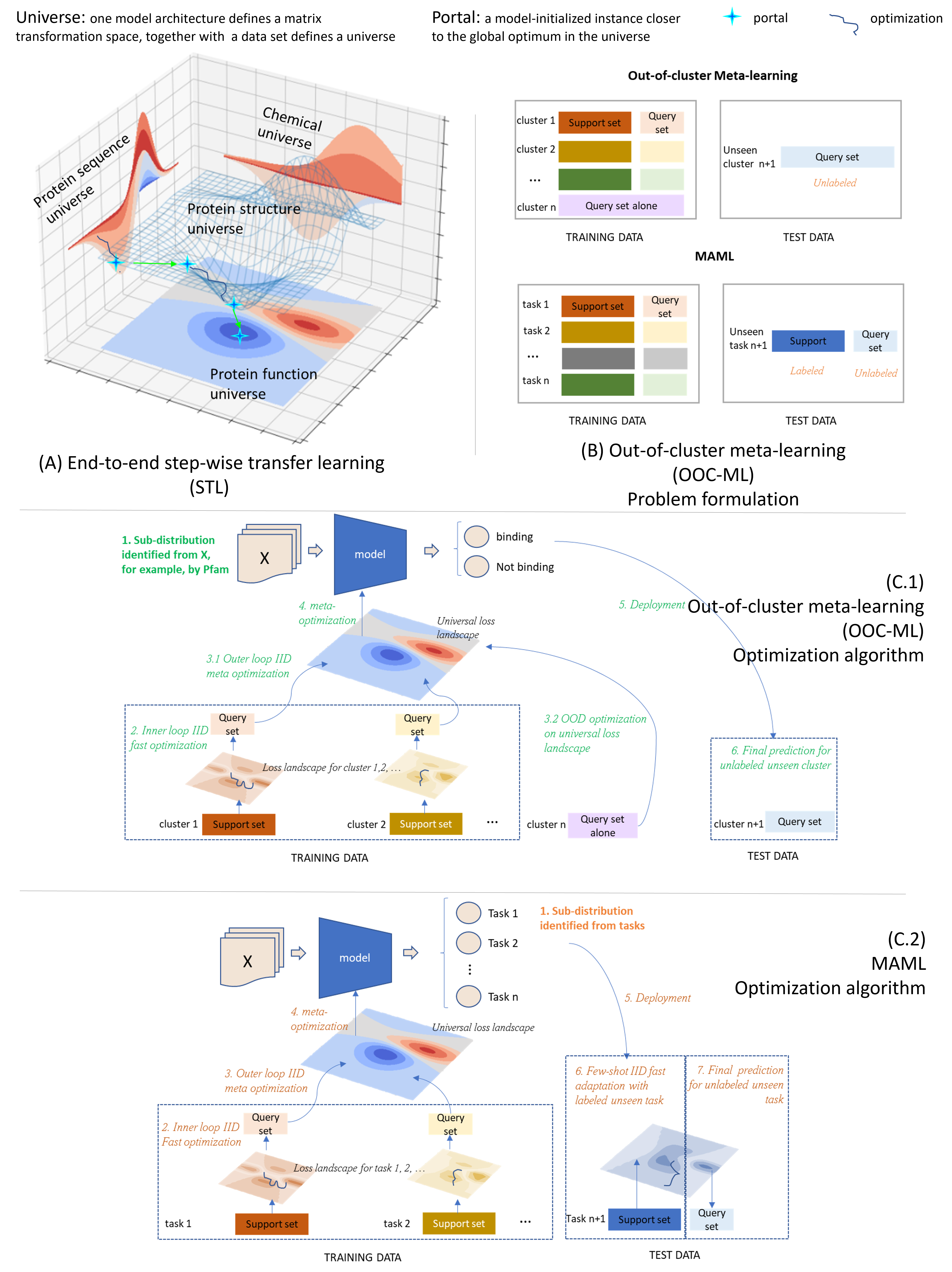}
    \caption{Illustration of two of the three major Portal Learning components for OOD problems, End-to-end step-wise transfer learning (STL) and out-of-cluster meta-learning (OOC-ML), using the prediction of out-of-gene family chemical-protein interactions (CPIs) as an example: 
    \textbf{A. STL}: 3D structure of protein ligand binding site is in the center connecting protein sequences to CPIs. There are two portals, the first traveling from the protein sequence universe to the binding site structure universe by pre-training a protein language model that is optimal in the protein sequence universe and leads to a model initialization instance closer to the global optimum in the binding site structure universe. The optimization based on this initialized instance leads to the discovery of the second portal through which protein function universe gets a model initialization instance closer to its own global optimum.
    \textbf{B. Problem formulation of OOC-ML in comparison with MAML}: Different from MAML where training data is grouped based on the task, the training data in OOC-ML is clustered in the instance space. Instead of decomposing the data in all clusters into support and query set like MAML, there is only a query set in certain training clusters and all testing clusters in OOC-ML to simulate OOD scenario.  
    \textbf{C.  Optimization of OOC-ML in comparison with MAML}: Intuitively, OOC-ML first performs local optimizations on each cluster of training data with the support/query decomposition, then meta optimizations on the training set that has only query sets by ensembling the knowledge learned from the local optimization. The optimized model is applied to the test data in a zero-shot learning setting. In contrast, the meta-optimization in MAML requires query sets in the setting of few-shot learning. 
    }
    \label{fig:wholepipeline}
\end{figure}

\begin{table}[ht]
\centering
\resizebox{\textwidth}{!}{%
\begin{tabular}{|c|c|c|c|l|}
\hline
\textbf{data split} &
  \textbf{Common practice} &
  \textbf{\begin{tabular}[c]{@{}c@{}}classic   scheme\\       applied in OOD\end{tabular}} &
  \textbf{Portal learning} &
  \multicolumn{1}{c|}{\textbf{specification}} \\ \hline
\multirow{2}{*}{train} & IID train & IID train & /           & each batch is from   the same distribution                              \\ \cline{2-5} 
                       & /         & /         & OOD   train & differentiate sub-distributions in each  batch                          \\ \hline
\multirow{2}{*}{dev}   & IID-dev   & IID-dev   & /           & from   the same distribution as the train set                     \\ \cline{2-5} 
                       & /         & /         & OOD-dev     & from   a different distribution from the training set                  \\ \hline
\multirow{2}{*}{test}  & IID-test  & /         & /           & from   the same distribution as the training set                     \\ \cline{2-5} 
                       & /         & OOD-test  & OOD-test    & from   a different distribution from both OOD-dev and training set \\ \hline
\end{tabular}%
}
\caption{Data split for stress model instance selection}
\label{tab:4split}
\end{table}

 To enable the exploration of dark regions of chemical and biological space, Portal Learning rests upon a systematic, well-principled training strategy, the underpinnings of which are shown in Figure 1. In Portal Learning, a model architecture together with a data set and a task defines a \textbf{universe}. Each universe has some global optimum with respect to the task based on a pre-defined loss function. The model-initialized instance in a universe---which could be a local optimum in the current universe, but which facilitates moving the model to the global optimum in the ultimately targeted universe---is called a \textbf{portal}. The portal is similar to a catalyst that lows the energy barrier via a transition state for a chemical reaction to occur. The dark chemical genomics space cannot be explored effectively if the learning process is confined only to the observed universe of protein sequences that have known ligands, as the known data are highly sparse and biased (details in Result section). Hence, it is critical to successfully identify portals into the dark chemical genomics universe starting from the observed protein sequence and structure universe. For clarity and ease of reference, key terms related to Portal Learning are given in the Supplemental Materials.

The remainder of this section describes the three key components of the Portal Learning approach---namely, end-to-end step-wise transfer learning (STL), out-of-cluster meta-learning (OOC-ML), and stress model selection. 

\textbf{End-to-end step-wise transfer learning (STL)}. Information flow in biological systems generally involves multiple intermediate steps, from a source instance to a target. For example, a discrete genotype (source) ultimately yields a downstream phenotype (target) via many steps of gene expression, in some environmental context. For predicting genotype-phenotype associations, explicit machine learning models that represent information transmission from DNA to RNA to cellular phenotype are more powerful than those that ignore the intermediate steps \cite{di-cleit}. In Portal Learning, transcriptomics profiles can be used as a portal to link the source genetic variation (e.g., variants, SNPs, homologs, etc.) and target cellular phenotype (e.g., drug sensitivity). Using deep neural networks, this process can be modeled in an end-to-end fashion.

\textbf{Out-of-cluster meta-learning (OOC-ML)}. Even if we can successfully transfer the information needed for the target through intermediate portals from the source universe, we still need additional portals to reach those many sparsely-populated regions of the dark universe that lack labeled data in the target. Inspired by Model Agnostic Meta-Learning (MAML)\cite{maml}, we designed a new OOC-ML approach to explore the dark biological space. MAML cannot be directly applied to Portal Learning in the context of the OOD problem because it is designed for few-shot learning under a multi-task formulation. Few-shot learning expects to have a few labeled samples from the test data set to update the trained model during inference for a new task. This approach cannot be directly applied to predicting gene functions of dark gene families where the task (e.g., binary classification of ligand binding) is unchanged, but rather there are no labeled data for a unseen distribution that may differ significantly from the training data. In a sense, rather than MAML's "few-shot/multi-task" problem context, mapping dark chemical/biological space is more of a "zero-shot/single-task" learning problem. A key insight of OOC-ML is to define sub-distributions (clusters) for the labeled data in the source instance universe. An example demonstrated in this paper is to define sub-distributions using Pfam families when the source instance is a protein sequence. Intuitively, OOC-ML involves a two-stage learning process. In the first stage, a model is trained using each individual labeled cluster (e.g., a given Pfam ID), thereby learning whatever knowledge is (implicitly) specific to each cluster. In the second stage, all trained models from the first stage are combined and a new ensemble model is trained, using labeled clusters that were not used in the first stage. In this way, we may extract common intrinsic patterns shared by all clusters and apply the learned essential knowledge to dark ones.

\textbf{Stress model selection}. Finally, training should be stopped at a suitable point in order to avoid overfitting. This was achieved by stress model selection. Stress model selection is designed to basically recapitulate an OOD scenario by splitting the data into OOD train, OOD development, and OOD test sets as listed in Table \ref{tab:4split}; in this procedure, the data distribution for the development set differs from that of the training data, and the distribution of the test data set differs from both the training and development data.

For additional details and perspective, the conceptual and theoretical basis of Portal Learning is further described in the Methods section of the Supplemental Materials.


\section{Results and Discussion}
\subsection{Overview of PortalCG}

\begin{figure}
    \centering
    \includegraphics[width=\textwidth]{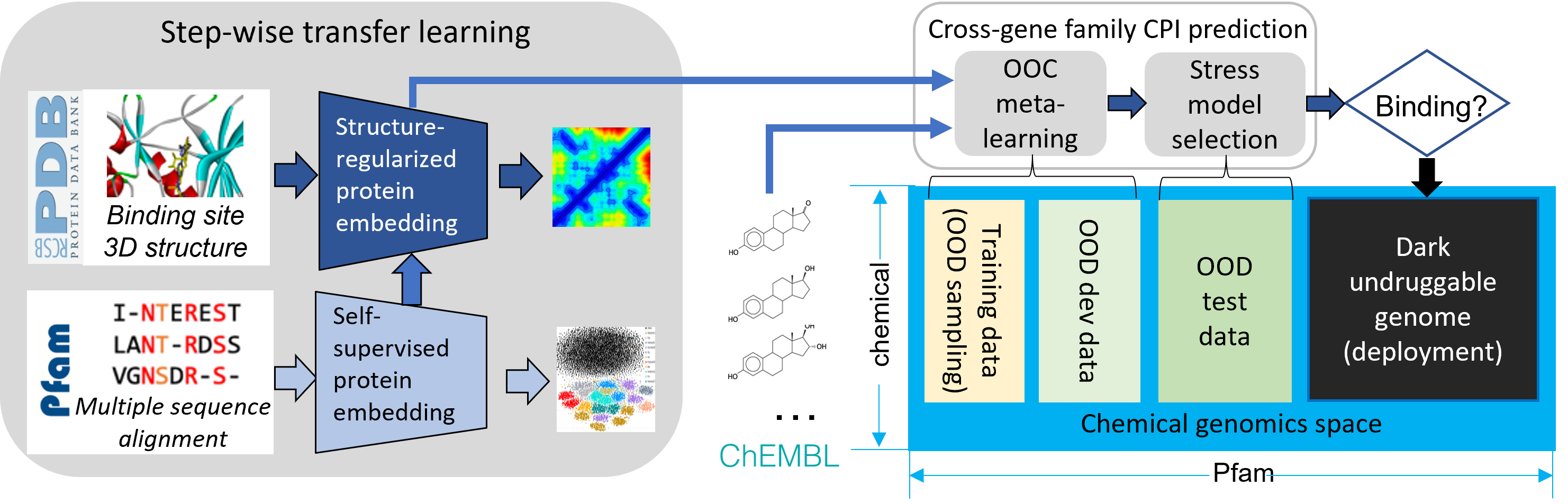}
    \caption{Scheme of PortalCG. PortalGC enables to predict chemical protein interactions (CPIs) for dark genes across gene families. It includes three key components: end-to-end transfer learning following sequence-structure-function paradigm, Out-of-cluster (OOC) meta-learning, and stress model selection.}
    \label{fig:in-a-nutshell}
\end{figure}

We implemented the Portal Learning concept as a concrete model, PortalCG, for exploring the dark chemical genomics space. In terms of Portal Learning's three key components (STL, OOC-ML, and stress model selection), PortalCG makes the following design choices (see also Figure \ref{fig:in-a-nutshell}).

\textbf{End-to-end sequence-structure-function STL}. The function of a protein---e.g., serving as a target receptor for ligand binding---stems from its three-dimensional (3D) shape and dynamics which, in turn, is ultimately encoded in its primary amino acid sequence. In general, information about a protein's structure is more powerful than purely sequence-based information for predicting its molecular function because sequences drift/diverge far more rapidly than do 3D structures on evolutionary timescales. Although the number of experimentally-determined structures continues to exponentially increase, and now AlphaFold2 can reliably predict 3D structures of most single-domain proteins, it nevertheless remains quite challenging to directly use protein structures as input for predicting ligand-binding properties of dark proteins. In PortalCG, protein structure information is used as a portal to connect a source protein sequence and a corresponding target protein function (Figure \ref{fig:wholepipeline}A). We begin by performing self-supervised training to map tens of millions of sequences into a universal embedding space, using our recent {\textit{distilled sequence alignment embedding}} (DISAE) algorithm \cite{DISAE}. Then, 3D structural information about the ligand-binding site is  used to fine-tune the sequence embedding. Finally, this structure-regularized protein embedding was used as a hidden layer for supervised learning of cross-gene family CPIs, following an end-to-end sequence-structure-function training process. By encapsulating the role of structure in this way, inaccuracies and  uncertainties in structure prediction are `insulated' and will not propagate to the function prediction. 

\textbf{Out-of-cluster meta-learning}. In the OOC-ML framework, Pfam gene families provide natural clusters as sub-distributions. In each Pfam family, the data is split into support set and query set as shown in Figure \ref{fig:wholepipeline}(B). Specifically, a model is trained for a single Pfam family independently to reach a local minimum  using the support set of the Pfam family as shown in the inner loop IID optimization in Figure \ref{fig:wholepipeline}(C.1). Then a query set from the same Pfam family is used on the locally optimized model to get a loss from the local loss landscape, i.e. outer loop IID meta optimization in Figure \ref{fig:wholepipeline}(C.1). Local losses from the query sets of multiple Pfam families will be aggregated to calculate the loss on a global loss landscape, i.e. meta optimization in Figure \ref{fig:wholepipeline}(C.1). For some cluster with very limited number of data, they don't have a support set hence will only participate in the optimization on the global loss landscape. There could be many choices of aggregations. A simple way is to calculate the average loss. The aggregated loss will be used to optimize the model on the global loss landscape. Note that weights learned on each local loss landscape will be memorized during the global optimization. In our implementation, it is realized by creating a copy of the model trained from the each family's local optimization. In this way, the local knowledge learned is ensured to be only passed to the global loss landscape by the query set loss.

\textbf{Stress model selection}. The final model was selected using Pfam families that were not used in the training stage (Figure \ref{fig:in-a-nutshell}, right panel). 

The Supplemental Materials provide further methodological details, covering data pre-processing, the core algorithm, model configuration, and implementation details.

\subsection{There are significantly unexplored dark spaces in chemical genomics}

\begin{figure}
    \centering
        \includegraphics[width=\textwidth]{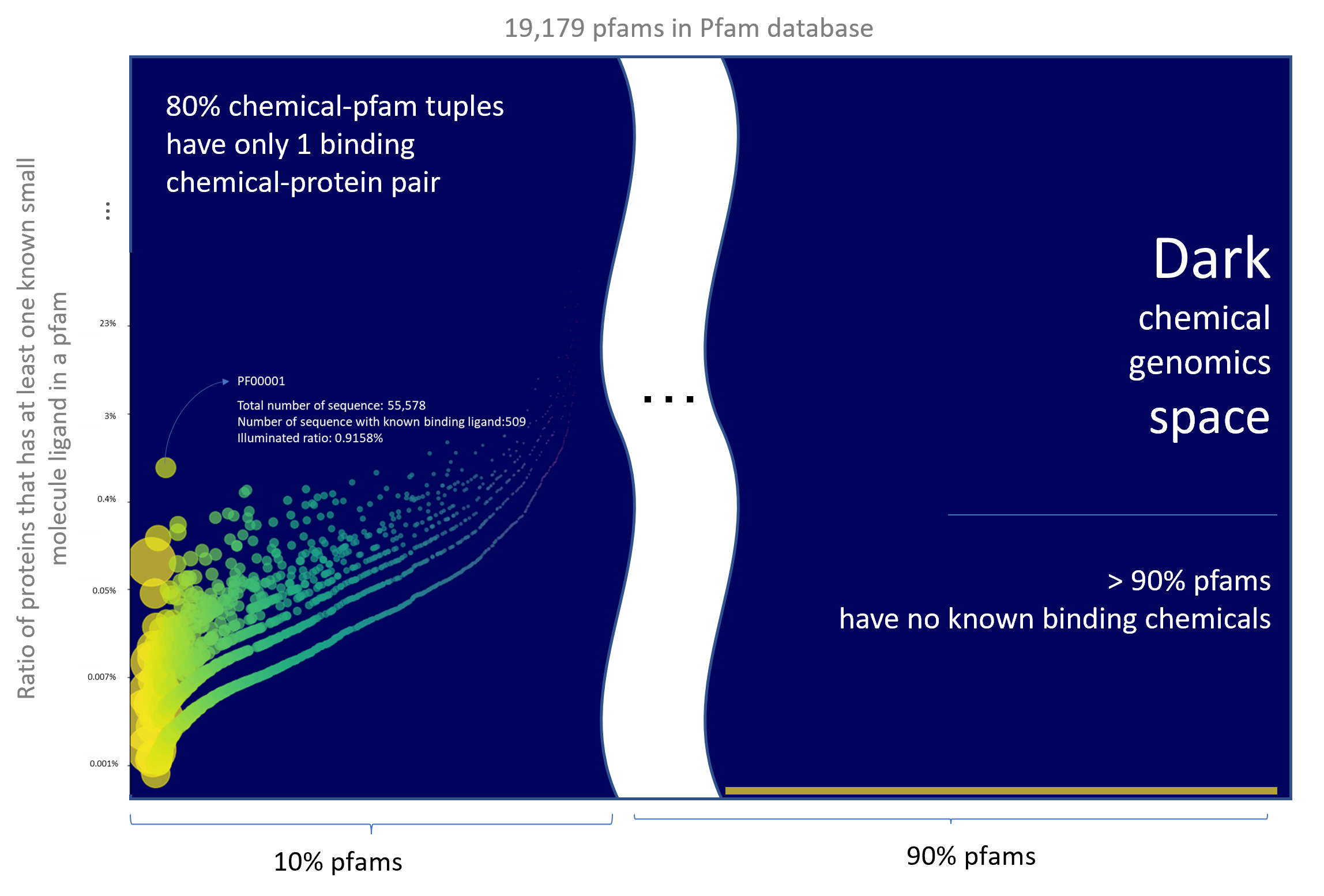}
        
        \caption{\textbf{Chemical genomics space in statistics: The ratio of proteins that have at least a known ligand in each Pfam family.} Each color bubble represents a Pfam family. The size of a bubble is proportional to the total number of proteins in the Pfam family. Y-axis is the ratio of proteins with known ligand(s) in a Pfam family. Around $2,000$ Pfam families have at least one known small molecule ligand. Most of these Pfam families have less than $1\%$ proteins with known ligands. Furthermore, around 90\% of total $19,179$ Pfam families are in the dark chemical genomics space without any known ligand information. }
    
\label{fig:darkspace}
\end{figure}

We inspected the known CPIs between (i) molecules in the manually-curated ChEMBL database, which consists of only a small portion of all chemical space, and (ii) proteins annotated in Pfam-A \cite{Pfam}, which represents only a narrow slice of the whole protein sequence universe. The ChEMBL26\cite{chembl26} database supplies $1,950,765$ chemicals paired to $13,377$ protein targets, constituting $15,996,368$ known interaction pairs. Even for just this small portion of chemical genomics space, unexplored CPIs are enormous, can be seen in the dark region in Figure \ref{fig:darkspace}. Approximately 90\% of Pfam-A families do not have any known small-molecule binder. Even in Pfam families with annotated CPIs (e.g., GPCRs), there exists a significant number of `orphan' receptors with unknown cognate ligands (Figure \ref{fig:darkspace}). Fewer than $1\%$ of chemicals bind to more than two proteins, and $<0.4\%$ of chemicals bind to more than five proteins, as shown in Supplemental Figures S1, S2 and S3. Because protein sequences and chemical structures in the dark chemical genomics space could be significantly different from those for the known CPIs, predicting CPIs in the dark space is an archetypal, unaddressed OOD problem. 

\subsection{Portal Learning significantly outperforms state-of-the-art approaches to predicting dark CPIs}

\begin{figure}[ht]
    \centering

    \begin{subfigure}[b]{0.8\textwidth}
        \centering
        \includegraphics[scale=0.55]{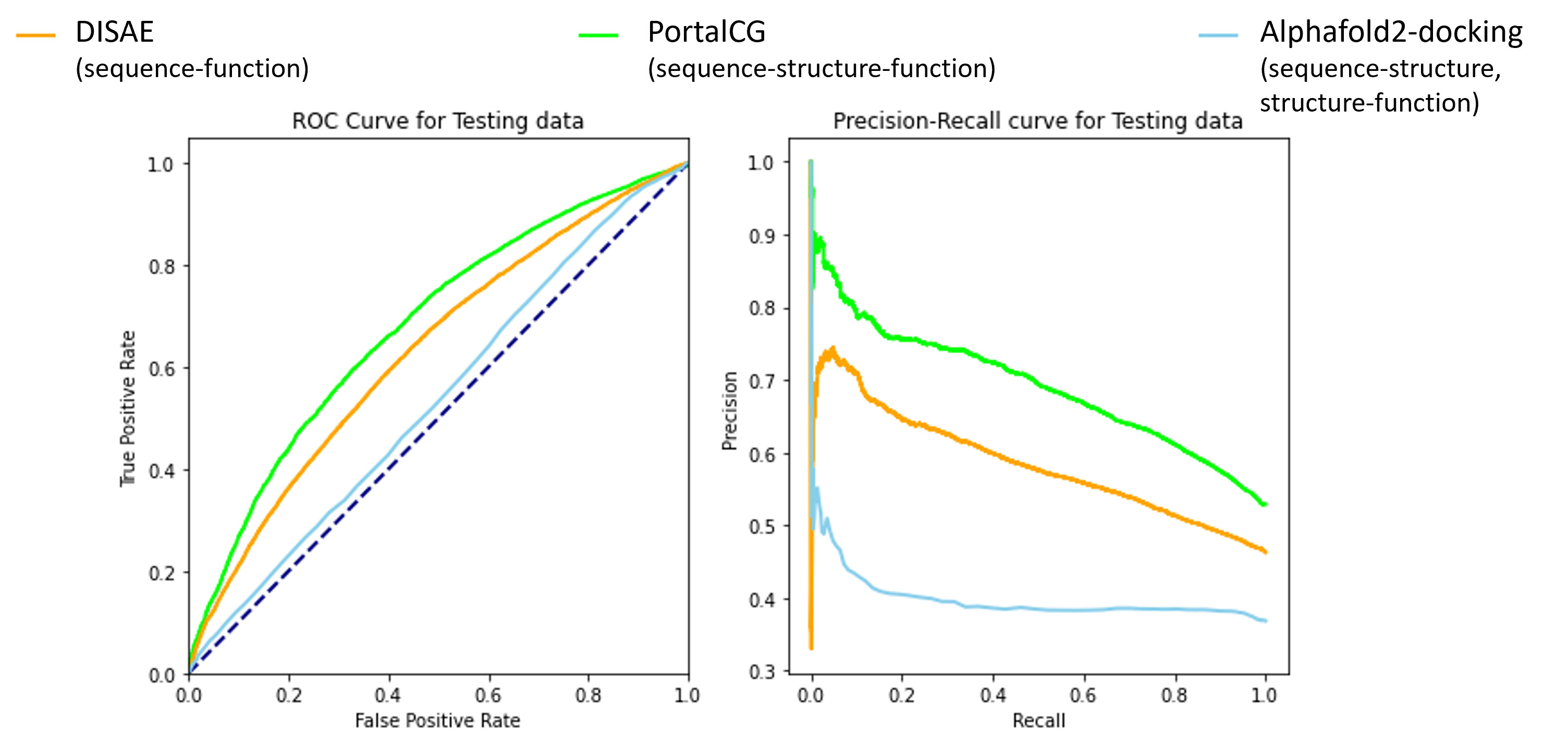}{(a)}
    \end{subfigure}
    \vfill
    \begin{subfigure}[b]{0.8\textwidth}
        \centering
        \includegraphics[scale=0.7]{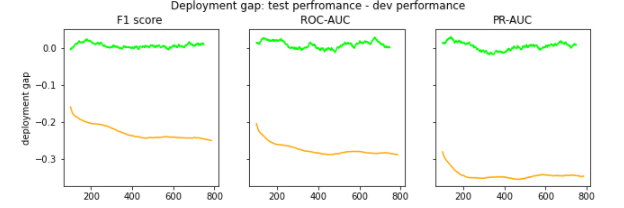}{(b)}
    \end{subfigure}
    \caption{Comparison of PortalCG with the state-of-the-art method DISAE as baseline using the shifted evaluation test. (a)  ROC and Precision-Recall curves for the ``best'' model instance selected by stress test; (b) 
    Deployment gaps where the deployment gap of PortalCG is steadily around zero as training step increases while the deployment performance of DISAE deteriorates.}
        \label{fig:auc-curve}
\end{figure}

When compared with the state-of-the-art method DISAE\cite{DISAE}, which already was shown to outperform other leading methods for predicting CPIs of orphan receptors, PortalCG demonstrates superior performance in terms of both Receiver Operating Characteristic (ROC) and Precision-Recall (PR) curves, as shown in Figure \ref{fig:auc-curve}(a). Because the ratio of positive and negative cases is imbalanced, the PR curve is more informative than the ROC curve. The PR-AUC of PortalCG and DISAE is 0.714 and 0.603, respectively. In this regard, the performance gain of Portal Learning (18.4\%) is significant (p-value < $1e-40$). Performance breakdowns for binding and non-binding classes can be found in Supplemental Figure S4.

PortalCG exhibits much higher recall and precision scores for positive cases (i.e., a chemical-protein pair that is predicted to bind) versus negative, as shown in Supplemental Figure S4; this is a highly encouraging result, given that there are many more negative (non-binding) than positive cases. The deployment gap, shown in Figure \ref{fig:auc-curve}(b), is steadily around zero for PortalCG; this promising finding means that we can expect that, when applied to the dark genomics space, the performance will be similar to that measured using the development data set.

With the advent of high-accuracy protein structural models, predicted by AlphaFold2 \cite{alphafold2}, it now becomes feasible to use reversed protein-ligand docking (RPLD)\cite{huang2018reverse} to predict ligand-binding sites and poses on dark proteins, on a genome-wide scale. In order to compare our method with the RPLD approach, blind docking to putative targets was performed via Autodock Vina\cite{autodock}. After removing proteins that failed in the RPLD experiments (mainly due to extended structural loops), docking scores for 28,909 chemical-protein pairs were obtained. The performance of RPLD was compared with that of PortalGC and DISAE. As shown in Figure \ref{fig:auc-curve}(a), both ROC and PR for RPLD are significantly worse than for PortalGC and DISAE. It is well known that PLD suffers from a high false-positive rate due to poor modeling of protein dynamics, solvation effects, crystallized waters, and other challenges \cite{DTI-challenge-structure}; often, small-molecule ligands will indiscriminately `stick' to concave, pocket-like patches on protein surfaces. For these reasons, although AlphaFold2 can accurately predict many protein structures, the relatively low reliability of PLD still poses a significant limitation, even with a limitless supply of predicted structures \cite{virtual-screen-performance}. Thus, the direct application of RPLD remains a challenge for predicting ligand binding to dark proteins. PortalCG's  end-to-end sequence-structure-function learning could be a more effective strategy: protein structure information is not used as a fixed input, but rather as an intermediate layer that can be tuned using various structural and functional information. From this perspective, again the role of protein structure in PortalCG can be seen as that of a portal (sequence$\rightarrow$function; Figure \ref{fig:wholepipeline}) and a regularizer (Figure \ref{fig:in-a-nutshell}).

\subsection{Both the STL and OOC-ML stages contribute to the improved performance of PortalCG}

\begin{table}[ht]
\centering
\resizebox{\textwidth}{!}{%
\begin{tabular}{|l|l|c|c|c|c|}
\hline
\multicolumn{1}{|c|}{} &
  \multicolumn{1}{c|}{\textbf{models}} &
  \textbf{\begin{tabular}[c]{@{}c@{}}PR-AUC\\ (OOD-test set)\end{tabular}} &
  \textbf{\begin{tabular}[c]{@{}c@{}}ROC-AUC\\ (OOD-test set)\end{tabular}} &
  \textbf{\begin{tabular}[c]{@{}c@{}}ROC-AUC\\ Deployment Gap\end{tabular}} &
  \textbf{\begin{tabular}[c]{@{}c@{}}PR-AUC\\ Deployment Gap\end{tabular}} \\ \hline
DIASE &
  PortalCG w/o STL \& OOC-ML &
  0.603±0.005 &
  0.636±0.004 &
  -0.275±0.016 &
  -0.345±0.012 \\ \hline
variant 1 &
  \begin{tabular}[c]{@{}l@{}}PortalCG w/o OOC-ML\end{tabular} &
  0.629±0.005 &
  0.661±0.004 &
  --- &
  --- \\ \hline
variant 2 &
  \begin{tabular}[c]{@{}l@{}}PortalCG w/o STL\end{tabular} &
  0.698±0.015 &
  0.654±0.062 &
  --- &
  --- \\ \hline
PortalCG &
  \begin{tabular}[c]{@{}l@{}}Portal learning\end{tabular} &
  0.714±0.010 &
  0.677±0.010 &
  0.010±0.009 &
  0.005±0.010 \\ \hline
\end{tabular}%
}
\caption{Ablation study of PortalCG.}
\label{tab:PRAUC}
\end{table}

To gauge the potential contribution of each component of PortalCG to the overall system effectiveness in predicting dark CPIs, we systematically compared the four models shown in Table \ref{tab:PRAUC}. Details of the exact model configurations for these experiments can be found in the Supplemental Materials Table S10 and Figure S13. As shown in Table \ref{tab:PRAUC}, Variant 1, with a higher PR-AUC compared to the DISAE baseline, is the direct gain from transfer learning through 3D binding site information, all else being equal; yet, with transfer learning alone and without OOC-ML as an optimization algorithm in the target universe (i.e., Variant 2 versus Variant 1), the PR-AUC gain is minor. Variant 2 yields a 15\% improvement while Variant 1 achieves  only a 4\% improvement. PortalCG (i.e., full Portal Learning), in comparison, has the best PR-AUC score. With all other factors held constant, the advantage of PortalCG appears to be the synergistic effect of both STL and OOC-ML. The performance gain measured by PR-AUC under a shifted evaluation setting is significant (p-value < 1e-40), as shown in Supplemental Figure S5. 

We find that stress model selection is able to mitigate potential overfitting problems, as expected. Training curves for the stress model selection are in Supplemental Figures S4 and S6. As shown in Supplemental Figure S6, the baseline DISAE approach tends to over-fit with training, and IID-dev performances are all higher than PortalCG but deteriorate in OOD-test performance. Hence, the deployment gap for the baseline is -0.275 and -0.345 on ROC-AUC and PR-AUC, respectively, while PortalCG deployment is around 0.01 and 0.005, respectively.

\subsection{Application of PortalCG to explore dark chemical genomics space}
A production-level model using PortalCG was trained with ensemble methods for the deployment. Details are in the Supplemental Methods section. The trained PortalCG model was applied to two case-studies in order to assess  its utility in the exploration of dark space. As long as a protein and chemical pair was presented to this model with their respective sequence and SMILES string, a prediction could be made, along with a corresponding prediction score. To select high confidence predictions, a histogram of prediction scores was built based on known pairs (Supplemental Figure S7). A threshold of $0.67$, corresponding to a false positive rate of 2.18e-05, was identified to filter out high-confidence positive predictions. Around 6,000 drugs from the Drug Repurposing Hub\cite{clue} were used in the screening. The remainder of this section describes the two case-studies that were examined with PortalCG, namely (i) COVID-19 polypharmacology and (ii) the `undruggable' portion of the human genome.

\subsubsection{COVID-19 polypharmacology}
In order to identify lead compounds that may disrupt SARS-CoV-2-Human interactions, we screened 5,886 approved and investigational drugs against the 332 human proteins known to interact with SARS-CoV-2. We considered a drug-protein pair as a positive hit and selected it for further analysis only when all models in an ensemble vote as positive and the false positive rate does not exceed \sout{is} 2.18e-05. Drugs involved in these positive pairs were ranked according to the number of proteins to which they are predicted to bind. Detailed information is given in Supplemental Table S1. Most of these drugs are protein kinase inhibitors and are already in Phase 2 clinical trials. Among them, Fenebrutinib and NMS-P715 are predicted to bind to seven human SARS-CoV-2 interactors, as shown in Table \ref{tab:docking}.      
In order to elucidate how these drug molecules might associate with a SARS-CoV-2 interactor partner, we performed molecular docking for Fenebrutinib and NMS-P715. Structures of two SARS-CoV-2 interactors were obtained from the Protein Data Bank; the remaining five proteins do not have experimentally solved structures so  their predicted structures (via AlphaFold2) were used for docking. For most of these structures, the binding pockets are unknown. Therefore, blind docking was employed, using Autodock Vina\cite{autodock} to search the full surfaces (the accessible molecular envelope) and identify putative binding sites of Fenebrutinib and NMS-P715 on these interactors. Docking conformations with the best (lowest) predicted binding energies were selected for each protein; the respective binding energies are listed in Table \ref{tab:docking}.

\begin{figure}[ht]
    \centering
    \includegraphics[width=\textwidth]{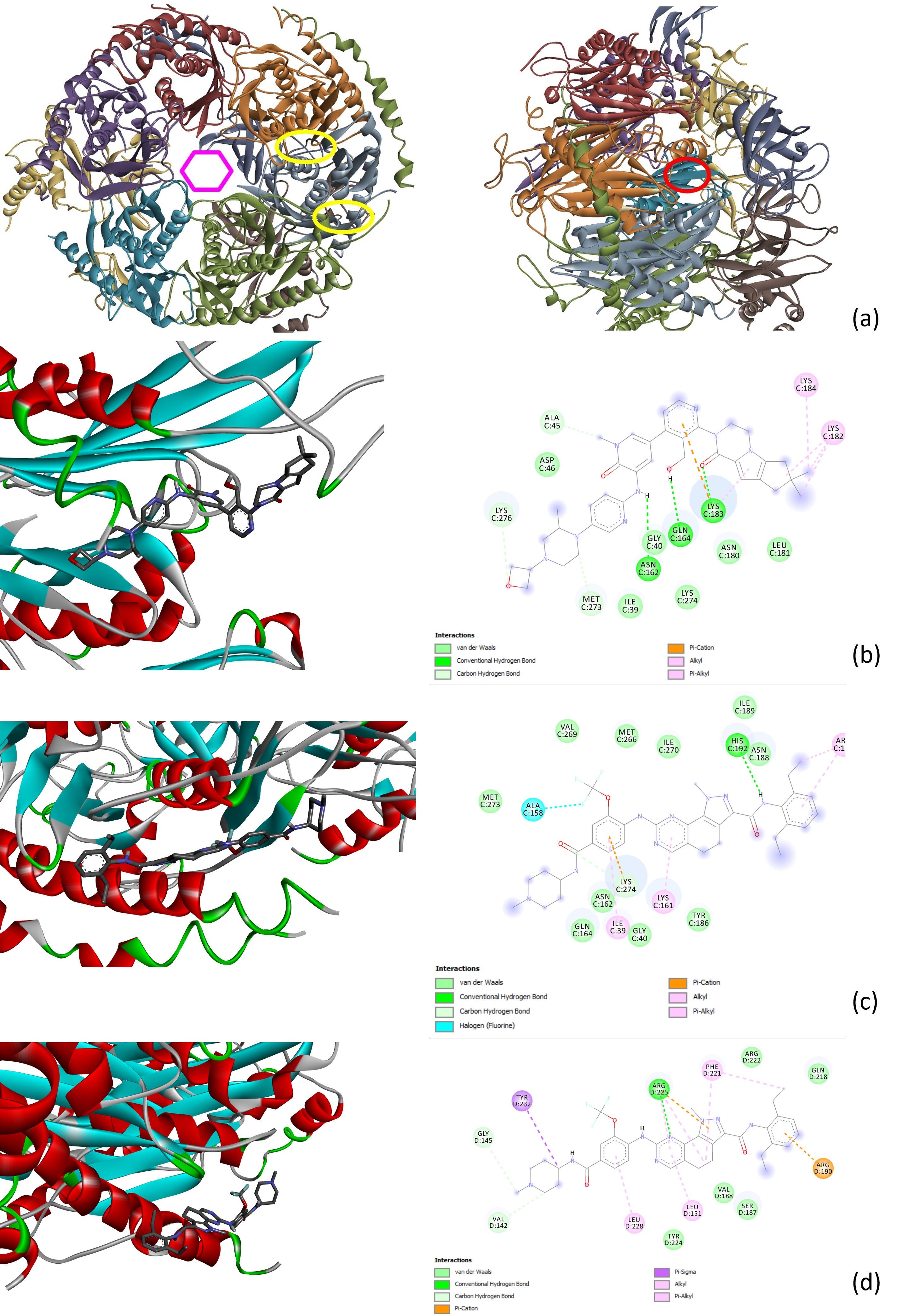}
    \caption{The 3D structure of the exosome complex and the binding conformations of Fenebrutinib and NMS-P715 on the complex components predicted by using Autodock: (a) The exosome complex structure; Left: yellow circles shows the binding pocket of NMS-P715 on RRP43 and RRP46, purple hexagon shows the gate; Right: red circle shows the binding pocket of Fenebrutinib on RRP43. (b) Fenebrutinib on RRP43. (c) NMS-P715 on RRP43. (d) NMS-P715 on RRP46.}
    \label{fig:docking conformation}
\end{figure}

Components of the exosome complex are predicted targets for both Fenebrutinib and NMS-P715. The exosome complex is a multi-protein, intracellular complex which is involved in degradation of many types of RNA molecules (e.g., via 3'$\rightarrow$5' exonuclease activities). As shown in Figure \ref{fig:docking conformation}, the subunits of the exosomal assembly form a central channel; RNA passes through this region as part of the degradation/processing. Intriguingly, SARS-CoV-2's genomic RNA has been found to be localized in the exosomal cargo, suggesting a key mechanistic role for the channel region in SARS-CoV-2 virion infectivity pathways \cite{Exosomes}. Fenebrutinib and NMS-P715 were also predicted to bind to a specific exonuclease, RRP43, of the exosome complex, while NMS-P715 was also predicted to  bind yet another exonuclease, RRP46.

The predicted binding poses for Fenebrutinib and NMS-P715 with the exosomal complex components are shown in Figure \ref{fig:docking conformation}. The physicochemical/interatomic interactions between these two drugs and the exosome complex components are also schematized as a 2D layout in this figure. The favorable hydrogen bond, pi-alkyl, pi-cation and Van der Waals interactions provide additional support that  Fenebrutinib and NMS-P715 do indeed bind to these components of the exosome complex. The predicted binding poses and 2D interactions maps for Fenebrutinib and NMS-P715 with other targeted proteins are shown in Supplementary Figures S8, S9, and S10.

\begin{table}[ht]
\centering
\resizebox{\textwidth}{!}{%

\begin{tabular}{|c|c|c|c|}
\hline
\multicolumn{4}{|c|}{Docking scores of \textbf{Fenebrutinib} binding to predicted targets}                                                                                                \\ \hline
Uniprot ID & Protein name                                    & PDB ID                                      & \begin{tabular}[c]{@{}c@{}}Docking score \\ (kcal/mol)\end{tabular} \\ \hline
Q96B26     & Exosome complex component RRP43                 & 2NN6\_C                                     & -7.9                                                                \\ \hline
Q5JRX3     & Presequence protease, mitochondrial             & 4L3T\_A                                     & -10.8                                                               \\ \hline
Q99720     & Sigma non-opioid intracellular receptor 1       & 5HK1\_A                                     & -9.6                                                                \\ \hline
Q5VT66     & Mitochondrial amidoxime-reducing component 1    & 6FW2\_A                                     & -10.4                                                               \\ \hline
P29122     & Proprotein convertase subtilisin/kexin type 6   & \multicolumn{1}{l|}{AF-P29122-F1 (157-622)} & -8.5                                                                \\ \hline
Q96K12     & Fatty acyl-CoA reductase 2                      & \multicolumn{1}{l|}{AF-Q96K12-F1 (1-478)}   & -10.1                                                               \\ \hline
O94973     & AP-2 complex subunit alpha-2                    & \multicolumn{1}{l|}{AF-O94973-F1 (3-622)}   & -8.6                                                                \\ \hline
\multicolumn{4}{|c|}{Docking scores of \textbf{NMS-P715} binding to predicted targets}                                                                                                    \\ \hline
Uniprot ID & Protein name                                    & PDB ID                                      & \begin{tabular}[c]{@{}c@{}}Docking score\\ (kcal/mol)\end{tabular}  \\ \hline
Q9UN86     & Ras GTPase-activating protein-binding protein 2 & 5DRV\_A                                     & -9.5                                                                \\ \hline
P67870     & Casein kinase II subunit beta                   & 1QF8\_A                                     & -8.6                                                                \\ \hline
Q96B26     & Exosome complex component RRP43                 & 2NN6\_C                                     & -9.3                                                                \\ \hline
P62877     & E3 ubiquitin-protein ligase RBX1                & 2HYE\_D                                     & -7.9                                                                \\ \hline
P61962     & DDB1- and CUL4-associated factor 7              & \multicolumn{1}{l|}{AF-P61962-F1 (9-341)}   & -8.7                                                                \\ \hline
Q9NXH9     & tRNA (guanine(26)-N(2))-dimethyltransferase     & \multicolumn{1}{l|}{AF-Q9NXH9-F1 (53-556)}  & -9.0                                                                \\ \hline
Q9NQT4     & Exosome complex component RRP46                 & 2NN6\_D                                     & -8.6                                                                \\ \hline
\end{tabular}%
}
           
\caption{Docking scores for Fenebrutinib and NMS-P715} 
    \label{tab:docking}

\end{table}

\subsubsection{Illuminating the undruggable human genome}

\begin{figure}[ht]
   
        \includegraphics[width=1.0\linewidth]{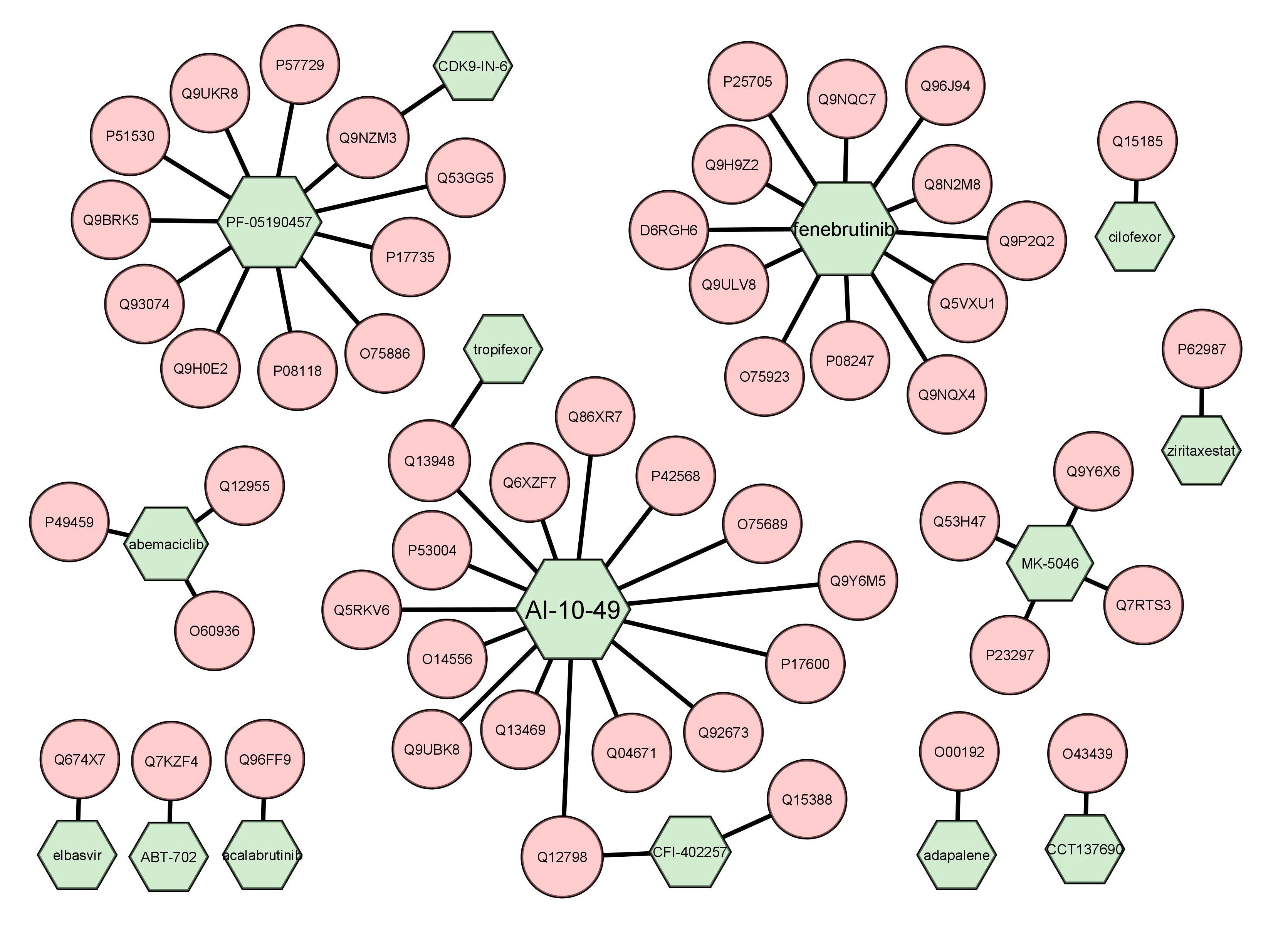}

    \caption {Drug-target interaction network for proteins associated with Alzheimer's disease. Green represents drugs and pink represents targets. }
    \label{fig:AD-drug-target}
\end{figure}

It is well known that only a small subset of the human genome is considered druggable \cite{finan2017druggable}. Many proteins are deemed ``undruggable'' because there is no information on their ligand-binding properties or other interactions with small-molecule compounds (be they endogenous or exogenous ligands). Here, we built an ``undruggable'' human disease protein database by removing the druggable proteins in Pharos \cite{pharos} and Casas's druggable proteins \cite{cesas} from human disease associated genes \cite{disgenet} and applied PortalCG to predict the probability for these ``undruggable'' proteins to bind to drug-like molecules.  A total of 12,475 proteins were included in our disease-associated undruggable human protein list. These proteins were ranked according to their probability scores, and 267 of them have a false positive rate lower than 2.18e-05, as listed in the supplementary material  Table S2. Table \ref{tab:enrichment} shows the statistically significantly enriched functions of these top ranked proteins as determined by DAVID \cite{David}. The most enriched proteins are involved in alternative splicing of mRNA transcripts. Malfunctions in alternative splicing are linked to many diseases, including several cancers \cite{alternative-splicing}\cite{alternative-splicing2} and Alzheimer's disease \cite{love2015alternative}. However, pharmaceutical modulation of alternative splicing process is a challenging task. Identifying new drug targets and their lead compounds for targeting alternative splicing pathways may open new doors to developing novel therapeutics for complex diseases with few treatment options. Diseases associated with these 267 human proteins were also listed in Table \ref{tab:disease}. Since one protein is always related to multiple diseases, these diseases are ranked by the number of their associated proteins. Most of top ranked diseases are related with cancer development. 21 drugs that are approved or in clinical development are predicted to interact with these proteins as shown in Table S3. Several of these drugs are highly promiscuous. For example, AI-10-49, a molecule that disrupts protein-protein interaction between CBFb-SMMHC and tumor suppressor RUNX1, may bind to more than 60 other proteins. The off-target binding profile of these proteins may provide invaluable information on potential side effects and opportunities for drug repurposing and polypharmacology. The drug-target interaction network built for predicted positive proteins associated with Alzheimer's disease was shown in Figure \ref{fig:AD-drug-target}. Functional enrichment, disease associations, and top ranked drugs for the undruggable proteins with well-studied biology (classified as Tbio in Pharos) and those excluding Tbio are list in Supplemental Table S4-S9. 

\begin{table}[ht]
\centering
\resizebox{\textwidth}{!}{%

\begin{tabular}{|c|c|c|c|c|}

\hline
\multicolumn{5}{|c|}{David Functional Annotation enrichment analysis}    

\\ \hline
\begin{tabular}[c]{@{}c@{}}Enriched terms in \\ UniProtKB keywords\end{tabular} & \begin{tabular}[c]{@{}c@{}}Number of \\ proteins involved\end{tabular} & \begin{tabular}[c]{@{}c@{}}Percentage of \\ proteins involved\end{tabular} & P-value  & \begin{tabular}[c]{@{}c@{}}Modified \\ Benjamini p-value\end{tabular} \\ \hline
{\color[HTML]{080808} Alternative splicing}                                     & 171                                                                    & 66.5                                                                       & 7.70E-07 & 2.00E-04                                                              \\ \hline
{\color[HTML]{080808} Phosphoprotein}                                           & 140                                                                    & 54.5                                                                       & 2.60E-06 & 3.40E-04                                                              \\ \hline
{\color[HTML]{080808} Cytoplasm}                                                & 91                                                                     & 35.4                                                                       & 1.30E-05 & 1.10E-03                                                              \\ \hline
{\color[HTML]{080808} Nucleus}                                                  & 93                                                                     & 36.2                                                                       & 1.20E-04 & 8.10E-03                                                              \\ \hline
{\color[HTML]{080808} Metal-binding}                                            & 68                                                                     & 26.5                                                                       & 4.20E-04 & 2.20E-02                                                              \\ \hline
{\color[HTML]{080808} Zinc}                                                     & 48                                                                     & 18.7                                                                       & 6.60E-04 & 2.90E-02                                                              \\ \hline
\end{tabular}%
}
\caption{Functional Annotation enrichment for undruggable human disease proteins selected by PortalCG} 
    \label{tab:enrichment}
\end{table}

\begin{table}[ht]
\resizebox{\textwidth}{!}{%
\begin{tabular}{|c|c|}

\hline
DiseaseName                   & \# of undruggable proteins associated with disease \\ \hline
Breast Carcinoma              & 90                                                 \\ \hline
Tumor Cell Invasion           & 86                                                 \\ \hline
Carcinogenesis                & 83                                                 \\ \hline
Neoplasm Metastasis           & 75                                                 \\ \hline
Colorectal Carcinoma          & 73                                                 \\ \hline
Liver carcinoma               & 66                                                 \\ \hline
Malignant neoplasm of lung    & 56                                                 \\ \hline
Non-Small Cell Lung Carcinoma & 56                                                 \\ \hline
Carcinoma of lung             & 54                                                 \\ \hline
Alzheimer's Disease           & 54                                                 \\ \hline
\end{tabular}%
}    
\caption{Top ranked diseases associated with the undruggable human disease proteins selected by PortalCG} 
    \label{tab:disease}
\end{table}

\section{Conclusion }
This paper confronts the challenge of exploring dark chemical genomics space by recognizing it as an OOD generalization problem in machine learning, and by developing a new learning framework to treat this type of problem. We propose Portal Learning as a general framework that enables systematic control of the OOD generalization risk. As a concrete algorithmic example and use-case, PortalCG was implemented under the Portal Learning framework. Systematic examination of the PortalCG method revealed its superior performance compared to (i) a state-of-the-art deep learning model (DISAE), and (ii) an AlphaFold2-enabled, structure-based reverse docking approach. PortalCG showed significant improvements in terms of both sensitivity and specificity, as well as close to zero deployment performance gap. With this approach, we were able to explore the dark regions of the druggable genome. Applications of PortalCG to COVID-19 polypharmacology and to the targeting of hitherto undruggable human proteins affords novel new directions in drug discovery. 

\section{Methods}
\subsection{Full algorithm details }
Portal learning as a system level framework involves collaborative new design from data preprocessing, data splitting to model architecture, model initialization, and model optimization and evaluation. The main illustrations are Figure \ref{fig:wholepipeline} and Figure \ref{fig:in-a-nutshell}. Extensive explanation of each of the component and their motivations are available in Supplemental Materials section Methods with Figure S11, and Algorithm1. 

\subsection{Data}
PortalCG uses three database, Pfam\cite{Pfam}, Protein Data Bank (PDB)\cite{pdb} and ChEMBL\cite{chembl26}. Two applications are demonstrated, COVID-19 polypharmacology and undruggable human proteins, for which known approved drugs are collected from CLUE\cite{clue}, 332 human proteins interacting SARS-CoV-2 are listed in recent publication\cite{nature-covid-target}, 12,475 undruggable proteins are collected by removing the druggable proteins in Pharos \cite{pharos} and Casas's druggable proteins \cite{cesas} from human disease associated genes \cite{disgenet}. Detailed explanation of how each data set is used can be found in Supplemental Materials Methods section.

Major data statistics are demonstrated in Figure \ref{fig:darkspace} and Supplemental Materials Figure S1, S2, and S3.

\subsection{Experiment implementation}
Experiments are first organized to test PortalCG performance against baseline models, DISAE\cite{DISAE} and AlphFold2\cite{alphafold2}. DISAE is a protein language which predicts protein function based on protein sequence information alone. AlphaFold2 uses protein sequence information to predict protein structure, combing docking methods, can be used to predict protein function. Main results are shown with Table \ref{tab:PRAUC} and Figure \ref{fig:auc-curve}. Ablation studies is also performed mainly to test some variants of PortalCG components such as binding site distance prediction as shown in Supplemental Figure S12. Since Portal Learning is a general framework, there could be many interesting variants to pursue in future studies. To enhance application accuracy, a production level model is built with ensemble learning, and high confidence predictions are selected as demonstrated in Supplemental Material Figure S7. Evaluation metrics used are F1, ROC-AUC and PR-AUC.

Extensive details can be found in Supplemental Materials Methods section.

\subsection{Related works}
A literature review of related works could be found in Supplemental Materials section Related Works.
\clearpage

\addcontentsline{toc}{section}{Author Contributions}
\section*{Author Contributions}
TC conceived the concept of Portal Learning, implemented the algorithms, performed the experiments, and wrote the manuscript; Li Xie prepared data, performed the experiments, and wrote the manuscript; MC implemented algorithms; YL implemented algorithms; SZ prepared data; CM and PEB refined the concepts and wrote the manuscript; Lei Xie conceived and planned the experiments, wrote the manuscript. 
\addcontentsline{toc}{section}{Data and software availability}
\section*{Data and software availability}
Data, a pre-trained PortalCG model, and PortalCG codes can be found in the following link: \url{ https://github.com/XieResearchGroup/PortalLearning}

\addcontentsline{toc}{section}{Acknowledgement}
\section*{Acknowledgement}
This project has been funded with federal funds from the National Institute of General Medical Sciences of National Institute of Health (R01GM122845) and the National Institute on Aging of the National Institute of Health (R01AD057555). 
We appreciate that Hansaim Lim helped with proof reading and provided constructive suggestions.


\clearpage
\addcontentsline{toc}{section}{Reference}
\bibliography{refer} 
\bibliographystyle{ieeetr}
\clearpage
\appendix
\renewcommand\thefigure{S\arabic{figure}}    
\setcounter{figure}{0}
\setcounter{table}{0}
\renewcommand{\thetable}{S\arabic{table}}
\tableofcontents
\clearpage

\section{Key terms}
To provide common ground for discussion with readers of various backgrounds, we list the specification of key terms related to the methods in Supplemental materials. The following list provides explanations at an intuition level without the attempt to establish formal definitions. Readers could refer to referenced materials for more formal definitions.

\textbf{Deep learning specific}
\begin{enumerate}
   
    \item model architecture: the design of a model as a set of trainable parameters without specification on the exact weight of the parameters \cite{Goodfellow-et-al-2016}.
    
    \item loss landscape: the geometry of the global loss associated with a model architecture\cite{visual-loss-landscape}. 
    
    \item model instance: given a model architecture with certain amount of trainable parameters, a set of weights associated to the trainable parameters defines an instance of model; during the training process, each optimization step leads to a model instance.
    
    \item  optimization\cite{Goodfellow-et-al-2016}: neural network is trained by optimizing a object function, usually in the form of minimizing a loss function.
    
    \item global/local optimum: under the optimization formalization, the optimum point is at the end of the optimization process. As explained in \cite{bootstrap-generalization}, in an ideal world where the complete distribution of data is available to train a model, the optimum is global; while any stop point for any  certain subdistribution is a local optimum.
    
    \item model initialization: optimization always starts with an initialization of an model instance.
    
    \item pretraining\cite{albert}: train a model on a pretext task before training on the target task; the trained model instances will become the initialization of the target task.
    
    \item finetuning\cite{albert}: train a model on a target task with initialization pretrained task. 

    \item Independent and Identifically distributed (IID)\cite{iid-def}: if given a set of data ${x_i}$, each of these $x_i$  observations is an independent draw from a fixed (``stationary'') probability distribution.
    
    \item Out-of-distribution (OOD) generalization\cite{ood-thesis}: Generalization consists in reducing the gap in performance between training data and testing data. When the data generating process from the training data is indistinguishable from that of the test data, it's ``in-distribution'', if not, it's called out-of-distribution generalization problem \cite{ood-thesis}. As in \cite{invariant-risk-minimization}, considering datasets $Data_{e}:=\{(x_{i}^{e},y_{i}^{e} )\}_{i=1}^{n_{e}}$ collected under multiple domains $e$, each containing samples IID with a probability distribution $D(X^{e},Y^{e})$. The goal of OOD generalization is to use these datasets to learn a predictior $Y\approx f(X)$, which performs well across a large set of unseen but related domains $e\in  \mathcal{E}_{all}$. Namely, the goal is to  minimize
\begin{equation}
    R^{OOD}(f) = \max_{e\in  \mathcal{E}_{all}} R^{e}(f)
\end{equation}

, where 
\begin{equation}
    R^{e}(f) := E_{X^{e}, Y^{e}}[l(f(X^{e}),Y^{e})]
\end{equation}

is the risk under domain $e$. Here the set $\mathcal{E}_{all}$ contains all possible domains.

    \item generalization\cite{Goodfellow-et-al-2016}: the most general goal of generalization is to enable the model make reliable prediction on unseen data; out-of-distribution prediction (OOD) is a more challenging type of generalization problem which requires the model to be generalizable to unseen data distribution.
    
     \item mini-batch\cite{Goodfellow-et-al-2016}: as a common practice for robustness and computation memory concerns, no matter how large a data set is available, only a sub set of data is sampled to train a model at each optimization step.
     
    \item representation\cite{Goodfellow-et-al-2016}: as coined by the line of work named ``representation learning'', the word representation is interachangable with ``embedding'', referring to a vector/matrix of learned features.
    
    \item transfer model parameter: a technique that is related to pretraining-finuetuning; for the implementation, a simple initialization of part of the target model using the pretrained model could serve the purpose.
 
\end{enumerate}
\textbf{Chemical-protein interaction (CPI) specific}
\begin{enumerate}
  \item CPI prediction: formulated as a binary classification task, to predict whether or not a pair of protein and chemical will bind given only protein sequence and chemical SMILES string.
  
    \item protein descriptor, chemical descriptor: the modules of a whole CPI prediction model to extract protein/chemical embedding in a euclidean space.
\end{enumerate}
\textbf{Portal learning specific}
\begin{enumerate}
    
    \item universe: a model architecture that defines a data transformation space, together with  a data set.
    
    \item portal: a model instance in a universe---which could be a local optimum in the current universe, but which facilitates moving the model to the global optimum in the ultimately targeted universe.
    
    \item local loss landscape: optimize a model on a sub-distribution of the complete underlying distribution of the whole data set.
    
    \item global loss landscape: the direction of gradient search towards global optimum for all sub-distributions.
    
    \item stress test: a technique\cite{Underspecification} to evaluate a predictor by observing its outputs on specifically designed inputs; three common types are stratified performance evaluation, shifted evaluation and contrastive evaluation.
    
    \item shifted evaluation\cite{Underspecification}: the stress test employed in this paper, which splits train/test data set by Pfam families, i.e. proteins in the testing and training sets come from different Pfam families. This is a simple simulation off dark space model deployment.
    
    \item deployment gap: the difference between the performance evaluated by the test set and that evaluated by the development set.
    
    \item classic deep learning training scheme: randomly split a whole data set into train/dev/test set; optimize model on a randomly sampled mini-batch data; choose a final trained model instance based on the best test evaluation metrics; usually adopts empirical risk minimization \cite{erm} formulation.
    
\end{enumerate}

\clearpage


\section{Methods: PortalCG in four-universe configuration for dark chemical genomics space}
In this section, we present the detailed methodology used in portal learning in the context of a four-universe configuration.

\subsection{Data preprocessing }
The four-universe configuration is built on three major databases, Pfam\cite{Pfam}, Protein Data Bank (PDB)\cite{pdb}, BioLp\cite{Biolip} and ChEMBL\cite{chembl26}. The data were preprocessed as follows.

\begin{itemize}
    \item Protein sequence universe. All sequences from Pfam-A families are used to pretrain the protein descriptor following the same setting in DISAE \cite{DISAE} that highlights a MSA-based distillation process.
    
    \item Protein structure universe. In our protein structure data set, there are 30,593 protein structures, 13,104 ligands, and 91,780 ligand binding sites. Binding sites were selected according to the annotation from BioLip (updated to the end of 2020). Binding sites which contact with DNA/RNA and metal ions were not included. If a protein has more than one ligand, multiple binding pockets were defined for this protein. For each binding pocket, the distances between $C\alpha$ atoms of amino acid residues of the binding pocket were calculated. In order to obtain the distances between the ligand and its surrounding binding site residues, the distances between atom i in the ligand and each atom in the residue j of the binding pocket were calculated and the smallest distance wa selected as the distance between atom i and residue j.  In order to get the sequence feature of the binding site residues in the DISAE protein sequence representation\cite{DISAE}, binding site residues obtained from PDB structures (queries) were mapped onto the multiple sequence alignments of its corresponding Pfam family. First, a profile HMM database was built for the whole Pfam families. hmmscan \cite{Hmmer} was applied to search the query sequence against this profile database to decide which Pfam family it belongs to. For those proteins with multiple domains, more than one Pfam families were identified. Then the query sequence was aligned to the most similar sequence in the corresponding Pfam family by using phmmer. Aligned residues on the query sequence were mapped to the multiple sequence alignments of this family according to the alignment between the query sequence and the most similar sequence.   
    
    \item Chemical universe.  All chemicals in the ChEMBL26 database consists of the chemical universe.
    
    \item Protein function universe. CPI classification prediction data is the whole ChEMBL26\cite{chembl26} database where the same threshold for defining positive and negative labels creating as that in DISAE \cite{DISAE} was used.

\end{itemize}

\subsection{Algorithm }

In the four-universe configuration, portal learning starts with portal identification in the protein sequence universe, then travels into protein structure universe for portal calibration before finally comes into the target protein function universe, where OOC-ML  will be invoked for model optimization. Along the way, shifted evaluation, one type of stress model selection, is used to select the ``best'' model instance, which splits train/test based on Pfam families, i.e. training set and testing set have proteins come from different Pfam families. Each phase will be specified in the following sections. 


\subsubsection{Chemical representation}
A chemical was represented as a graph and its embedding was learned using GIN\cite{gin}. 

\subsubsection{Protein sequence pre-training}
Protein descriptor is pretrained from scratch following exactly DISAE \cite{DISAE} on whole Pfam families, making it a universal protein language model. With standard Adam optimization, shifted evaluation is used to select the ``best'' instance.

\subsubsection{Protein structure regularization}
With the protein descriptor pretrained using the sequences from the whole Pfam, chemical descriptors and a distance learner were plugged in to fine-tune the protein representation. The distance learner follows Alphafold\cite{alphafold1} which formulates a multi-way classification on a distrogram. Based on the histogram of binding site distances, a histogram equalization\footnote{Histogram equalization: \url{https://en.wikipedia.org/wiki/Histogram_equalization}} was applied to formulate a 10-way classification on our binding site structure data as in Supplemental material Figure S \ref{fig:hist-equal-10class}. Since protein and chemical descriptors output position-specific embeddings of a distilled protein sequence and all atoms of a chemical, pair-wise interaction features on the binding sites were created with a simple vector operation: a matrix multiplication was used to select embedding vectors of each binding residue and atom; multiply and broadcast the selected embedding vectors into a symmetric tensor as shown  in the following, where $H$ is embedding matrix of size $(number\_of\_residues, embedding_dimension)$ or $(number\_of\_atoms, embedding_dimension)$ and $A$ is selector matrix\cite{linear-alg},
\[
H_{binding\_site}^{protein} = A^{protein}*H_{full\_distilled}^{protein} 
\]
\[
H_{binding\_site}^{chemical} = A^{chemical}*H_{full\_chemical\_graph}^{chemical} 
\]
\[
H_{binding\_site}^{interaction} = (H_{binding\_site}^{protein})^{T} *H_{binding\_site}^{chemical} 
\]
This pair-wise interaction feature tensor $H_{binding\_site}^{interaction}$ was fed into a Attentive Pooling\cite{santos2016attentive} layer followed by feed-forward layer for final 10-way classification. Detailed model architecture configuration could be found in Table S \ref{tab:config} and Figure S\ref{fig:architecture-details} .The intuition for the simplest form of distance learner is to put all stress of learning on the shared protein and chemical descriptors which will carry information across universes. Again, with standard Adam optimization, shifted evaluation was used to select the ``best'' instance. Two versions of distance structure prediction were implemented, one formulated as a binary classification, i.e. contact prediction, one formulated as a multi-way classification, i.e. distogram prediction. The performance of the two version are similar, as shown in Figure S \ref{fig:ablation-contact-distogram}.

\subsubsection{Out-of-cluster Meta Learning (OOC-ML) in protein function universe}
With fine-tuned protein descriptor in the protein function universe, a binary classifier is plugged on, which is a ResNet\cite{resnet} layered with two linear layers as shown in Table S \ref{tab:config} and Figure S\ref{fig:architecture-details}. What plays the major role in this phase is the optimization algorithm OOC-ML as shown in pseudocode \textbf{Algorithm1} and main content Figure 1(B),(C.1). The local loss landscape exploration is reflected in line 4-9, and line 10 shows ensemble of global loss landscape. Note that more variants could be derived from changing sampling rule (line 3 and 5) and global loss ensemble rule.

OOC-ML is built on MAML\cite{maml} but has significant differences. Echoing to steps illustrated in the Figure 1 of the main text:
\begin{enumerate}
    \item As shown in main content Figure1 (B), OOC-ML has a sub-distribution data split into support set and query set, or as MAML named it, meta-train and meta-test within training set and test set. However, MAML sub-distributions are identified from the label space $\{Y\}$ while OOC-ML identifies sub-distributions, i.e. clusters, from input feature space $\{X\}$. In PortalCG, the clusters are identified by Pfam. Further, OOC-ML allows the utilization of very small clusters where very limited known data points are available for training. For example, in PortalCG, some Pfam families with too few samples to be split into support and query set are organized as query-set-alone, which participate only in the global loss optimization, as detailed below. 
    
    \item In each mini-batch, a few sub-distributions are sampled. The whole optimization has two layers, inner loop and outer loop. At the inner loop, each sub-distribution data has its own local loss landscape. The support set is used for in-distribution optimization on the local loss landscape. 
    
    \item The locally optimized model is then used on query set to get a query set loss, which will be fed to the global loss landscape. Each sub-distribution is independently optimized. This step is the same as MAML. What is different is that OOC-ML also calculates query-set without local in-distribution optimization for the small clusters. 
    
    \item Local query set losses are pooled together and the model will be optimized on the global loss landscape as meta-optimization defined in MAML.
    
    \item After finishing train, the model will be deployed.
    
    \item MAML is designed for multi-class classification in few-shot learning, at deployment stage, it's expected to meet new unseen class. And it's assumed that there are \textbf{a few labelled sample available} as support set, hence named as \textbf{few-shot learning}.  For each unseen class, the trained model will carry out a fast in-distribution adaptation using support set before final prediction on the query set. However, this is impossible in the context of dark space illumination. Portal learning trained model has to make robust predictions without any chance of in-distribution adaptation. 
\end{enumerate}

\IncMargin{1em}
\begin{algorithm}[ht]
\SetKwData{Left}{left}\SetKwData{This}{this}\SetKwData{Up}{up}
\SetKwFunction{Union}{Union}\SetKwFunction{FindCompress}{FindCompress}
\SetKwInOut{Input}{input}\SetKwInOut{Output}{output}

\Input{$p(\textbf{D})$, CPI data distribution over whole pfams, where each $D_i\in \textbf{D}$ is a set of CPI pairs for the $pfam_i$;\\

$\alpha,\beta$, learning step size hyperparameters;\\
$L$, number of optimization steps in  each round of local exploration;\\
$T$, number of global training steps;\\
$K$, number of points sampled from a local neighborhood 
}
\Output{$\theta$ the whole model weights}
\BlankLine
\SetAlgoLined initialization whole model weights $\theta$ (with weight transfered from portal for protein and chemical descriptors and random initiliazed weights for binary classifier)

\For{$t$ in $T$}{ 
Sample a $D_i \sim p(D)$;

\For{$l$ in $L$}{
  Sample a positive-negative balanced mini-batch of $K$ pairs $neighborhood_m \sim D_i$;
  
  \For{$point_j$ in $neighborhood_m$ }{
   Evaluate $\nabla_{\theta} L_{point_j}(f_\theta)$ with respect to $K$ examples\;
   Compute adapted parameters with gradient decent:$\theta_{i}^{'}= \theta - \alpha \nabla_{\theta}L_{point_j}(f_\theta)$\;
   }{
   Update $\theta \leftarrow \theta - \beta\nabla_{\theta}\sum_{D_{i}\sim p(D)}L_{point_j}(f_{\theta}^{'})$\;
  }
 }}

 \caption{Portal Learning Optimization Algorithm: Out-of-cluster Meta-learning}
\end{algorithm}

\subsubsection{Stress model instance selection}

 In classic training scheme common practice, there are 3-split data sets, ``train set'', ``dev set'' and ``test set''.  Train set as the name suggested is used to train model. Test set as commonly expected is used to set an expectation of performance when applying the trained model to unseen data. Dev set is to select the preferred model instance. In OOD setting, data is split (main content Table 1) such that dev set is a OOD from train set and test set is a OOD from both train and dev set. Deployment gap is calculated by deducting ODD-dev performance with OOD-test performance.

\subsection{Implementation details}
With portal learning being a framework, all experiments are based on the configuration of a four-universe design. Four major variants of models are trained as shown in main content Table 2  for controlled factor experiments to verify the contribution of key components of Portal Learning. In this section we present implementation details.

Due to the large number of total samples, all training are carried out under global step-based formalization instead of epoch-based. Typically, a deep learning model is trained for numerous epochs, in each epoch the model will loop over all training data. Evaluation will be carried out once on the whole test data set at the end of each epoch. In the global step formalization, a mini-batch is sampled at uniform random from pre-split training data set. For a pre-defined total number of global steps, this mini-batch sampling will be repeated. Training is stopped when loss decreases are within a pre-defined error margin. To evaluate along the way of training, for every $m$ global steps of training, a subset of test data is sampled uniformly randomly from a pre-split test set. To compute generalization gaps, in addition to evaluate on test set split according to the shifted evaluation, a dev set is held out from the train set for the evaluation as well. In this way, dev set and train set are iid. The performance difference between dev and train is the \textit{observed space generalization gap} while the performance difference between dev and test is the \textit{dark space generalization gap}.

\subsection{Evaluation metrics}
Distogram prediction uses an average accuracy on the distogram. CPI binary classification uses F1, ROC-AUC and PR-AUC for overall evaluation with breakdown by class F1, recall and precision scores. 


\subsection{Docking as baseline}
Protein-ligand docking was performed using Autodock Vina\cite{autodock}. The whole protein surface search implemented in the Autodock Vina was applied to identify the ligand binding pocket. The center of each protein was set as the center of the binding pocket. The largest distance of the protein atoms to the center of the protein is calculated for each x, y, and z direction to define the edge of each protein. 10 Angstrom of extra space was added to the protein edge to set up the search space for the docking.

\subsection{Production level for deployment}

To create a production level model, three models were trained in PortalCG with only difference in data split. Dev set was OOD in respect of training set to make sure there was no overlapped Pfam families between them. By rotating Pfam families between training set and OOD-dev set in the fashion of a cross-validation, each of the three models was trained on different train set in light of Pfam families involved. Then a voting mechanism was used to make the final prediction.


\subsection{Dark space exploration from a theoretical lens}
A neural network classifier is trained by minimizing a loss function with a standard form as the following:
\[L(\theta) = \frac{1}{|D_t|} \sum_{(x,y)\in D_t} -\log p_{\theta}(x,y)\] where $p_{\theta}(x,y)$ is the probability that a sample $x$ belongs to the class $y$ according to the trained neural network with parameters $\theta$, and $D_t$ is the training data set with the number of samples $|D_t|$. As laid out in the recent framework in  \cite{bootstrap-generalization} that reasons about generalization in  deep learning, the test error of a model $f_{t}$ could be decomposed as follows,

\[
TestError(f_{t}) = \underbrace{TestError(f_{t}^{iid})}_\text{ideal world} +
\underbrace{
[TestError(f_{t}) - TestError(f_{t}^{iid})]}_\text{real world generalization gap}
\]

When  data are sampled as independent and identically distributed (iid) random variables, ``ideal world'' is a scenario where the complete data distribution is available with infinite data and optimization is on a population loss landscape. By contrast, ``real world'' has only finite data, where optimization is on an empirical loss landscape.
In the dark space context of the OOD setting, this decomposition needs to be changed to
\[
TestError(f_{t}^{OOD}) = \underbrace{TestError(f_{t}^{iid})}_\text{observed space} +
\underbrace{
[TestError(f_{t}^{OOD}) - TestError(f_{t}^{iid})]}_\text{dark space generalization gap}
\]
and
\[TestError(f_{t}^{iid}) = TrainError(f_{t}^{iid}) +
\underbrace{[TestError(f_{t})^{iid} - TrainError(f_{t}^{iid})]}_\text{observed space generalization gap}
\]
This explains that the effort could be devoted to decrease the observed space error and/or the dark space generalization gap to reduce $TestError(f_{t}^{OOD})$.

\[\nabla_{\theta}J(\theta) \approx g=  \frac{1}{m}\sum_{i=1}^{m}\nabla_{\theta}L(x^{(i)} , y^{(i)},\theta)
\]

\[\theta \leftarrow \theta - \epsilon g
\]

When stochastic gradient descent (SDG) is applied to the optimization, it approximately estimates $\nabla_{\theta}J(\theta)$, the expectation of gradient, using a small set of sample of size $m$, i.e., the mini-batch drawn uniformly from the training set. When all data are IID, this approximation works fine to update $\theta$ with $g$. However, for the ODD with unknown distribution, this $\theta$ updating function could easily fall into a local minimum based on the $m$ mini-batch samples.

The test error for a trained model in the OOD setting includes two parts: test errors in the observed IID space and a generalized gap when stepping into the OOD space. Furthermore, as discussed and proved in \cite{ood-thesis}, \cite{invariant-risk-minimization}, not all OOD tasks are equal. Depending on how different the OOD data set is from the train set, some OOD task could be more challenge. It is true for predicting ligand binding to dark proteins.  It is impossible for training data to provide sufficient coverage of the whole distribution in the dark chemical genomics space. The motivation of Portal Learning for exploring the dark space follows:  one model architecture defines a functional mapping space, together with  a data set defines a \textbf{universe}. The model initialized instance in a universe closer to the global optimum  universe is a \textbf{portal} that is transferred from an associated universe. CPI dark space is impossible to be explored if the learning is confined only in the observed protein function, i.e. CPI universe since the known data are far sparse as shown in main content Figure 3. Hence STL is important to identify portals. The model optimization on a loss function can decrease IID training errors but will not help with the observed IID space generalization gap $TestError(f_{t})^{iid} - TrainError(f_{t}^{iid})$ or the dark space generalization gap $TestError(f_{t}^{OOD}) - TestError(f_{t}^{iid})$. With Portal Learning, stress model instance selection can narrow the first gap and OOC-ML can narrow the second gap.





\clearpage
\section{Additional tables}

\addcontentsline{toc}{subsection}{Table  \ref{tab:drug-covid-interactors}: Drugs predicted to interact with SARS-COVID-2 interactors}

\begin{table}[ht]
\resizebox{\textwidth}{!}{%
\begin{tabular}{|c|c|c|c|c|l}
\cline{1-5}
IhChIKey                    & Number of hits & Drug name           & Clinical trail  & Mechanism of Action                                &  \\ \cline{1-5}
WNEODWDFDXWOLU-QHCPKHFHSA-N & 7              & fenebrutinib        & phase 2         & Bruton's tyrosine kinase (BTK) inhibitor           &  \\ \cline{1-5}
JFOAJUGFHDCBJJ-UHFFFAOYSA-N & 7              & NMS-P715            & preclinical     & protein kinase inhibitor                           &  \\ \cline{1-5}
QHLVBNKYJGBCQJ-UHFFFAOYSA-N & 4              & NMS-1286937         & phase 2         & PLK inhibitor                                      &  \\ \cline{1-5}
FUXVKZWTXQUGMW-FQEVSTJZSA-N & 4              & 9-aminocamptothecin & phase 2         & topoisomerase inhibitor                            &  \\ \cline{1-5}
DKZYXHCYPUVGAF-JCNLHEQBSA-N & 2              & OTS167              & phase 1/phase 2 & maternal embryonic leucine zipper kinase inhibitor &  \\ \cline{1-5}
VYLOOGHLKSNNEK-PIIMJCKOSA-N & 1              & tropifexor          & phase 2         & FXR agonist                                        &  \\ \cline{1-5}
TZKBVRDEOITLRB-UHFFFAOYSA-N & 1              & GZD824              & preclinical     & Bcr-Abl kinase inhibitor                           &  \\ \cline{1-5}
KZSKGLFYQAYZCO-UHFFFAOYSA-N & 1              & cilofexor           & phase 3         & FXR agonist                                        &  \\ \cline{1-5}
\end{tabular}%
}
\caption{Drugs predicted to interact with SARS-COVID-2 interactors} 
    \label{tab:drug-covid-interactors}
\end{table}

\addcontentsline{toc}{subsection}{Table  \ref{tab:portalcg}: Undruggable human disease-associated proteins selected by ProtalCG}

\newcolumntype{C}[1]{>{\raggedright \arraybackslash\hspace{0pt}}m{#1}} 
\begin{longtable}{@{}|l|C{6.5cm}|l|l|@{}}
\hline
Uniprot & Protein name                                                              & Drug name             & Probscore  \\ \hline
\endfirsthead
\multicolumn{4}{c}%
{{\bfseries Table \thetable\ continued from previous page}} \\
\hline
Uniprot & Protein name                                                              & Drug name             & Probscore  \\ \hline
\endhead
Q8TEX9  & Importin-4                                                                & fenebrutinib          & 0.68089414 \\ \hline
Q8NB66  & Protein unc-13 homolog C                                                  & AI-10-49              & 0.67951804 \\ \hline
Q9NZF1  & Placenta-specific gene 8 protein                                          & AI-10-49              & 0.6778485  \\ \hline
Q96G03  & Phosphoglucomutase-2                                                      & fenebrutinib          & 0.677423   \\ \hline
Q8IWU4  & Zinc transporter 8                                                        & fenebrutinib          & 0.677423   \\ \hline
P53004  & Biliverdin reductase A                                                    & AI-10-49              & 0.6769635  \\ \hline
P40879  & Chloride anion exchanger                                                  & fenebrutinib          & 0.67688245 \\ \hline
Q8IYL2  & Probable tRNA                                                             & fenebrutinib          & 0.6768823  \\ \hline
Q9ULL4  & Plexin-B3                                                                 & AI-10-49              & 0.67652583 \\ \hline
Q96SZ6  & Mitochondrial tRNA methylthiotransferase CDK5RAP1                         & fenebrutinib          & 0.6763639  \\ \hline
Q9BRT9  & DNA replication complex GINS protein SLD5                                 & fenebrutinib          & 0.6763639  \\ \hline
Q8N2U0  & Transmembrane protein 256                                                 & CCT137690             & 0.67634773 \\ \hline
Q9UC06  & Zinc finger protein 70                                                    & fenebrutinib          & 0.6761195  \\ \hline
Q9BPX5  & Actin-related protein 2/3 complex subunit 5-like protein                  & Q-203                 & 0.67571765 \\ \hline
Q8TDF6  & RAS guanyl-releasing protein 4                                            & NMS-1286937           & 0.6756182  \\ \hline
Q9P2G3  & Kelch-like protein 14                                                     & NMS-1286937           & 0.6756182  \\ \hline
Q9HBT7  & Zinc finger protein 287                                                   & CCT137690             & 0.6752016  \\ \hline
Q9UKR8  & Tetraspanin-16                                                            & PF-05190457           & 0.6751168  \\ \hline
P59044  & NACHT, LRR and PYD domains-containing protein 6                           & MK-5046               & 0.67498016 \\ \hline
Q16774  & Guanylate kinase                                                          & MK-5046               & 0.6749801  \\ \hline
Q9HCE5  & N6-adenosine-methyltransferase non-catalytic subunit                      & PF-05190457           & 0.67497396 \\ \hline
Q15185  & Prostaglandin E synthase 3                                                & cilofexor             & 0.6745854  \\ \hline
Q8WUA7  & TBC1 domain family member 22A                                             & fenebrutinib          & 0.67439413 \\ \hline
Q17RS7  & Flap endonuclease GEN homolog 1                                           & CGM097                & 0.67434615 \\ \hline
Q14244  & Ensconsin                                                                 & PF-05190457           & 0.67431164 \\ \hline
Q9BRK5  & 45 kDa calcium-binding protein                                            & PF-05190457           & 0.67431164 \\ \hline
Q9UKJ5  & Cysteine-rich hydrophobic domain-containing protein 2                     & CCT137690             & 0.67429835 \\ \hline
Q6ZWJ1  & Syntaxin-binding protein 4                                                & AI-10-49              & 0.6742442  \\ \hline
Q9HAR2  & Adhesion G protein-coupled receptor L3                                    & AI-10-49              & 0.67414063 \\ \hline
P98161  & Polycystin-1                                                              & AI-10-49              & 0.67414063 \\ \hline
Q92673  & Sortilin-related receptor                                                 & AI-10-49              & 0.6740716  \\ \hline
D6RGH6  & Multicilin                                                                & fenebrutinib          & 0.67397785 \\ \hline
Q8NHH9  & Atlastin-2                                                                & fenebrutinib          & 0.67395777 \\ \hline
Q9H0E2  & Toll-interacting protein                                                  & PF-05190457           & 0.6739072  \\ \hline
O15145  & Actin-related protein 2/3 complex subunit 3                               & PF-05190457           & 0.6739072  \\ \hline
P51116  & Fragile X mental retardation syndrome-related protein 2                   & abemaciclib           & 0.6738332  \\ \hline
Q9BR09  & Neuralized-like protein 2                                                 & elbasvir              & 0.67369795 \\ \hline
P42568  & Protein AF-9                                                              & AI-10-49              & 0.67364717 \\ \hline
P17600  & Synapsin-1                                                                & AI-10-49              & 0.67364717 \\ \hline
P48553  & Trafficking protein particle complex subunit 10                           & AI-10-49              & 0.6736086  \\ \hline
Q12955  & Ankyrin-3                                                                 & abemaciclib           & 0.6735056  \\ \hline
O60936  & Nucleolar protein 3                                                       & abemaciclib           & 0.6735056  \\ \hline
Q02575  & Helix-loop-helix protein 1                                                & AI-10-49              & 0.6734969  \\ \hline
P49640  & Homeobox even-skipped homolog protein 1                                   & CFI-402257            & 0.67348343 \\ \hline
P22670  & MHC class II regulatory factor RFX1                                       & PF-05190457           & 0.67346805 \\ \hline
Q8IUF8  & Ribosomal oxygenase 2                                                     & NMS-1286937           & 0.67343307 \\ \hline
P22681  & E3 ubiquitin-protein ligase CBL                                           & NMS-1286937           & 0.67343307 \\ \hline
Q7RTS3  & Pancreas transcription factor 1 subunit alpha                             & MK-5046               & 0.67334837 \\ \hline
Q9Y5L4  & Mitochondrial import inner membrane translocase subunit Tim13             & NMS-P715              & 0.6733338  \\ \hline
P17735  & Tyrosine aminotransferase                                                 & PF-05190457           & 0.673238   \\ \hline
O95294  & RasGAP-activating-like protein 1                                          & PF-05190457           & 0.673224   \\ \hline
Q8NCD3  & Holliday junction recognition protein                                     & PF-05190457           & 0.673224   \\ \hline
Q86W28  & NACHT, LRR and PYD domains-containing protein 8                           & MK-5046               & 0.6731598  \\ \hline
Q04671  & P protein                                                                 & AI-10-49              & 0.67310166 \\ \hline
Q9C035  & Tripartite motif-containing protein 5                                     & AI-10-49              & 0.6731016  \\ \hline
Q66K64  & DDB1- and CUL4-associated factor 15                                       & AI-10-49              & 0.672944   \\ \hline
Q5T9A4  & ATPase family AAA domain-containing protein 3B                            & CFI-402257            & 0.67286545 \\ \hline
Q96MP8  & BTB/POZ domain-containing protein KCTD7                                   & CFI-402257            & 0.6728654  \\ \hline
Q9UIE0  & Zinc finger protein 230                                                   & CFI-402257            & 0.672853   \\ \hline
Q3SYG4  & Protein PTHB1                                                             & abemaciclib           & 0.67280704 \\ \hline
Q8IUY3  & GRAM domain-containing protein 2A                                         & AI-10-49              & 0.67273754 \\ \hline
O75689  & Arf-GAP with dual PH domain-containing protein 1                          & AI-10-49              & 0.67273754 \\ \hline
Q9NZP8  & Complement C1r subcomponent-like protein                                  & fenebrutinib          & 0.6727066  \\ \hline
Q9NQX4  & Unconventional myosin-Vc                                                  & fenebrutinib          & 0.6727064  \\ \hline
Q15388  & Mitochondrial import receptor subunit TOM20 homolog                       & CFI-402257            & 0.6726911  \\ \hline
Q9NXL9  & DNA helicase MCM9                                                         & piperaquine-phosphate & 0.6726752  \\ \hline
P53370  & Nucleoside diphosphate-linked moiety X motif 6                            & piperaquine-phosphate & 0.6726752  \\ \hline
Q8N7C0  & Leucine-rich repeat-containing protein 52                                 & PF-05190457           & 0.67262363 \\ \hline
O95231  & Homeobox protein VENTX                                                    & NMS-P715              & 0.67261267 \\ \hline
Q9HC56  & Protocadherin-9                                                           & AI-10-49              & 0.6725752  \\ \hline
A6NNW6  & Enolase 4                                                                 & AI-10-49              & 0.6725752  \\ \hline
P35612  & Beta-adducin                                                              & CFI-402257            & 0.67255336 \\ \hline
Q9Y6M5  & Zinc transporter 1                                                        & AI-10-49              & 0.67255074 \\ \hline
Q01804  & OTU domain-containing protein 4                                           & AI-10-49              & 0.6725507  \\ \hline
Q9Y678  & Coatomer subunit gamma-1                                                  & MK-5046               & 0.67254347 \\ \hline
Q9BTY7  & Protein HGH1 homolog                                                      & AI-10-49              & 0.67243195 \\ \hline
Q8NEM0  & Microcephalin                                                             & CFI-402257            & 0.6724283  \\ \hline
Q7LGA3  & Heparan sulfate 2-O-sulfotransferase 1                                    & AI-10-49              & 0.67242616 \\ \hline
O75956  & Cyclin-dependent kinase 2-associated protein 2                            & ziritaxestat          & 0.6723895  \\ \hline
P62987  & Ubiquitin-60S ribosomal protein L40                                       & ziritaxestat          & 0.6723893  \\ \hline
Q14257  & Reticulocalbin-2                                                          & CDK9-IN-6             & 0.6723883  \\ \hline
Q9H0I3  & Coiled-coil domain-containing protein 113                                 & tropifexor            & 0.6723477  \\ \hline
P32019  & Type II inositol 1,4,5-trisphosphate 5-phosphatase                        & fenebrutinib          & 0.6723469  \\ \hline
O95409  & Zinc finger protein ZIC 2                                                 & fenebrutinib          & 0.6723469  \\ \hline
P56179  & Homeobox protein DLX-6                                                    & fenebrutinib          & 0.67233485 \\ \hline
Q12798  & Centrin-1                                                                 & AI-10-49              & 0.6722851  \\ \hline
Q14714  & Sarcospan                                                                 & AI-10-49              & 0.6722851  \\ \hline
Q96LI5  & CCR4-NOT transcription complex subunit 6-like                             & abemaciclib           & 0.6722749  \\ \hline
Q86XR7  & TIR domain-containing adapter molecule 2                                  & AI-10-49              & 0.6722719  \\ \hline
P53672  & Beta-crystallin A2                                                        & CCT137690             & 0.6722543  \\ \hline
O43439  & Protein CBFA2T2                                                           & CCT137690             & 0.6722543  \\ \hline
Q9H0E3  & Histone deacetylase complex subunit SAP130                                & fenebrutinib          & 0.6722474  \\ \hline
Q8NFZ0  & F-box DNA helicase 1                                                      & fenebrutinib          & 0.6722474  \\ \hline
Q8N0S2  & Synaptonemal complex central element protein 1                            & CFI-402257            & 0.67223567 \\ \hline
Q92667  & A-kinase anchor protein 1, mitochondrial                                  & fenebrutinib          & 0.67222863 \\ \hline
O60884  & DnaJ homolog subfamily A member 2                                         & PF-05190457           & 0.6721347  \\ \hline
P41214  & Eukaryotic translation initiation factor 2D                               & AI-10-49              & 0.6721203  \\ \hline
Q6R327  & Rapamycin-insensitive companion of mTOR                                   & AI-10-49              & 0.67211396 \\ \hline
O75886  & Signal transducing adapter molecule 2                                     & PF-05190457           & 0.67206836 \\ \hline
Q9UMX6  & Guanylyl cyclase-activating protein 2                                     & PF-05190457           & 0.6720683  \\ \hline
Q99447  & Ethanolamine-phosphate cytidylyltransferase                               & CCT137690             & 0.67205685 \\ \hline
Q6FI13  & Histone H2A type 2-A                                                      & PF-05190457           & 0.6720233  \\ \hline
Q969S2  & Endonuclease 8-like 2                                                     & fenebrutinib          & 0.67200994 \\ \hline
Q6XZF7  & Dynamin-binding protein                                                   & AI-10-49              & 0.67196065 \\ \hline
Q96EX1  & Small integral membrane protein 12                                        & fenebrutinib          & 0.6719141  \\ \hline
Q8TD57  & Dynein axonemal heavy chain 3                                             & fenebrutinib          & 0.6719141  \\ \hline
O15375  & Monocarboxylate transporter 6                                             & PF-05190457           & 0.6719058  \\ \hline
Q9H7E2  & Tudor domain-containing protein 3                                         & NMS-1286937           & 0.6718896  \\ \hline
A6NI73  & Leukocyte immunoglobulin-like receptor subfamily A member 5               & AI-10-49              & 0.671879   \\ \hline
Q9NZM3  & Intersectin-2                                                             & CDK9-IN-6             & 0.6718554  \\ \hline
O75145  & Liprin-alpha-3                                                            & CDK9-IN-6             & 0.6718554  \\ \hline
Q9NSD9  & Phenylalanine--tRNA ligase beta subunit                                   & abemaciclib           & 0.6718319  \\ \hline
O15195  & Villin-like protein                                                       & abemaciclib           & 0.6718319  \\ \hline
P05538  & HLA class II histocompatibility antigen, DQ beta 2 chain                  & cilofexor             & 0.671822   \\ \hline
Q8TD57  & Dynein axonemal heavy chain 3                                             & cilofexor             & 0.671822   \\ \hline
Q9NRD9  & Dual oxidase 1                                                            & abemaciclib           & 0.6718094  \\ \hline
Q8WUQ7  & Cactin                                                                    & AI-10-49              & 0.67178315 \\ \hline
Q9H7C4  & Syncoilin                                                                 & AI-10-49              & 0.67178315 \\ \hline
Q70EK9  & Ubiquitin carboxyl-terminal hydrolase 51                                  & PF-05190457           & 0.671742   \\ \hline
O43837  & Isocitrate dehydrogenase {[}NAD{]} subunit beta, mitochondrial            & PF-05190457           & 0.67174196 \\ \hline
Q9UKY1  & Zinc fingers and homeoboxes protein 1                                     & CFI-402257            & 0.6717407  \\ \hline
Q6ZMJ2  & Scavenger receptor class A member 5                                       & CFI-402257            & 0.6717407  \\ \hline
P61371  & Insulin gene enhancer protein ISL-1                                       & fenebrutinib          & 0.6717256  \\ \hline
Q6NSZ9  & Zinc finger and SCAN domain-containing protein 25                         & fenebrutinib          & 0.6717256  \\ \hline
Q92754  & Transcription factor AP-2 gamma                                           & IACS-10759            & 0.67171496 \\ \hline
Q5RKV6  & Exosome complex component MTR3                                            & AI-10-49              & 0.6716919  \\ \hline
Q9BRA0  & N-alpha-acetyltransferase 38, NatC auxiliary subunit                      & MK-5046               & 0.6716785  \\ \hline
Q9BSF8  & BTB/POZ domain-containing protein 10                                      & CCT137690             & 0.67165    \\ \hline
Q6PJ69  & Tripartite motif-containing protein 65                                    & AI-10-49              & 0.6716369  \\ \hline
Q9NRP0  & Oligosaccharyltransferase complex subunit OSTC                            & AI-10-49              & 0.6716369  \\ \hline
Q8TBM7  & Transmembrane protein 254                                                 & PF-05190457           & 0.6716311  \\ \hline
Q674X7  & Kazrin                                                                    & elbasvir              & 0.6716015  \\ \hline
Q9Y3C4  & EKC/KEOPS complex subunit TPRKB                                           & abemaciclib           & 0.6715913  \\ \hline
P59190  & Ras-related protein Rab-15                                                & abemaciclib           & 0.6715913  \\ \hline
Q9NZM3  & Intersectin-2                                                             & PF-05190457           & 0.6715465  \\ \hline
P05455  & Lupus La protein                                                          & AI-10-49              & 0.6715072  \\ \hline
Q96KN3  & Homeobox protein PKNOX2                                                   & abemaciclib           & 0.67148525 \\ \hline
P54278  & Mismatch repair endonuclease PMS2                                         & abemaciclib           & 0.6714852  \\ \hline
P49069  & Guided entry of tail-anchored proteins factor CAMLG                       & fenebrutinib          & 0.6714454  \\ \hline
Q8IYT4  & Katanin p60 ATPase-containing subunit A-like 2                            & AI-10-49              & 0.6714413  \\ \hline
Q8WVI0  & Small integral membrane protein 4                                         & AI-10-49              & 0.67144126 \\ \hline
Q9Y6X6  & Unconventional myosin-XVI                                                 & MK-5046               & 0.6713607  \\ \hline
Q8N357  & Solute carrier family 35 member F6                                        & MK-5046               & 0.6713607  \\ \hline
Q96BP2  & Coiled-coil-helix-coiled-coil-helix domain-containing protein 1           & abemaciclib           & 0.671357   \\ \hline
Q9UJQ4  & Sal-like protein 4                                                        & MK-5046               & 0.67135173 \\ \hline
Q9HCQ5  & Polypeptide N-acetylgalactosaminyltransferase 9                           & fenebrutinib          & 0.67132807 \\ \hline
Q9P2Q2  & FERM domain-containing protein 4A                                         & fenebrutinib          & 0.67132807 \\ \hline
Q7Z7M8  & UDP-GlcNAc:betaGal beta-1,3-N-acetylglucosaminyltransferase 8             & NMS-P715              & 0.6713157  \\ \hline
Q9HCK5  & Protein argonaute-4                                                       & PF-05190457           & 0.6713007  \\ \hline
O95405  & Zinc finger FYVE domain-containing protein 9                              & AI-10-49              & 0.67127687 \\ \hline
Q92911  & Sodium/iodide cotransporter                                               & PF-05190457           & 0.67127484 \\ \hline
Q8WWI1  & LIM domain only protein 7                                                 & fenebrutinib          & 0.6712641  \\ \hline
Q8N3C7  & CAP-Gly domain-containing linker protein 4                                & fenebrutinib          & 0.671264   \\ \hline
Q9UBS8  & E3 ubiquitin-protein ligase RNF14                                         & PF-05190457           & 0.671259   \\ \hline
Q9BY66  & Lysine-specific demethylase 5D                                            & PF-05190457           & 0.671259   \\ \hline
Q9UPM8  & AP-4 complex subunit epsilon-1                                            & abemaciclib           & 0.6712124  \\ \hline
Q9BYE0  & Transcription factor HES-7                                                & MK-5046               & 0.6711568  \\ \hline
Q9BSD7  & Cancer-related nucleoside-triphosphatase                                  & AI-10-49              & 0.67114973 \\ \hline
Q9NXH9  & tRNA                                                                      & tropifexor            & 0.6711371  \\ \hline
Q13948  & Protein CASP                                                              & tropifexor            & 0.6711371  \\ \hline
Q9C019  & Tripartite motif-containing protein 15                                    & abemaciclib           & 0.67111254 \\ \hline
Q5TA50  & Ceramide-1-phosphate transfer protein                                     & ABT-702               & 0.6711099  \\ \hline
Q7KZF4  & Staphylococcal nuclease domain-containing protein 1                       & ABT-702               & 0.6711099  \\ \hline
P08247  & Synaptophysin                                                             & fenebrutinib          & 0.6711092  \\ \hline
A4IF30  & Solute carrier family 35 member F4                                        & AI-10-49              & 0.67109805 \\ \hline
Q9NRA8  & Eukaryotic translation initiation factor 4E transporter                   & AI-10-49              & 0.67109805 \\ \hline
Q96FF9  & Sororin                                                                   & acalabrutinib         & 0.6710895  \\ \hline
P10073  & Zinc finger and SCAN domain-containing protein 22                         & PF-05190457           & 0.67107534 \\ \hline
P0CW19  & LIM and senescent cell antigen-like-containing domain protein 3           & PF-05190457           & 0.67107534 \\ \hline
Q9UPM8  & AP-4 complex subunit epsilon-1                                            & fenebrutinib          & 0.6710706  \\ \hline
P08118  & Beta-microseminoprotein                                                   & PF-05190457           & 0.6710468  \\ \hline
Q96P16  & Regulation of nuclear pre-mRNA domain-containing protein 1A               & fenebrutinib          & 0.6710315  \\ \hline
P62280  & 40S ribosomal protein S11                                                 & piperaquine-phosphate & 0.6710226  \\ \hline
Q8TE49  & OTU domain-containing protein 7A                                          & AI-10-49              & 0.6710194  \\ \hline
Q53HI1  & Protein unc-50 homolog                                                    & AI-10-49              & 0.6710194  \\ \hline
Q8WW32  & High mobility group protein B4                                            & PF-05190457           & 0.6709929  \\ \hline
P57729  & Ras-related protein Rab-38                                                & PF-05190457           & 0.67099285 \\ \hline
Q8N2M8  & CLK4-associating serine/arginine rich protein                             & fenebrutinib          & 0.6709883  \\ \hline
P25705  & ATP synthase subunit alpha, mitochondrial                                 & fenebrutinib          & 0.6709781  \\ \hline
Q9Y5H3  & Protocadherin gamma-A10                                                   & PF-05190457           & 0.6709645  \\ \hline
Q9NU63  & Zinc finger protein 57 homolog                                            & fenebrutinib          & 0.67096215 \\ \hline
P43363  & Melanoma-associated antigen 10                                            & AI-10-49              & 0.6709133  \\ \hline
Q13469  & Nuclear factor of activated T-cells, cytoplasmic 2                        & AI-10-49              & 0.670904   \\ \hline
Q8NG77  & Olfactory receptor 2T12                                                   & AI-10-49              & 0.670904   \\ \hline
Q9Y2G8  & DnaJ homolog subfamily C member 16                                        & PF-05190457           & 0.67090386 \\ \hline
Q9ULV8  & E3 ubiquitin-protein ligase CBL-C                                         & fenebrutinib          & 0.67089945 \\ \hline
Q96EF0  & Myotubularin-related protein 8                                            & elbasvir              & 0.67085487 \\ \hline
Q8TD16  & Protein bicaudal D homolog 2                                              & abemaciclib           & 0.6708334  \\ \hline
P49459  & Ubiquitin-conjugating enzyme E2 A                                         & abemaciclib           & 0.6708334  \\ \hline
Q6XZF7  & Dynamin-binding protein                                                   & AI-10-49              & 0.6708271  \\ \hline
Q9Y4I1  & Unconventional myosin-Va                                                  & AI-10-49              & 0.6708271  \\ \hline
Q96FL9  & Polypeptide N-acetylgalactosaminyltransferase 14                          & MK-5046               & 0.67080116 \\ \hline
Q9HD67  & Unconventional myosin-X                                                   & MK-5046               & 0.6708011  \\ \hline
Q13214  & Semaphorin-3B                                                             & MK-5046               & 0.6707926  \\ \hline
Q53H47  & Histone-lysine N-methyltransferase SETMAR                                 & MK-5046               & 0.6707926  \\ \hline
P46776  & 60S ribosomal protein L27a                                                & AI-10-49              & 0.67078424 \\ \hline
Q15306  & Interferon regulatory factor 4                                            & AI-10-49              & 0.6707842  \\ \hline
Q2TAA2  & Isoamyl acetate-hydrolyzing esterase 1 homolog                            & abemaciclib           & 0.67076474 \\ \hline
O43324  & Eukaryotic translation elongation factor 1 epsilon-1                      & PF-05190457           & 0.67074585 \\ \hline
Q96R27  & Olfactory receptor 2M4                                                    & fenebrutinib          & 0.6707435  \\ \hline
Q5VXU1  & Sodium/potassium-transporting ATPase subunit beta-1-interacting protein 2 & fenebrutinib          & 0.6707435  \\ \hline
Q5VWQ8  & Disabled homolog 2-interacting protein                                    & fenebrutinib          & 0.67072076 \\ \hline
Q9UJF2  & Ras GTPase-activating protein nGAP                                        & elbasvir              & 0.67071134 \\ \hline
O75923  & Dysferlin                                                                 & fenebrutinib          & 0.6707069  \\ \hline
O75592  & E3 ubiquitin-protein ligase MYCBP2                                        & PF-05190457           & 0.6706958  \\ \hline
Q14714  & Sarcospan                                                                 & CFI-402257            & 0.67069536 \\ \hline
Q12798  & Centrin-1                                                                 & CFI-402257            & 0.67069536 \\ \hline
Q9H9Z2  & Protein lin-28 homolog A                                                  & fenebrutinib          & 0.67065567 \\ \hline
Q8N2A8  & Mitochondrial cardiolipin hydrolase                                       & fenebrutinib          & 0.67065555 \\ \hline
Q5VV41  & Rho guanine nucleotide exchange factor 16                                 & abemaciclib           & 0.6706521  \\ \hline
Q96QT6  & PHD finger protein 12                                                     & fenebrutinib          & 0.67064524 \\ \hline
Q13948  & Protein CASP                                                              & AI-10-49              & 0.6706354  \\ \hline
Q9NXH9  & tRNA                                                                      & AI-10-49              & 0.6706354  \\ \hline
Q969E2  & Secretory carrier-associated membrane protein 4                           & fenebrutinib          & 0.67062545 \\ \hline
P58107  & Epiplakin                                                                 & fenebrutinib          & 0.67062545 \\ \hline
P55036  & 26S proteasome non-ATPase regulatory subunit 4                            & NMS-P715              & 0.67059726 \\ \hline
Q11128  & 4-galactosyl-N-acetylglucosaminide 3-alpha-L-fucosyltransferase FUT5      & AI-10-49              & 0.6705861  \\ \hline
Q86W28  & NACHT, LRR and PYD domains-containing protein 8                           & AI-10-49              & 0.6705852  \\ \hline
O43566  & Regulator of G-protein signaling 14                                       & Q-203                 & 0.67055416 \\ \hline
Q9BVI0  & PHD finger protein 20                                                     & Q-203                 & 0.67055416 \\ \hline
Q93074  & Mediator of RNA polymerase II transcription subunit 12                    & PF-05190457           & 0.67054135 \\ \hline
P51530  & DNA replication ATP-dependent helicase/nuclease DNA2                      & PF-05190457           & 0.6705413  \\ \hline
Q6NS38  & DNA oxidative demethylase ALKBH2                                          & AI-10-49              & 0.67052937 \\ \hline
B4DJY2  & Transmembrane protein 233                                                 & AI-10-49              & 0.67052513 \\ \hline
Q8N807  & Protein disulfide-isomerase-like protein of the testis                    & PF-05190457           & 0.67051905 \\ \hline
P13929  & Beta-enolase                                                              & AI-10-49              & 0.67051786 \\ \hline
Q9NRJ7  & Protocadherin beta-16                                                     & AI-10-49              & 0.67051774 \\ \hline
O95751  & Protein LDOC1                                                             & MK-5046               & 0.67051023 \\ \hline
Q92947  & Glutaryl-CoA dehydrogenase, mitochondrial                                 & fenebrutinib          & 0.67050344 \\ \hline
Q96J94  & Piwi-like protein 1                                                       & fenebrutinib          & 0.6704877  \\ \hline
O15400  & Syntaxin-7                                                                & adapalene             & 0.67046946 \\ \hline
O00192  & Armadillo repeat protein deleted in velo-cardio-facial syndrome           & adapalene             & 0.67046934 \\ \hline
Q17RD7  & Synaptotagmin-16                                                          & CFI-402257            & 0.6704517  \\ \hline
A0PJZ3  & Glucoside xylosyltransferase 2                                            & MK-5046               & 0.67042667 \\ \hline
Q53GG5  & PDZ and LIM domain protein 3                                              & PF-05190457           & 0.6704057  \\ \hline
P20929  & Nebulin                                                                   & PF-05190457           & 0.6704056  \\ \hline
Q5NDL2  & EGF domain-specific O-linked N-acetylglucosamine transferase              & abemaciclib           & 0.67039376 \\ \hline
Q10570  & Cleavage and polyadenylation specificity factor subunit 1                 & fenebrutinib          & 0.6703786  \\ \hline
Q53FE4  & Uncharacterized protein C4orf17                                           & AI-10-49              & 0.67036396 \\ \hline
Q96T21  & Selenocysteine insertion sequence-binding protein 2                       & cilofexor             & 0.67036384 \\ \hline
P17980  & 26S proteasome regulatory subunit 6A                                      & PF-05190457           & 0.67036086 \\ \hline
Q8TE77  & Protein phosphatase Slingshot homolog 3                                   & tropifexor            & 0.67035466 \\ \hline
Q96LD4  & E3 ubiquitin-protein ligase TRIM47                                        & tropifexor            & 0.67035466 \\ \hline
P25205  & DNA replication licensing factor MCM3                                     & fenebrutinib          & 0.6703515  \\ \hline
Q9NQC7  & Ubiquitin carboxyl-terminal hydrolase CYLD                                & fenebrutinib          & 0.6703515  \\ \hline
Q71F23  & Centromere protein U                                                      & NMS-P715              & 0.6703202  \\ \hline
P46778  & 60S ribosomal protein L21                                                 & fenebrutinib          & 0.67028475 \\ \hline
Q04727  & Transducin-like enhancer protein 4                                        & AI-10-49              & 0.6702773  \\ \hline
O14556  & Glyceraldehyde-3-phosphate dehydrogenase, testis-specific                 & AI-10-49              & 0.6702773  \\ \hline
Q92994  & Transcription factor IIIB 90 kDa subunit                                  & fenebrutinib          & 0.67027074 \\ \hline
Q15646  & 2'-5'-oligoadenylate synthase-like protein                                & fenebrutinib          & 0.67027074 \\ \hline
O15131  & Importin subunit alpha-6                                                  & fenebrutinib          & 0.6702412  \\ \hline
Q14254  & Flotillin-2                                                               & fenebrutinib          & 0.6702412  \\ \hline
O75764  & Transcription elongation factor A protein 3                               & CFI-402257            & 0.67017365 \\ \hline
P40425  & Pre-B-cell leukemia transcription factor 2                                & MK-5046               & 0.67016166 \\ \hline
O43766  & Lipoyl synthase, mitochondrial                                            & PF-05190457           & 0.67014426 \\ \hline
Q969L2  & Protein MAL2                                                              & fenebrutinib          & 0.67013794 \\ \hline
Q9NSG2  & Uncharacterized protein C1orf112                                          & CCT137690             & 0.67013377 \\ \hline
Q9H892  & Tetratricopeptide repeat protein 12                                       & PF-05190457           & 0.67012644 \\ \hline
O95248  & Myotubularin-related protein 5                                            & CCT137690             & 0.6701216  \\ \hline
O60481  & Zinc finger protein ZIC 3                                                 & PF-05190457           & 0.670091   \\ \hline
Q9UBK8  & Methionine synthase reductase                                             & AI-10-49              & 0.6700795  \\ \hline
Q4G0J3  & La-related protein 7                                                      & CCT137690             & 0.67007875 \\ \hline
Q8WW32  & High mobility group protein B4                                            & CCT137690             & 0.67005986 \\ \hline
P23297  & Protein S100-A1                                                           & MK-5046               & 0.67002577 \\ \hline
Q8TCB7  & tRNA N                                                                    & elbasvir              & 0.6700179  \\ \hline
Q8IXH6  & Tumor protein p53-inducible nuclear protein 2                             & abemaciclib           & 0.67000955 \\ \hline
\caption{Undruggable human disease-associated proteins selected by ProtalCG} 
\label{tab:portalcg}

\end{longtable}

\addcontentsline{toc}{subsection}{Table  \ref{tab:chemicals}: Chemicals interacted with undruggable human proteins}
\begin{table}[ht]
\centering
\resizebox{\textwidth}{!}{%

\begin{tabular}{|c|c|c|c|}
\hline
Drug name                & Clinical phase  & Mechanism of Action                                  & \begin{tabular}[c]{@{}c@{}}Number of \\ targeted proteins\end{tabular} \\ \hline
AI-10-49                 & Preclinical     & core binding factor inhibitor                        & 63                                                                     \\ \hline
Fenebrutinib             & Phase 2         & Bruton's tyrosine kinase (BTK) inhibitor             & 56                                                                     \\ \hline
PF-05190457              & Phase 2         & growth hormone secretagogue receptor inverse agonist & 42                                                                     \\ \hline
Abemaciclib              & Launched        & CDK inhibitor                                        & 21                                                                     \\ \hline
MK-5046                  & Preclinical     & bombesin receptor agonist                            & 18                                                                     \\ \hline
CFI-402257               & Phase 1/Phase 2 & dual specificity protein kinase inhibitor            & 14                                                                     \\ \hline
CCT137690                & Preclinical     & Aurora kinase inhibitor                              & 11                                                                     \\ \hline
Tropifexor               & Phase 2         & FXR agonist                                          & 5                                                                      \\ \hline
NMS-1286937              & Phase 2         & PLK inhibitor                                        & 5                                                                      \\ \hline
NMS-P715                 & Preclinical     & protein kinase inhibitor                             & 5                                                                      \\ \hline
Elbasvir                 & Launched        & HCV inhibitor                                        & 5                                                                      \\ \hline
Cilofexor                & Phase 3         & FXR agonist                                          & 4                                                                      \\ \hline
CDK9-IN-6                & Preclinical     & CDK inhibitor       
                                 & 3                                                                      \\ \hline
piperaquine-phosphate    & Launched        & antimalarial agent                                   & 3                                                                      \\ \hline
Q-203                    & Phase 2         & ATP synthase inhibitor                               & 3                                                                      \\ \hline
ABT 702 dihydrochloride  & Preclinical     & adenosine kinase inhibitor                           & 2                                                                      \\ \hline
Ziritaxestat             & Phase 3         & autotaxin inhibitor                                  & 2                                                                      \\ \hline
Adapalene                & Launched        & retinoid receptor agonist                            & 1                                                                      \\ \hline
Acalabrutinib            & Launched        & Bruton's tyrosine kinase (BTK) inhibitor             & 1                                                                      \\ \hline
IACS-10759 Hydrochloride & Preclinical     & mitochondrial complex I inhibitor                    & 1                                                                      \\ \hline
NVP-CGM097               & Phase 1         & MDM inhibitor                                        & 1                                                                      \\ \hline
\end{tabular}%
}
\caption{Chemicals interacted with undruggable human proteins} 
    \label{tab:chemicals}
\end{table}


When we consider the proteins in Tbio, there are 9545 proteins which are not int Casas's druggable proteins. If 0.67 was used as the cutoff, 219 proteins were predicted as positive hits. The gene enrichment analysis result for these proteins was listed in Table S\ref{tab:tbioonly-enrichment}. Disease associated with these 219 human proteins were also listed in Table S\ref{tab:tbiodisease}. Since one protein is always related with multiple diseases, these diseases are ranked by the number of their associated proteins and the top 10 diseases were listed in the table. Most of top ranked diseases are related with cancer development. 21 drugs that are approved or in clinical trial are predicted to interact with these proteins as shown in Table S\ref{tab:Tbiochemicals}.

If the proteins in Tbio were removed from the undruggable list, only 2930 proteins were left. If 0.67 was used as the cutoff, there will be only 41 proteins predicted positive and no significant enrichment with David gene enrichment analysis. So 0.665 was used as a cutoff, and 348 proteins were predicted as positive hits. The gene enrichment analysis result for these proteins was listed in Table S\ref{tab:without-tbio-enrichment}. Disease associated with these 348 human proteins were also listed in Table S\ref{tab:undrugnoTbiodisease}. 42 drugs that are approved or in clinical trial are predicted to interact with these proteins as shown in Table S\ref{tab:UndrugnoTbiochemicals}. 

\addcontentsline{toc}{subsection}{Table  \ref{tab:tbioonly-enrichment}: Functional Annotation enrichment for human proteins in Tbio selected by PortalCG}
\begin{table}[ht]
\resizebox{\textwidth}{!}{%
\begin{tabular}{|c|c|c|c|c|}
\hline
\multicolumn{5}{|c|}{David Functional Annotation enrichment analysis}  

\\ \hline
\begin{tabular}[c]{@{}c@{}}Enriched terms in \\ UniProtKB keywords\end{tabular} & \begin{tabular}[c]{@{}c@{}}Number of \\ proteins involved\end{tabular} & \begin{tabular}[c]{@{}c@{}}Percentage of \\ proteins involved\end{tabular} & P-value  & \begin{tabular}[c]{@{}c@{}}Modified \\ Benjamini p-value\end{tabular} \\ \hline
{\color[HTML]{080808} Alternative splicing}                                     & 153                                                                    & 69.9                                                                       & 2.70E-08 & 6.80E-06                                                              \\ \hline
{\color[HTML]{080808} Phosphoprotein}                                           & 126                                                                    & 57.5                                                                       & 1.70E-07 & 2.20E-05                                                              \\ \hline
{\color[HTML]{080808} Cytoplasm}                                                & 82                                                                     & 37.4                                                                       & 3.20E-06 & 2.40E-04                                                              \\ \hline
{\color[HTML]{080808} Nucleus}                                                  & 87                                                                     & 39.7                                                                       & 3.70E-06 & 2.40E-04                                                              \\ \hline
{\color[HTML]{080808} Metal-binding}                                            & 61                                                                     & 27.9                                                                       & 2.00E-04 & 1.00E-02                            
                                  \\ \hline
\end{tabular}%
}
\caption{Functional Annotation enrichment for human proteins in Tbio selected by PortalCG} 
    \label{tab:tbioonly-enrichment}
\end{table}


\clearpage


\addcontentsline{toc}{subsection}{Table  \ref{tab:tbiodisease}: Top ranked diseases associated with proteins in Tbio selected by portal learning}
\begin{table}[ht]
\resizebox{\textwidth}{!}{%
\begin{tabular}{|c|c|}
\hline
Disease Name                  & \# of undruggable proteins associated with the disease \\ \hline
Breast Carcinoma              & 89                                                     \\ \hline
Tumor Cell Invasion           & 85                                                     \\ \hline
Carcinogenesis                & 83                                                     \\ \hline
Neoplasm Metastasis           & 73                                                     \\ \hline
Colorectal Carcinoma          & 68                                                     \\ \hline
Liver carcinoma               & 66                                                     \\ \hline
Non-Small Cell Lung Carcinoma & 56                                                     \\ \hline
Malignant neoplasm of lung    & 56                                                     \\ \hline
Carcinoma of lung             & 54                                                     \\ \hline
Alzheimer's Disease           & 54                                                     \\ \hline
\end{tabular}%
}
\caption{Top ranked diseases associated with proteins in Tbio selected by portal learning} 
    \label{tab:tbiodisease}
\end{table}
\addcontentsline{toc}{subsection}{Table  \ref{tab:Tbiochemicals}: Chemicals interacted with human proteins in Tbio}
\begin{table}[ht]
\resizebox{\textwidth}{!}{%
\begin{tabular}{|c|c|c|c|}
\hline

rug name                & Clinical phase  & Mechanism of Action                                  & \begin{tabular}[c]{@{}c@{}}Number of \\ targeted proteins\end{tabular} \\ \hline

AI-10-49              & Preclinical     & core binding factor inhibitor                        & 52                          \\ \hline
fenebrutinib          & Phase 2         & Bruton's tyrosine kinase (BTK) inhibitor             & 45                          \\ \hline
PF-05190457           & Phase 2         & growth hormone secretagogue receptor inverse agonist & 36                          \\ \hline
abemaciclib           & Launched        & CDK inhibitor                                        & 20                          \\ \hline
MK-5046               & Preclinical     & bombesin receptor agonist                            & 15                          \\ \hline
CFI-402257            & Phase 1/Phase 2 & dual specificity protein kinase inhibitor            & 14                          \\ \hline
CCT137690             & Preclinical     & Aurora kinase inhibitor                              & 7                           \\ \hline
NMS-1286937           & Phase 2         & PLK inhibitor                                        & 5                           \\ \hline
tropifexor            & Phase 2         & FXR agonist                                          & 4                           \\ \hline
NMS-P715              & Preclinical     & protein kinase inhibitor                             & 4                           \\ \hline
cilofexor             & Phase 3         & FXR agonist                                          & 3                           \\ \hline
CDK9-IN-6             & Preclinical     & CDK inhibitor                                        & 3                           \\ \hline
piperaquine-phosphate & Launched        & antimalarial agent                                   & 3                           \\ \hline
elbasvir              & Launched        & HCV inhibitor                                        & 3                           \\ \hline
ABT-702               & Preclinical     & adenosine kinase inhibitor                           & 2                           \\ \hline

adapalene             & Launched        & retinoid receptor agonist                            & 2                           \\ \hline
Q-203                 & Phase 2         & ATP synthase inhibitor                               & 2                           \\ \hline
ziritaxestat          & Phase 3         & autotaxin inhibitor                                  & 2                           \\ \hline
acalabrutinib         & Launched        & Bruton's tyrosine kinase
 (BTK) inhibitor             & 1                           \\ \hline
IACS-10759            & Preclinical     & mitochondrial complex I inhibitor                    & 1                           \\ \hline
CGM097                & Phase 1         & MDM inhibitor                                        & 1                           \\ \hline
\end{tabular}%
}
\caption{Chemicals interacted with human proteins in Tbio} 
    \label{tab:Tbiochemicals}
\end{table}

\addcontentsline{toc}{subsection}{Table  \ref{tab:without-tbio-enrichment}: Functional Annotation enrichment for undruggable human disease proteins without Tbio selected by PortalCG}
\begin{table}[ht]
\resizebox{\textwidth}{!}{%
\begin{tabular}{|c|c|c|c|c|}

\hline
\multicolumn{5}{|c|}{David Functional Annotation enrichment analysis}                                                                                                                                                                                                                                                 \\ \hline
\begin{tabular}[c]{@{}c@{}}Enriched terms \\ in UniProtKB keyword\end{tabular} & \begin{tabular}[c]{@{}c@{}}Number of \\ proteins involved\end{tabular} & \begin{tabular}[c]{@{}c@{}}Percentage of \\ proteins involved (\%)\end{tabular} & P-value  & \begin{tabular}[c]{@{}c@{}}Modified\\ Benjamini p-value\end{tabular} \\ \hline
{\color[HTML]{080808} Zinc-finger}                                             & 80                                                                     & 23                                                                              & 7.60E-16 & 1.20E-13                                                             \\ \hline
{\color[HTML]{080808} Zinc}                                                    & 85                                                                     & 24.4                                                                            & 1.20E-11 & 9.90E-10                                                             \\ \hline
{\color[HTML]{080808} Metal-binding}                                           & 96                                                                     & 27.6                                                                            & 4.30E-06 & 2.30E-04                                                             \\ \hline
{\color[HTML]{080808} DNA-binding}                                             & 62                                                                     & 17.8                                                                            & 7.90E-06 & 3.20E-04                                                             \\ \hline
{\color[HTML]{080808} Transcription regulation}                                & 61                                                                     & 17.5                                                                            & 5.60E-04 & 1.80E-02                                                             \\ \hline

{\color[HTML]{080808} Transcription}                                           & 61                                                                     & 17.5                                                                            & 1.10E-03 & 3.00E-02                                                             \\ \hline
\end{tabular}%
}
\caption{Functional Annotation enrichment for undruggable human disease proteins without Tbio selected by PortalCG} 
    \label{tab:without-tbio-enrichment}
\end{table}

\addcontentsline{toc}{subsection}{Table  \ref{tab:undrugnoTbiodisease}: Top ranked diseases associated with undruggable human proteins excluding Tbio selected by portal learning}
\begin{table}[ht]
\resizebox{\textwidth}{!}{%
\begin{tabular}{|c|c|}
\hline
Disease Name                   & \# of undruggable proteins associated with the disease \\ \hline
Body Height                    & 31                                                     \\ \hline
Colorectal Carcinoma           & 28                                                     \\ \hline
Malignant neoplasm of breast   & 26                                                     \\ \hline
Breast Carcinoma               & 18                                                     \\ \hline
Blood Protein Measurement      & 18                                                     \\ \hline
Leukemia, Myelocytic, Acute    & 17                                                     \\ \hline
Carcinogenesis                 & 17                                                     \\ \hline
Neoplasm Metastasis            & 17                                                     \\ \hline
Liver carcinoma                & 15                                                     \\ \hline
Malignant neoplasm of prostate & 15                                                     \\ \hline
\end{tabular}%
}
\caption{Top ranked diseases associated with undruggable human proteins excluding Tbio selected by portal learning} 
    \label{tab:undrugnoTbiodisease}
\end{table}

\addcontentsline{toc}{subsection}{Table  \ref{tab:UndrugnoTbiochemicals}: Chemicals interacted with undruggable human proteins excluding Tbio}
\begin{table}[ht]
\resizebox{\textwidth}{!}{%
\begin{tabular}{|c|c|c|c|}
\hline
Drug\_name             & Clinical phase  & Mechanism of Action                                       & Number of targeted proteins \\ \hline
fenebrutinib           & Phase 2         & Bruton's tyrosine kinase (BTK) inhibitor                  & 80                          \\ \hline
PF-05190457            & Phase 2         & growth hormone secretagogue receptor inverse agonist      & 50                          \\ \hline
MK-5046                & Preclinical     & bombesin receptor agonist                                 & 38                          \\ \hline
CCT137690              & Preclinical     & Aurora kinase inhibitor                                   & 36                          \\ \hline
AI-10-49               & Preclinical     & core binding factor inhibitor                             & 31                          \\ \hline
abemaciclib            & Launched        & CDK inhibitor                                             & 26                          \\ \hline
CFI-402257             & Phase 1/Phase 2 & dual specificity protein kinase inhibitor                 & 20                          \\ \hline
NMS-P715               & Preclinical     & protein kinase inhibitor                                  & 14                          \\ \hline
NMS-1286937            & Phase 2         & PLK inhibitor                                             & 11                          \\ \hline
elbasvir               & Launched        & HCV inhibitor                                             & 8                           \\ \hline
cilofexor              & Phase 3         & FXR agonist                                               & 7                           \\ \hline
ABBV-744               & Phase 1         & bromodomain inhibitor                                     & 7                           \\ \hline
tropifexor             & Phase 2         & FXR agonist                                               & 7                           \\ \hline
CDK9-IN-6              & Preclinical     & CDK inhibitor                                             & 6                           \\ \hline
adapalene              & Launched        & retinoid receptor agonist                                 & 4                           \\ \hline
Q-203                  & Phase 2         & ATP synthase inhibitor                                    & 4                           \\ \hline
PLX8394                & Phase 1/Phase 2 & serine/threonine kinase inhibitor                         & 4                           \\ \hline
ABT-702                & Preclinical     & adenosine kinase inhibitor                                & 4                           \\ \hline
NVP-BSK805             & Preclinical     & JAK inhibitor                                             & 3                           \\ \hline
OTS167                 & Phase 1/Phase 2 & maternal embryonic leucine zipper kinase inhibitor        & 3                           \\ \hline
CHIR-99021             & Preclinical     & glycogen synthase kinase inhibitor                        & 3                           \\ \hline
DBPR-211               & Preclinical     & cannabinoid receptor antagonist                           & 2                           \\ \hline
A-887826               & Preclinical     & sodium channel blocker                                    & 2                           \\ \hline
integrin-antagonist-1  & Phase 1         & integrin antagonist                                       & 2                           \\ \hline
piperaquine-phosphate  & Launched        & antimalarial agent                                        & 2                           \\ \hline
cenerimod              & Phase 2         & sphingosine 1-phosphate receptor modulator                & 1                           \\ \hline
peposertib             & Phase 1/Phase 2 & DNA dependent protein kinase inhibitor                    & 1                           \\ \hline
tezacaftor             & Launched        & CFTR channel agonist                                      & 1                           \\ \hline
cot-inhibitor-2        & Preclinical     & MAPK-interacting kinase inhibitor                         & 1                           \\ \hline
itacitinib             & Phase 3         & JAK inhibitor                                             & 1                           \\ \hline
10-hydroxycamptothecin & Preclinical     & topoisomerase inhibitor                                   & 1                           \\ \hline
alectinib              & Launched        & ALK tyrosine kinase receptor inhibitor                    & 1                           \\ \hline
adarotene              & Phase 1         & retinoid receptor agonist                                 & 1                           \\ \hline
acalabrutinib          & Launched        & Bruton's tyrosine kinase (BTK) inhibitor                  & 1                           \\ \hline
XL041                  & Preclinical     & LXR agonist                                               & 1                           \\ \hline
WAY-207024             & Preclinical     & gonadotropin releasing factor hormone receptor antagonist & 1                           \\ \hline
MK-5108                & Phase 1         & Aurora kinase inhibitor                                   & 1                           \\ \hline
CGM097                 & Phase 1         & MDM inhibitor                                             & 1                           \\ \hline
CD-437                 & Preclinical     & retinoid receptor agonist                                 & 1                           \\ \hline
AMG-925                & Phase 1         & CDK inhibitor|FLT3 inhibitor                              & 1                           \\ \hline
ACT-132577             & Launched        & endothelin receptor antagonist                            & 1                           \\ \hline
ziritaxestat           & Phase 3         & autotaxin inhibitor                                       & 1                           \\ \hline
\end{tabular}%
}
\caption{Chemicals interacted with undruggable human proteins excluding Tbio} 
    \label{tab:UndrugnoTbiochemicals}
\end{table}

\addcontentsline{toc}{subsection}{Table  \ref{tab:config}: Model architecture configuration}
\begin{table}[ht]
\centering
\resizebox{\textwidth}{!}{%
\begin{tabular}{|c|l|r|}
\hline
\multirow{2}{*}{Protein   descriptor}                                    & layers & Albert --\textgreater Resnet                                      \\ \cline{2-3} 
                                     & embedding dimension & 256             \\ \hline
\multirow{5}{*}{Chemical descriptor} & backbone            & GIN             \\ \cline{2-3} 
                                     & number of layers    & 5               \\ \cline{2-3} 
                                     & embedding dimension & 300             \\ \cline{2-3} 
                                     & aggregation methods & sum             \\ \cline{2-3} 
                                     & drop out ratio      & 0.5             \\ \hline
\multirow{2}{*}{Interaction learner}                                     & layers & Attentive pooling --\textgreater{}2 layers   of MLP               \\ \cline{2-3} 
                                     & embedding dimension & 128             \\ \hline
\multicolumn{1}{|l|}{Structure residue-atom pair wise feature   learner} & layers & matrix multiplication of protein   and chemical embedding vectors \\ \hline
\multirow{2}{*}{Classifier}          & layers              & 2 layers of MLP \\ \cline{2-3} 
                                     & embedding dimension & 64              \\ \hline
\end{tabular}%
}
\caption{Model architecture configuration}
\label{tab:config}
\end{table}

\clearpage
\section{Related works }

\subsection{OOD generalization in deep learning}

The recent work Invariant risk minimization (IRM)\cite{invariant-risk-minimization} is a dedicated algorithm to OOD generalization, which is under the goal of transformative solution for invariant representation. However, given its completeness in theory, many experiments  \cite{agianst-IRM} report IRM are not doing well in large real-world data set. 

Many deep learning tasks are inherently OOD generalization. Among those jargon, some are famous for defining a type of OOD scenario problem, for example, Domain generalization \cite{domain-genearalization-survey} can be taken as the equivalent of OOD generalization; domain shift \cite{domain-genearalization-survey} rephrases the fact of distribution change in terms of $D(X,y)$. Some jargon define a type of solution: domain alignment \cite{domain-align-assumption} minimizes the difference/distance between source domains and target domains distributions for an invariant representation where the distance between source and target domain distributions are measured by a wide variety of statistical distance metrics from simple $l_2$, $f-divergence$ to Wasserstein distance; domain adaptation \cite{domain-adapt-spare-label} is to leverage pretrained model on a different domain and is just one idea to achieve domain generalization, the more general term that is equivalent to OOD generalization in a more practical sense; causal learning is proved by  \cite{ood-thesis} to be equivalent to OOD generalization when causality makes senses (taking into consideration the existence of cases where causality is meaningless); robust optimization \cite{robust-opt} that focuses on worst-group performance instead of the average one in ERM;  although robust optimization has not quite been adapted to modern deep learning, its sub-field  distributional robust optimization \cite{distributional-opt-survey} has witnessed quite a few recent works adapted to be used in deep learning. 

Worth to be clarified that, many works that are solving the sub-group or sub-population shift problem is quite different from the OOD generalization problem as discussed in the setting of dark chemical genomics space. Sub-population shift is more like a imbalanced data problem where the test set has major resemblance with training data just the shift from a major class to a minor class or vise-versa. For example, GroupDRO \cite{groupdro-1} was published in 2018 to address this problem, proposing to incorporate structural assumptions on distribution, which could be straight forward in some data sets which has more meta-data or is a multi-label classification case where the label structure could be used as the structural assumption.

\subsection{Portal learning key components related}

\textit{(Model architecture)}  
Even since the debut of the survey \cite{representation-benjio} enlightening the perspective of \textit{representation learning}, enormous research passion is motivated for model architecture design, almost taken as equivalent to deep learning and overshadowing all other directions. A key idea that echos the demand of generalization is to learn \textit{global representation} which helps to decrease both $TrainError(f_{i}^{iid})$ and \textit{known space generalization gap}, denoising large data set. Hence, to solve OOD, good model architecture design is not enough.

All existing work in CPI is confined in the known space and limited works have concerned  generalization. Generally, proposed CPI deep learning models follow the same fashion: build model architecture of three key modules, protein descriptor, chemical descriptor and interaction learner formulating  a classification  problem with a few variants as regression problem. Innovation is seen mostly for model architecture, particularly active  for chemical descriptor, reflecting all milestones in recent years deep learning advancement from CNN, LSTM to Transformer and GNN as demonstrated in DeepPurpose \cite{huang2020deeppurpose}. Generalization has not shown in any previous work as a main goal of research except for DISAE\cite{DISAE} which proves generalizability to orphan GPCR protein drug screening mainly relying on a general purpose pretrained protein language model. It's fine-tuned on GPCR data set with shifted evaluation. Hence, DISAE becomes the baseline model in this work. 

\textit{(Model intialization)}
Although could be categorized as a type of representation learning,  \textit{transfer learning} became an iconic independent concept for its huge success with breakthroughs in many NLP and CV benchmark tasks. It features a \textit{pretraining-finetuning} procedure. An intuitive example is to pretrain a language model on large general English vocabulary with pretext task formulation such as predicting next word and  then to finetune the language model on specific downstreawm task such as machine translation in biology domain. Well-renowned Transformer based pretrained models starting from human language models are a combined success of attention based model architecture design and transfer learning. In the computation biology field, most eye-catching equivalent is protein language model, i.e. protein descripto, which inspired several similar works at the same time by different groups: TAPE and ESM showcases pretraining on large protein vocabulary could significantly improves downstream task such as protein-protein interaction prediction; MSA-based-tranformer and DISAE incorporates MSA in pretraining. From the perspective of the target downstrem task, the power of transfer learning comes from a better model initialization. This is a major breakthrough that could fill the gap of $TestError(f_{t}^{OOD}) - TestError(f_{t}^{iid})$ but not necessarily, depending on how it's incorporated into the whole training scheme at system level, particularly depending on data fed in. 

DISAE is used in our work here as a pretrained protein descriptor. This choice over other protein language models is due to the fact that DISAE is the smallest among other in terms of memory required to use and optimize with same level of performance. STL is a way to leverage transfer learning to fine better model initialization. The main difference and innovation is that transfer learning naively relies on the belief that more general knowledge transferred will bring better performance while STL in portal learning actively leverage biology endorsed biased when transferring general knowledge. Further, by defining the goal ``to learn the portal'', which will be closer to global optimum in target universe loss landscape, the whole training system is steered actively solve ODD. 

\textit{(STL )} Sparked by the breakthrough of Alphafold 1 \cite{alphafold1} and Alphafold 2\cite{alphafold2} in protein structure prediction, deep learning has been trusted in molecule interaction distance map prediction to learn structure information. The inclusion of CPI-structure, i.e. protein function prediction portal calibration is inspired by recent success in protein structure prediction led by the great work of AlphaFold1\cite{alphafold1} and AlphaFold2\cite{alphafold2}. Specifically, we pretrain the model to predict residue-residue contacts for a protein whose structure is solved and chemical atom-protein residue contacts given a known CPI complex structure. There are three popular ways of residue-residue pairwise distance matrix prediction depending on how to formulate it as a machine learning task. On the one end is to formulate it as a binary classification where a distance threshold is set defining whether a pair of residues are in contact or not, hence the name contact prediction. On the other end is to formulate it as a regression problem where the exact distance is used as a regression target, hence the name exact distance prediction. AlphaFold1 showed another way in between the two ends, which is to formulate it as a multi-class classification problem, where the distribution of pair-wise residue distances is broken down into multiple class labels according to a histogram, hence the name distogram prediction. We first focus on residue-atom pair wise distance at binding sites and then experiments contact prediction and distogram prediction. In our results, the two formulations have similar performance in light of the final CPI prediction through ablation study as shown in Supplemental Figure S\ref{fig:ablation-contact-distogram}. 

\textit{(OOC-ML)}  It's long be aware that the sequence order of training data exposed to the model has an impact on model generalizability. Active learning \cite{active-learning-drug-nature} emphasises to actively query data  in a iterative fashion  to only expose the model to data close to he decision boundary . Curriculum learning \cite{bengio-curriculum} emphasises to sort all training so that the model is exposed to challenges of increasing difficulty.  This element of data logistics has also been closely weaved into many optimization algorithms that aim to improve model generalizability. For example, contrastive loss\cite{siamese} requires certain ratio of positive v.s. negative samples in each mini-bath. Most related to portal learning is meta-learning which can be categorized into metric-based, model-based and optimizer based ``learn to learn'' algorithms \cite{DBLP:meta-survey} with application to few-shot learning and zero-shot learning. Meta-learning started for the data-efficiency challenge instead of generalization or OOD. Although meta-learning is defined very general, making many algorithms seem to be mere an variants falling under its umbrella, in practice, algorithms proposed bearing the name of meta-learning are defined on multi-class classification data set, typically image classification, where the main challenge is the huge number of classes while limited data points are known in each class. Because of this underlying motivation, meta-learning features a very involving data logistics with multiple layer of optimization each has its meta-train/meta-test set sampled based on label distribution. These unspecified facts reveal that there is no existing meta-learning algorithm fit into CPI data.

However, the idea of ``learn to learn'' is attractive. MAML\cite{maml} is the optimization based meta-learning work that inspired OOC-ML proposed as a major component of portal learning. The differences are major. OOC-ML algorithm expands on it by focusing on data feature distribution instead of label distribution, encouraging active sampling in local neighborhood, which simplifies the support/query meta-train/meta-test data logistics, and ensembling a few local loss directions to learn global gravity direction. 

\clearpage

\section{Additional figures}
\begin{figure}[ht]
    \centering
    \includegraphics[scale=0.7]{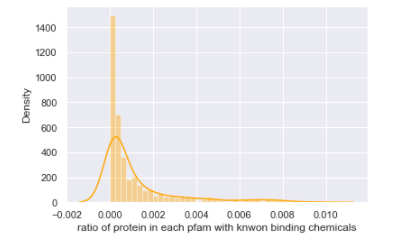}
    \caption{Dark space statistics histogram  based on known CPI pairs in ChEMBL26. $< 1\%$ proteins in each pfam invovled in ChEMBLE have known binding chemcials. }
    \label{fig:hist-darkspace}
\end{figure}

\begin{figure}[ht]
    \centering
    \includegraphics[width=\textwidth]{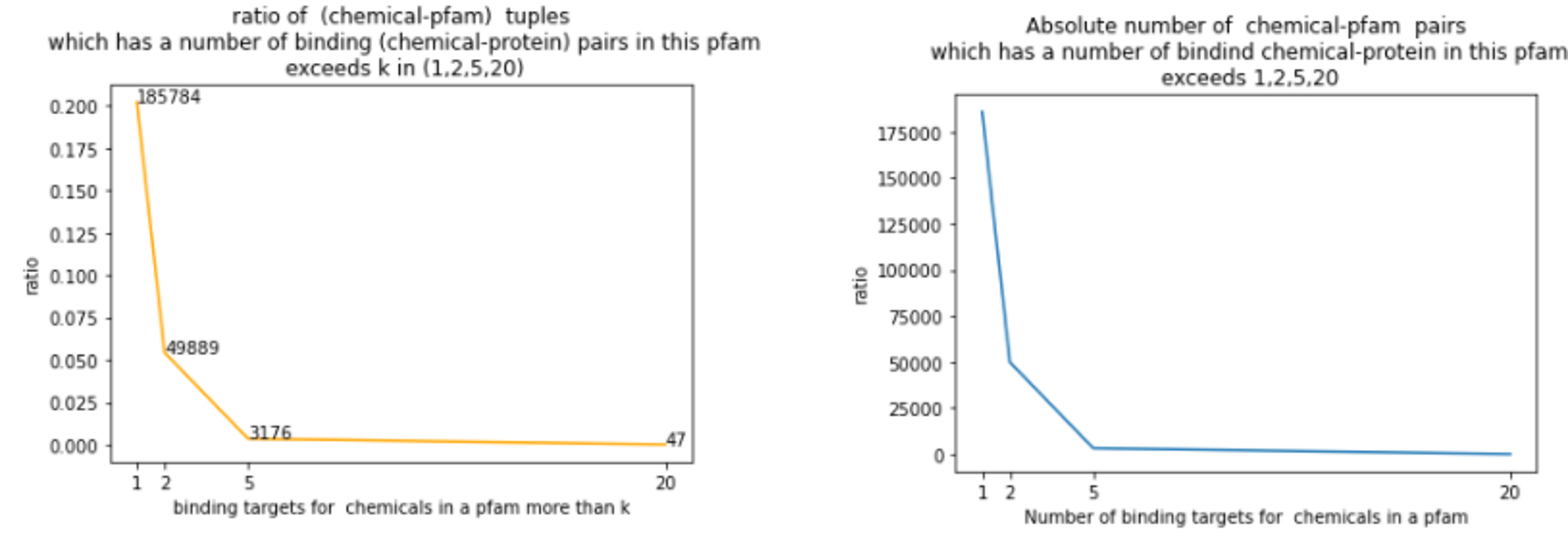}
    \caption{Dark space statistics histogram  based on known CPI pairs in ChEMBL26. $<1\%$ chemicals bind to more then 2 proteins; $<0.4\%$ chemicals bind to more than 5 proteins.}
    \label{fig:dark-dotted-dist}
\end{figure}

\addcontentsline{toc}{subsection}{Figure  \ref{fig:hist-darkspace}: Dark space statistics histogram}
\addcontentsline{toc}{subsection}{Figure  \ref{fig:dark-dotted-dist}: Dark space statistics trend line}

\begin{figure}[ht]
    \centering
    \includegraphics[width=\textwidth]{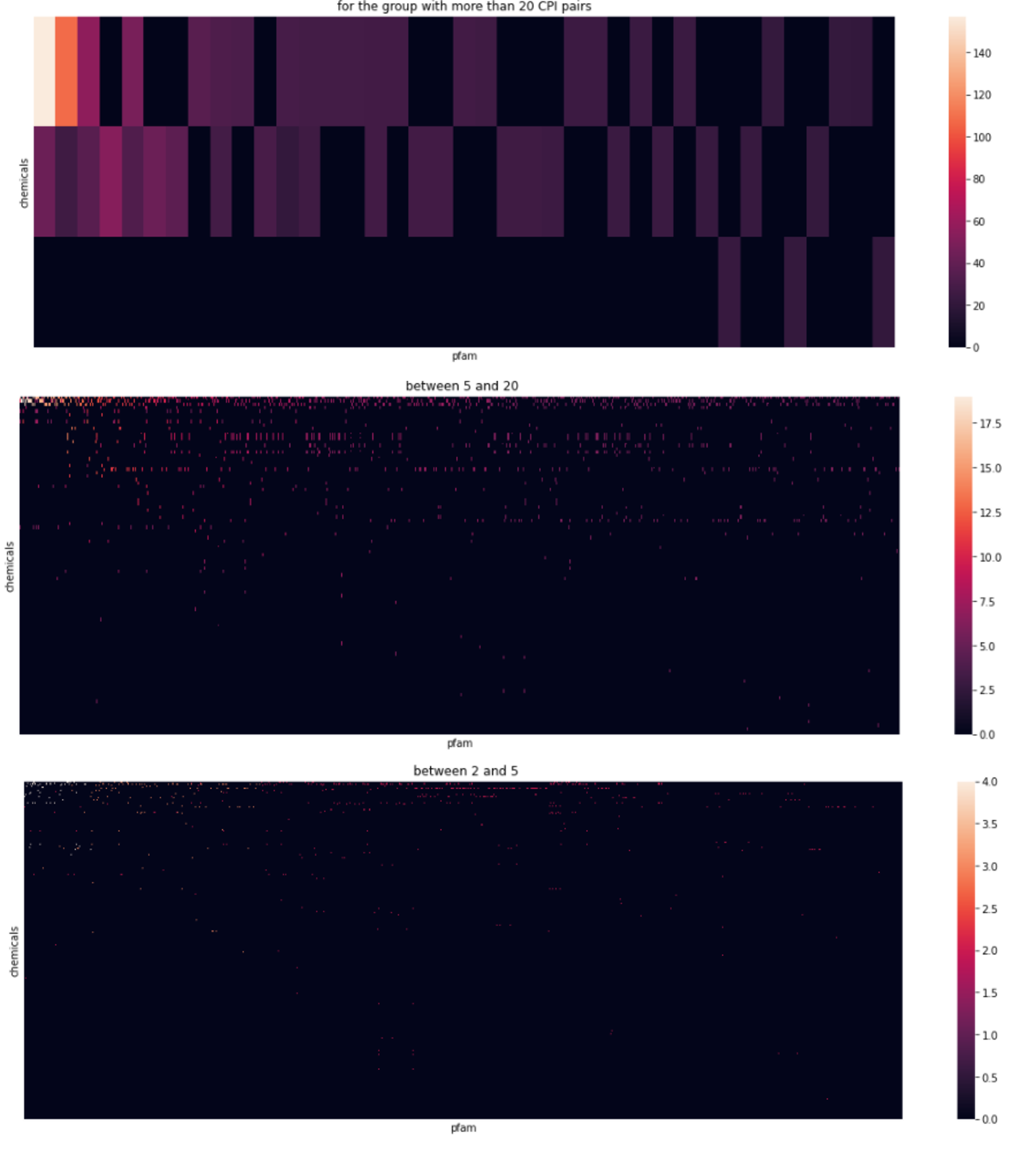}
    \caption{As shown in Figure S \ref{fig:dark-dotted-dist}, there are three main ranges in terms of the binding targets in a pfam for one chemical: [2,5],[5,20],[20,). For each of the range, a heatmap is shown with y axis representing each chemicals, x axis representing each pfam, each point representing the known binding pairs for one chemical and one pfam. As we can see, there is huge dark space. }
    \label{fig:dark-dotted}
\end{figure}

\addcontentsline{toc}{subsection}{Figure  \ref{fig:dark-dotted}: Dark space statistics heatmap}

\begin{figure}[ht]
    \centering
    \includegraphics[scale=0.8]{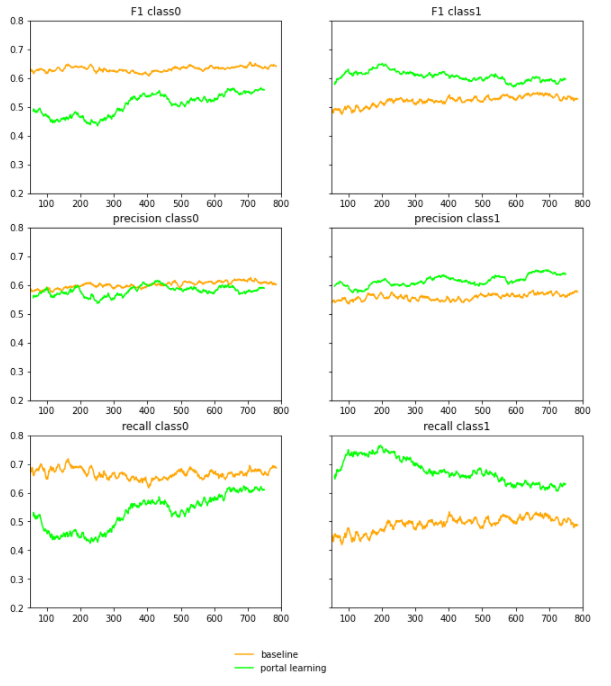}
    \caption{In the main text, overall evaluation across positive and negative classes are reported, such as F1, ROC-AUC, PR-AUC. Here is a breakdown of performance in each class, where class0 is negative, i.e. not binding, class1 is positive, i.e. binding.  against DISAE as baseline }
    \label{fig:beakdown}
\end{figure}

\addcontentsline{toc}{subsection}{Figure  \ref{fig:beakdown}: Model performance breakdown to each class}

\begin{figure}[ht]
    \centering
    \includegraphics[width=\textwidth]{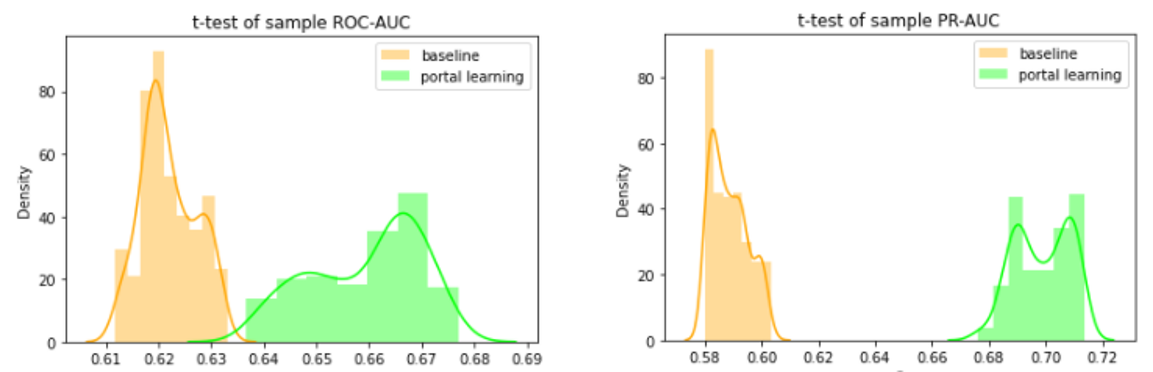}
    \caption{t-test comparison. The p-values for both ROC-AUC and PR-AUC are close to 0  against DISAE as baseline}
    \label{fig:t-test}
\end{figure}

\addcontentsline{toc}{subsection}{Figure  \ref{fig:t-test}: Model performance t-test}

\begin{figure}[ht]
    \centering
    \includegraphics[width=\textwidth]{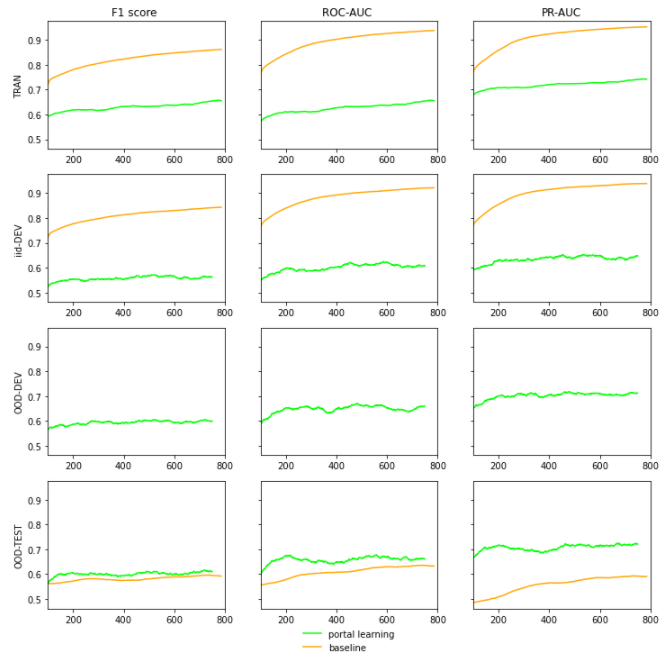}
    \caption{Stress model selection performance curves against DISAE as baseline}
    \label{fig:4split}
\end{figure}

\addcontentsline{toc}{subsection}{Figure \ref{fig:4split}: Stress model selection performance curves against DISAE as baseline}

\begin{figure}[ht]
    \centering
    \includegraphics[scale=0.8]{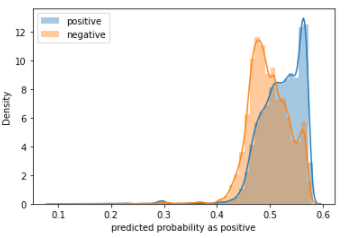}
    \caption{To select high confidence prediction, one additional procedure of filtering is to build a histogram of prediction scores based on known pairs. A threshold of $0.67$ is identified to filter out confident positive prediction.}
    \label{fig:false-positive}
\end{figure}

\addcontentsline{toc}{subsection}{Figure  \ref{fig:false-positive}: Model performance histogram}

\begin{figure}[ht]
   
    \begin{subfigure}[b]{0.5\textwidth}
        
        \includegraphics[width=0.9\textwidth, height=5cm]{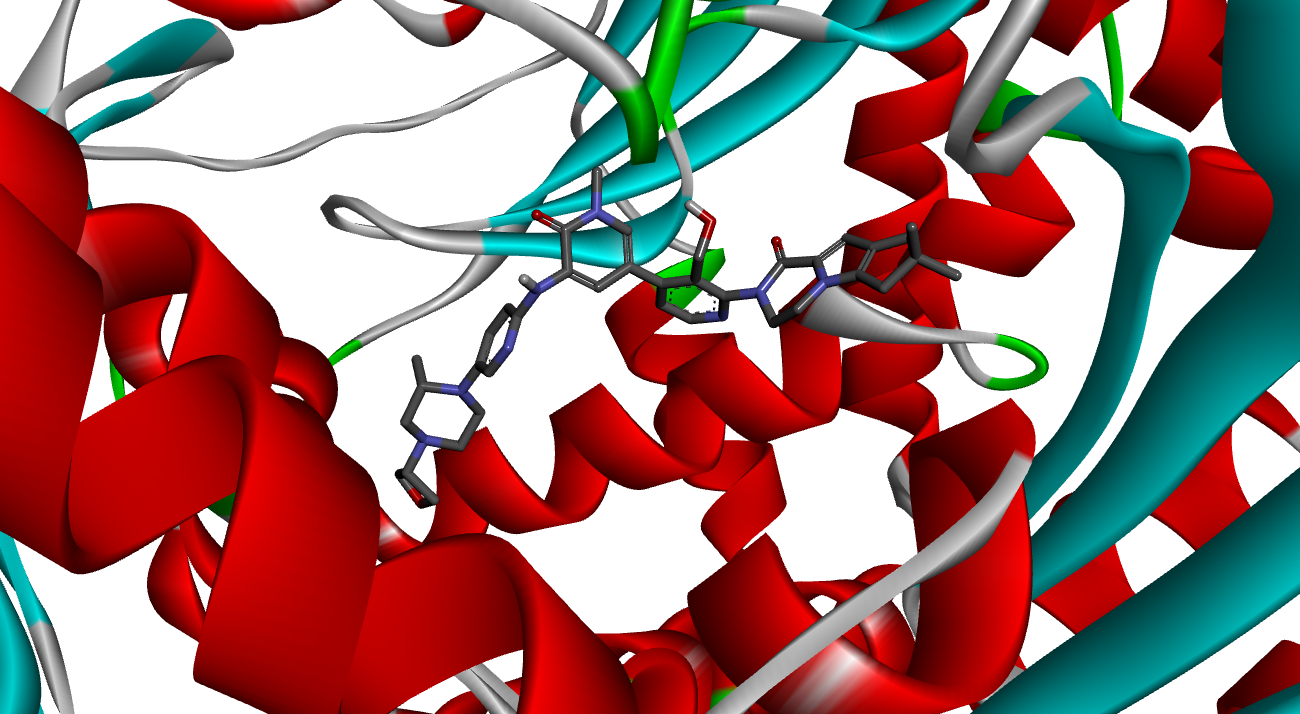}
    \end{subfigure}
    \hfill
    \begin{subfigure}[b]{0.5\textwidth}
        
        \includegraphics[width=0.9\textwidth, height=5cm]{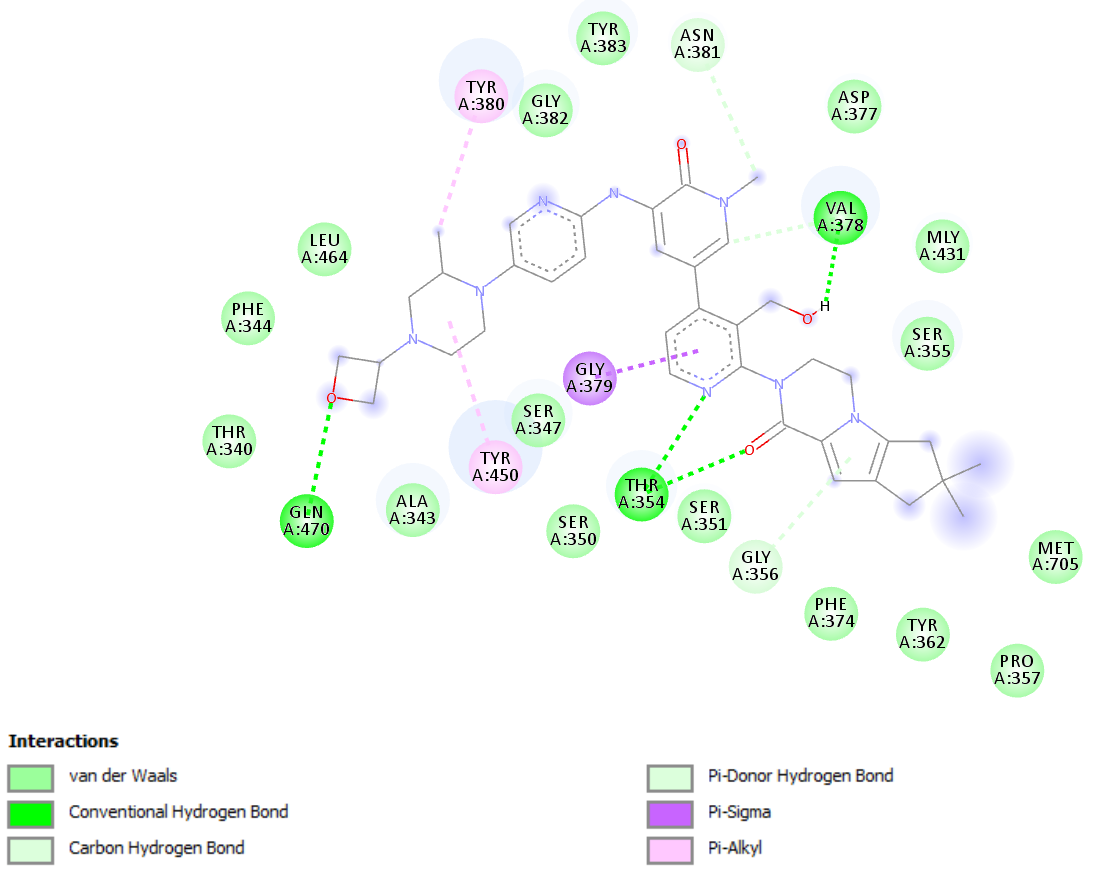}{(a)}
    \end{subfigure}
    
    \vfill
    
    \begin{subfigure}[b]{0.5\textwidth}
        \includegraphics[width=0.9\textwidth, height=5cm]{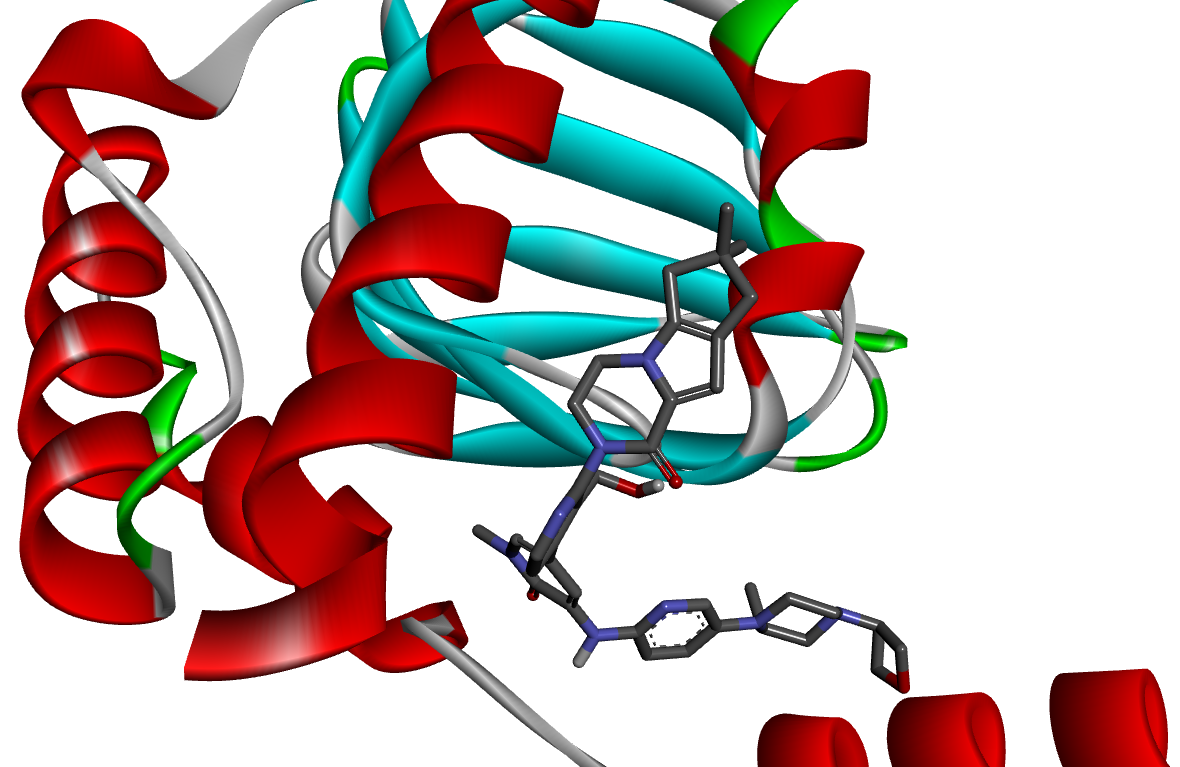}
    \end{subfigure}
    \hfill
    \begin{subfigure}[b]{0.5\textwidth}
        \includegraphics[width=0.9\textwidth, height=5cm]{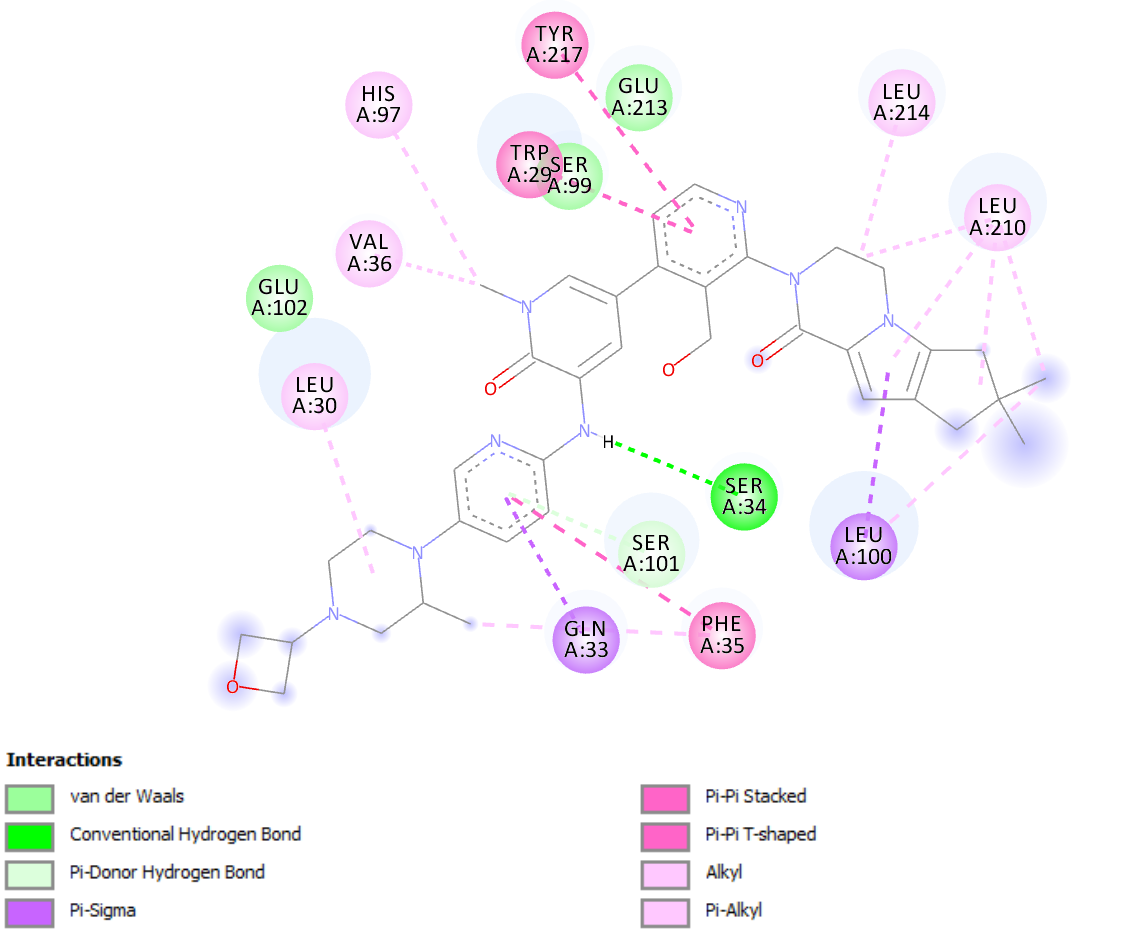}{(b)}
    \end{subfigure}
    
    \vfill
     \begin{subfigure}[b]{0.5\textwidth}
        \includegraphics[width=0.9\textwidth, height=5cm]{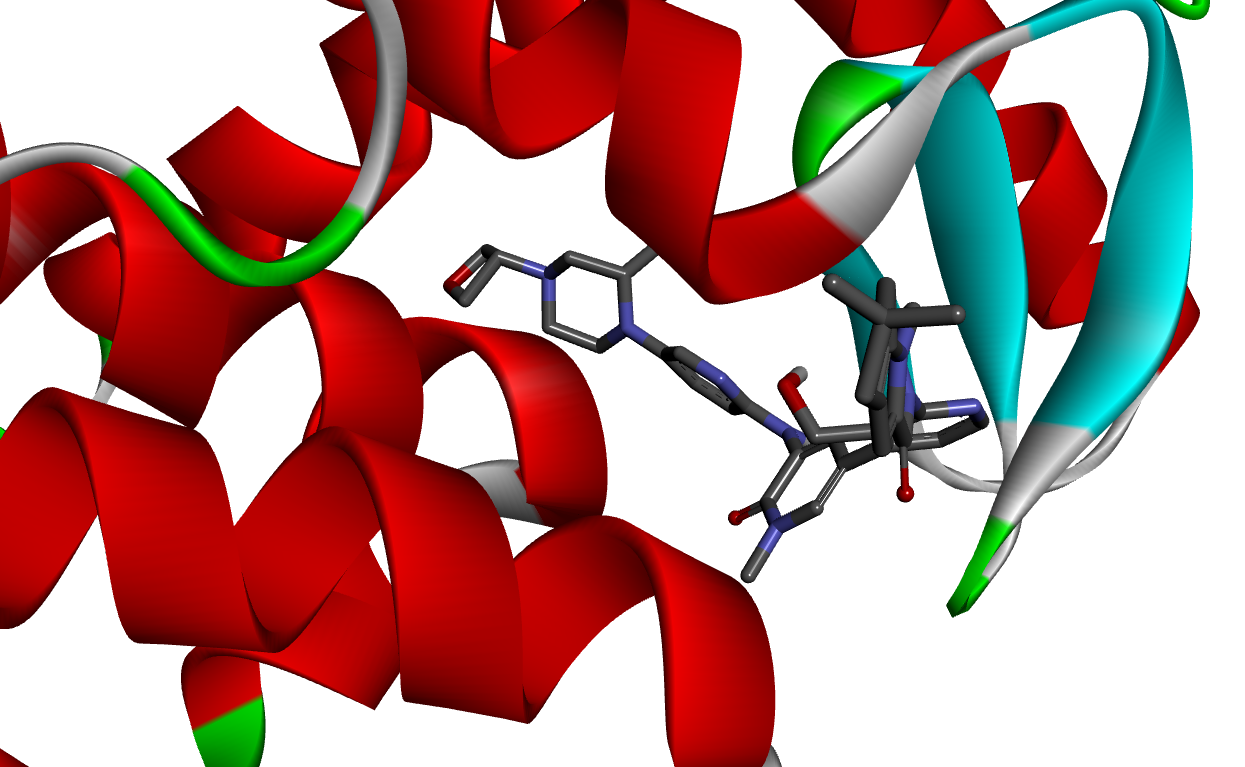}
    \end{subfigure}
    \hfill
    \begin{subfigure}[b]{0.5\textwidth}
        \includegraphics[width=0.9\linewidth, height=5cm]{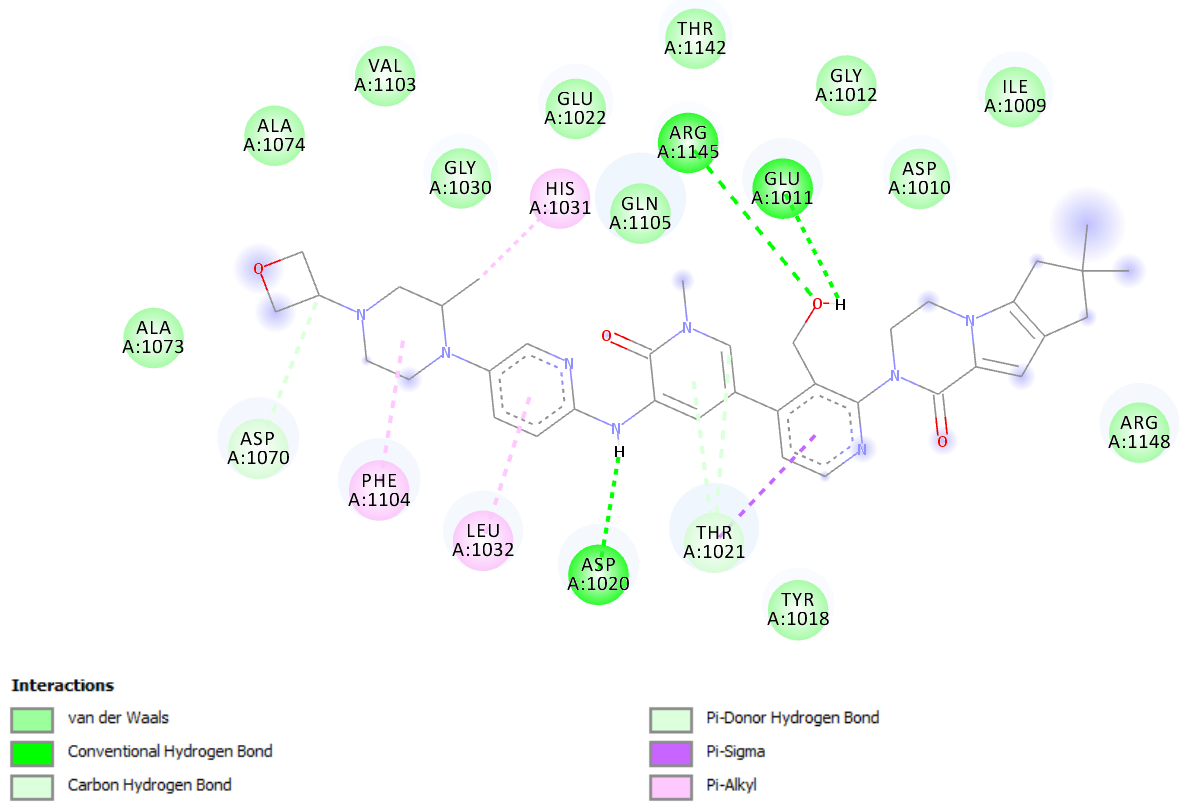}{(c)}
    \end{subfigure}
    \vfill

     \caption{The binding conformations and 2D-interactions of Fenebrutinib on the targeted proteins predicted by using Autodock: (a) Fenebrutinib on Presequence protease.  (b) Fenebrutinib on Sigma non-opioid intracellular receptor 1 . (c) Fenebrutinib on Mitochondrial amidoxime-reducing component 1. }
    \label{fig:fenebrutinib docking conformation 1}
\end{figure}

\addcontentsline{toc}{subsection}{Figure  \ref{fig:fenebrutinib docking conformation 1}: The binding conformations and 2D-interactions of Fenebrutinib on the targeted proteins predicted by using Autodock}

\begin{figure}[ht]    
    \begin{subfigure}[b]{0.5\textwidth}
        \includegraphics[width=0.9\linewidth, height=5cm]{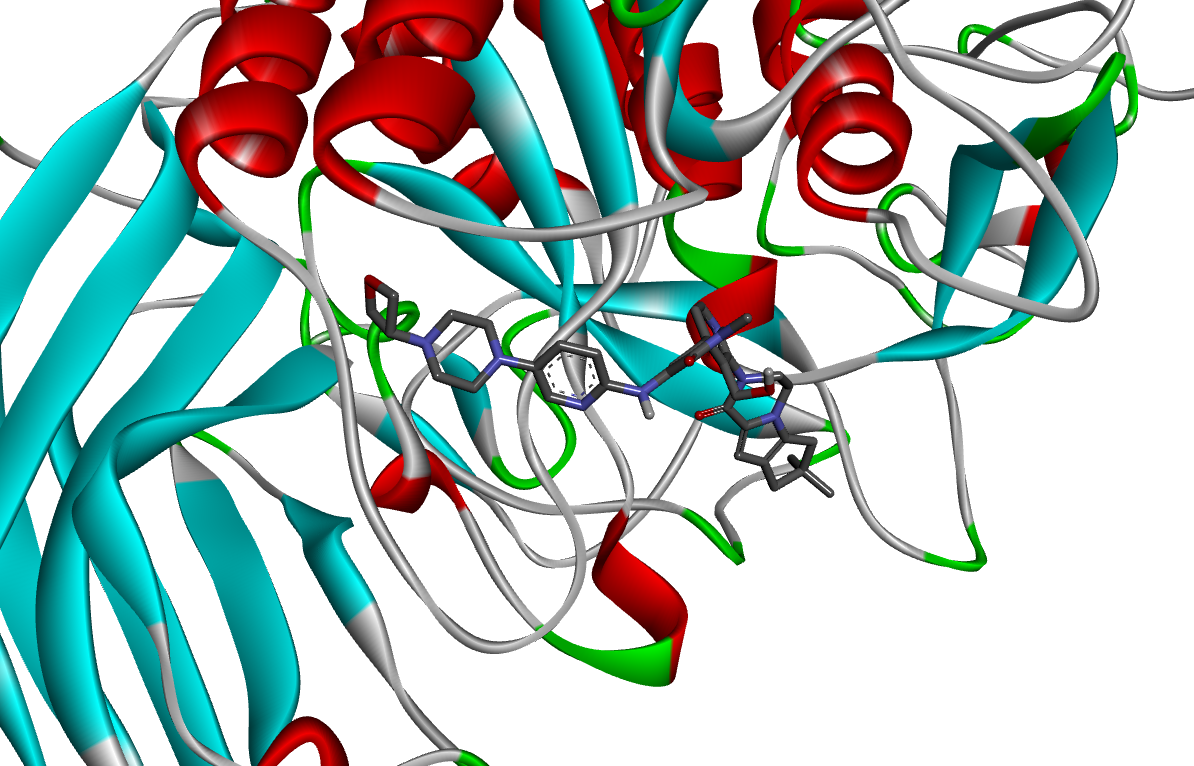}
    \end{subfigure}
    \hfill
    \begin{subfigure}[b]{0.5\textwidth}
        \includegraphics[width=0.9\linewidth, height=5cm]{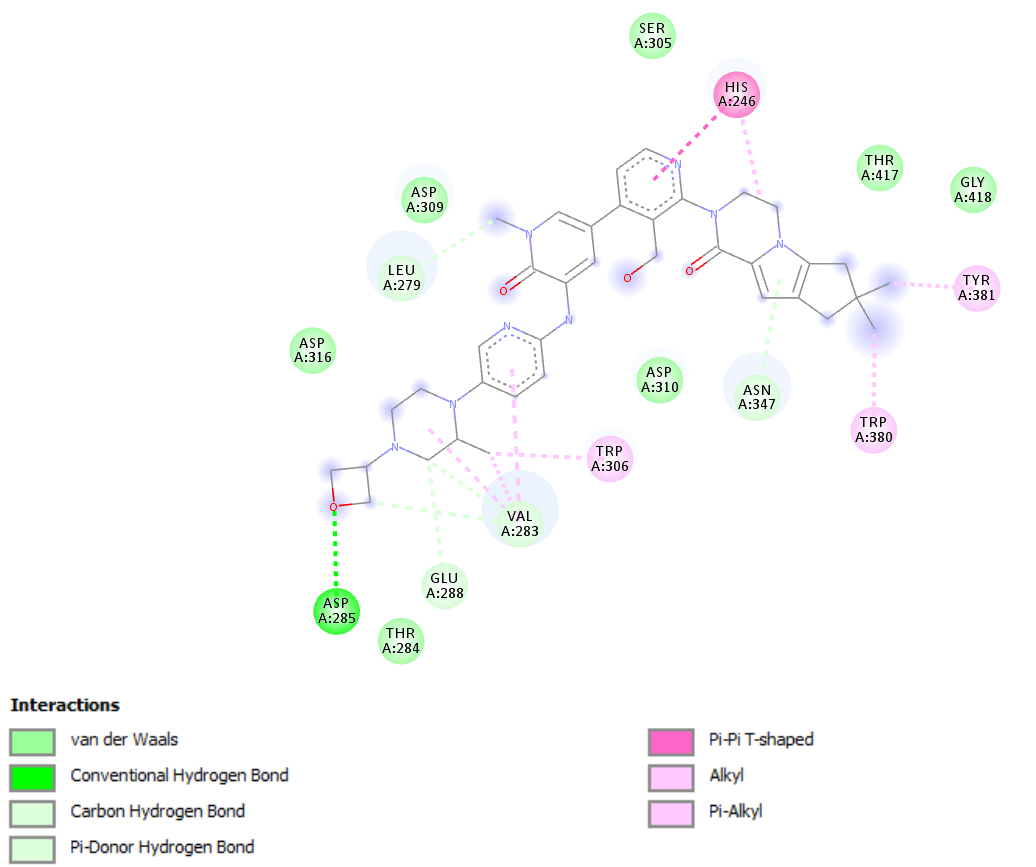}{(d)}
    \end{subfigure}
    
    \vfill
    \begin{subfigure}[b]{0.5\textwidth}
        \includegraphics[width=0.9\linewidth, height=5cm]{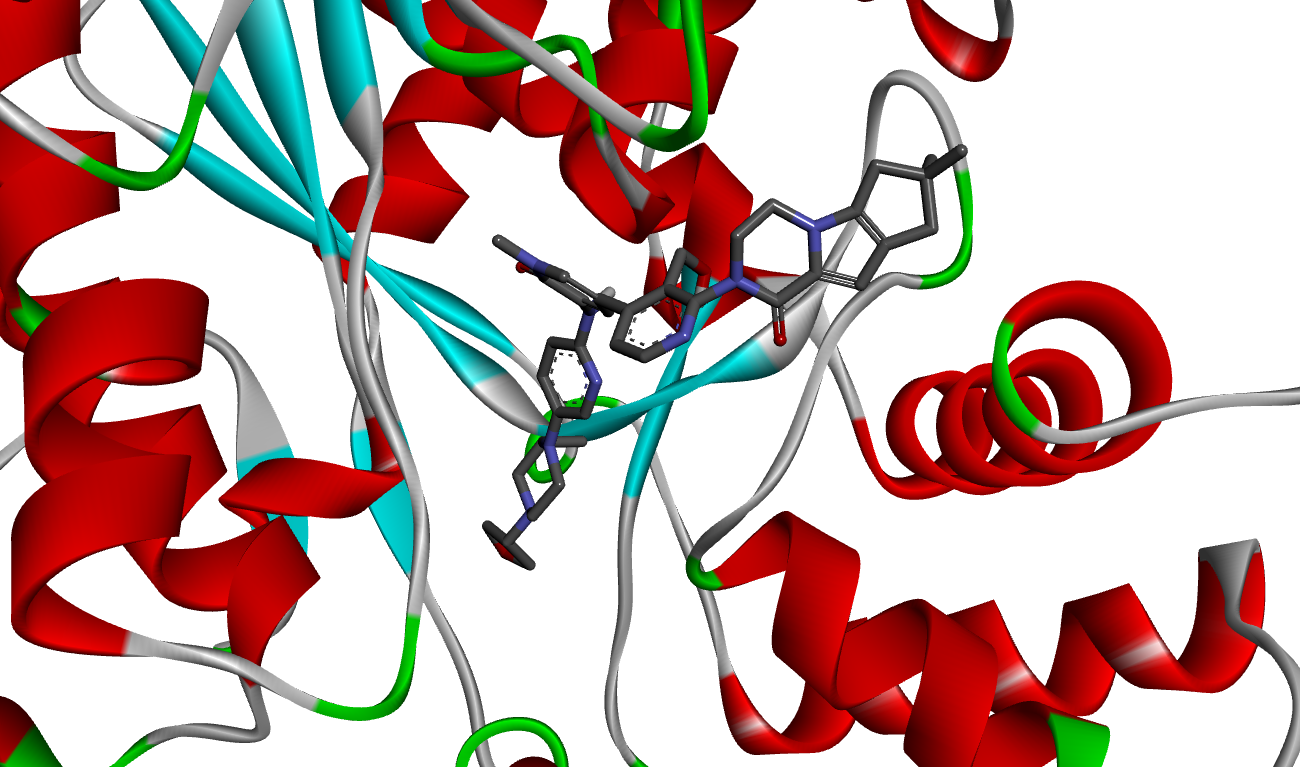}
    \end{subfigure}
    \hfill
    \begin{subfigure}[b]{0.5\textwidth}
        \includegraphics[width=0.9\linewidth, height=5cm]{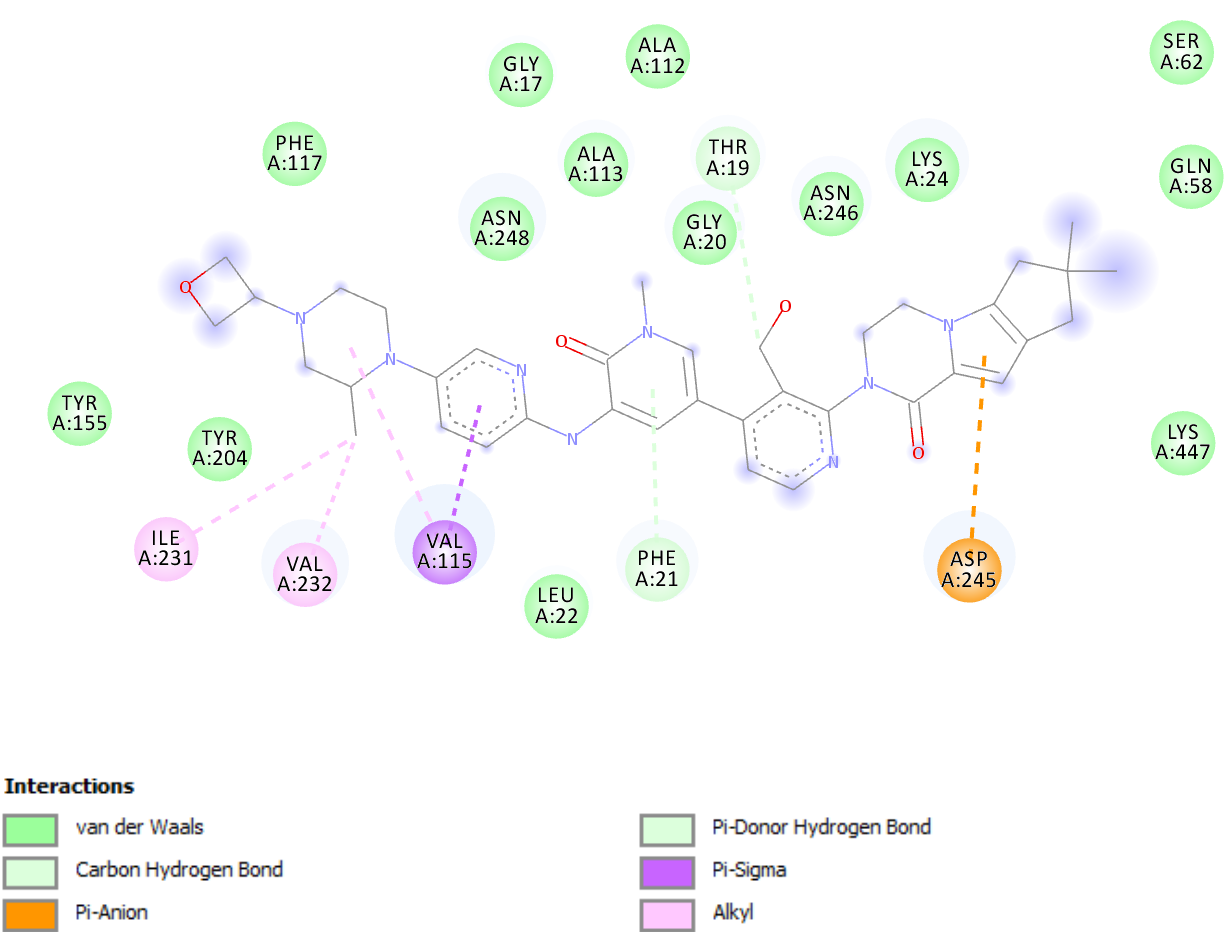}{(e)}
    \end{subfigure}
    
    \vfill
    \begin{subfigure}[b]{0.5\textwidth}
        \includegraphics[width=0.9\linewidth, height=5cm]{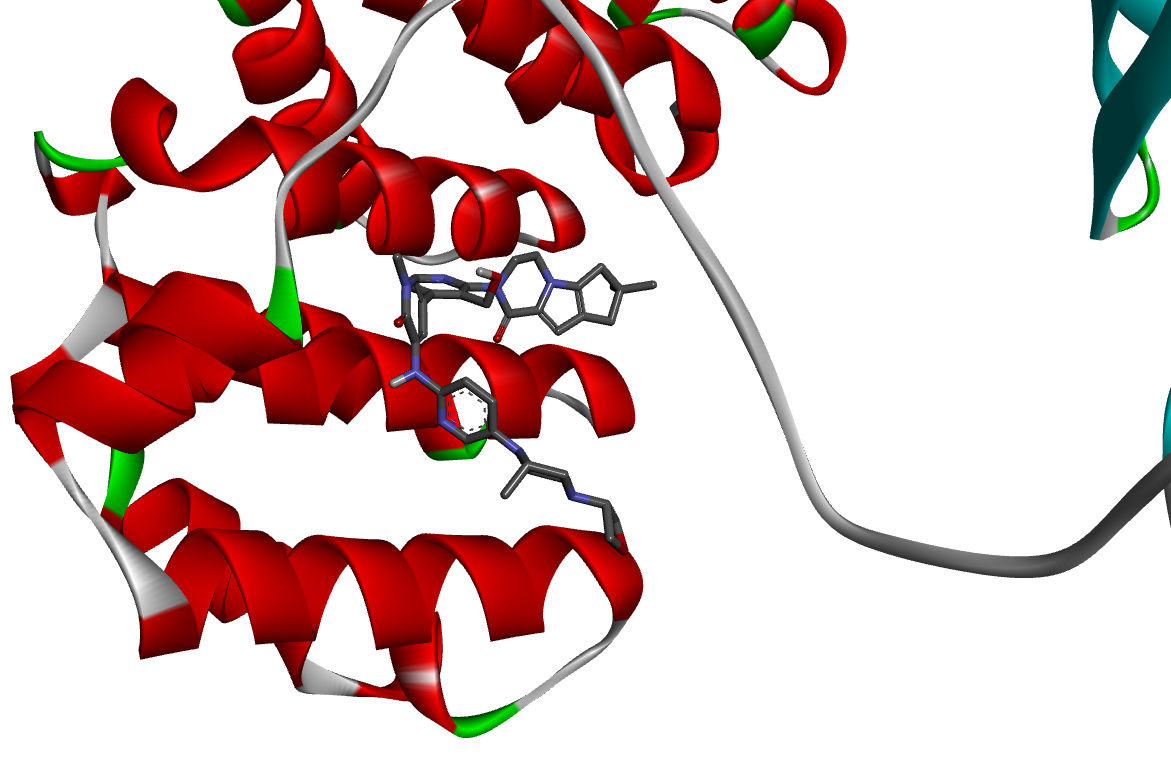}
    \end{subfigure}
    \hfill
    \begin{subfigure}[b]{0.5\textwidth}
        \includegraphics[width=0.9\linewidth, height=5cm]{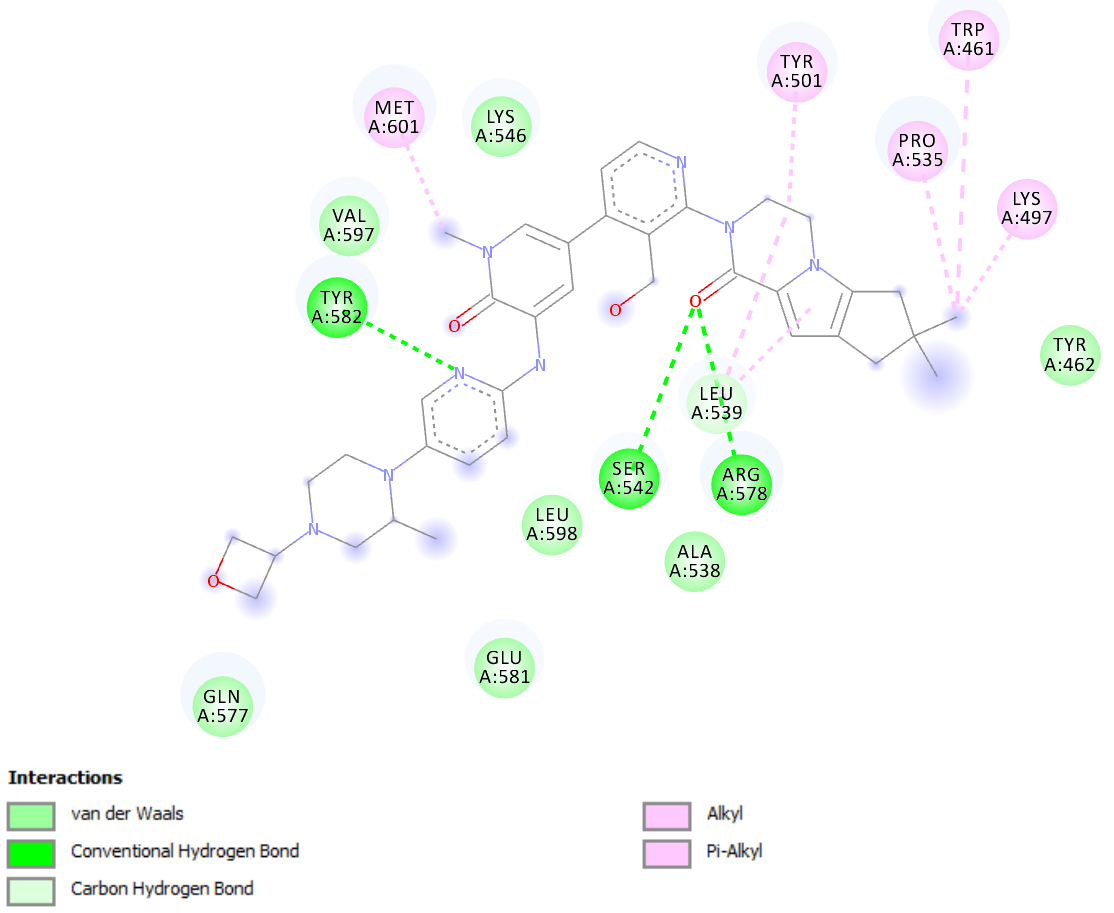}{(f)}
    \end{subfigure}
    
    \caption{The binding conformations and 2D-interactions of Fenebrutinib on the targeted proteins predicted by using Autodock: (d) Fenebrutinib on Proprotein convertase subtilisin/kexin type 6. (e) Fenebrutinib on Fatty acyl-CoA reductase 2. (f) Fenebrutinib on AP-2 complex subunit alpha-2. }
    \label{fig:fenebrutinib docking conformation 2}
\end{figure}

\addcontentsline{toc}{subsection}{Figure  \ref{fig:fenebrutinib docking conformation 2}: The binding conformations and 2D-interactions of Fenebrutinib on the targeted proteins predicted by using Autodock}
    
\begin{figure}[ht]
    \centering
    \includegraphics[scale=0.5]{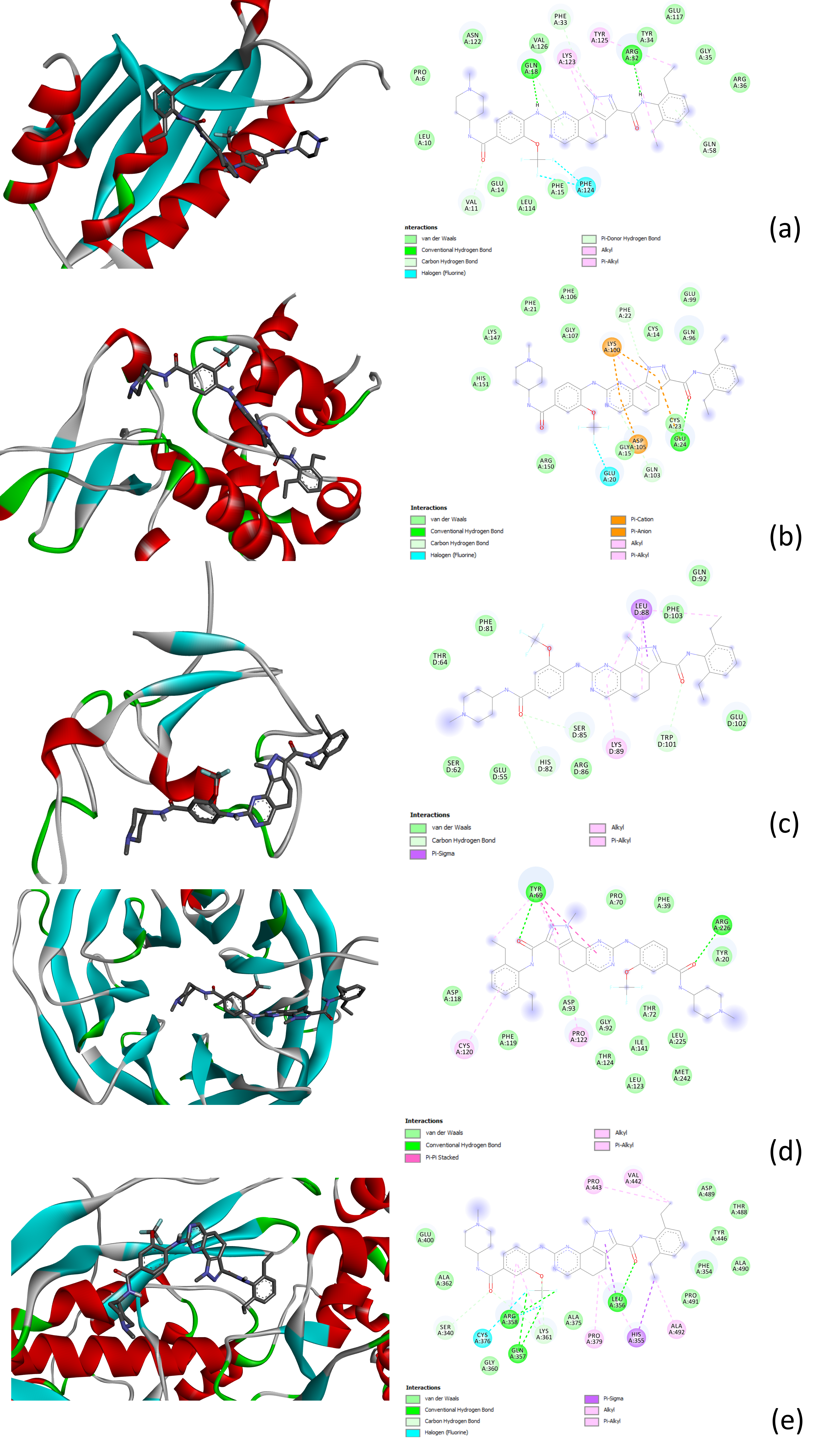}
    \caption{The binding conformations and 2D-interactions of NMS-P715 on the targeted proteins predicted by using Autodock: (a) NMS-P715 on Ras GTPase-activating protein-binding protein 2. (b) NMS-P715 on Casein kinase II subunit beta. (c) NMS-P715 on E3 ubiquitin-protein ligase RBX1. (d) NMS-P715 on DDB1- and CUL4-associated factor 7.(e) NMS-P715 on tRNA (guanine(26)-N(2))-dimethyltransferase. }
    \label{fig:NMS-P715 docking conformation}
\end{figure}

\addcontentsline{toc}{subsection}{Figure \ref{fig:NMS-P715 docking conformation}: The binding conformations and 2D-interactions of NMS-P715 on the targeted proteins predicted by using Autodock}
    
\begin{figure}[ht]
    \centering
    \includegraphics[width=\textwidth]{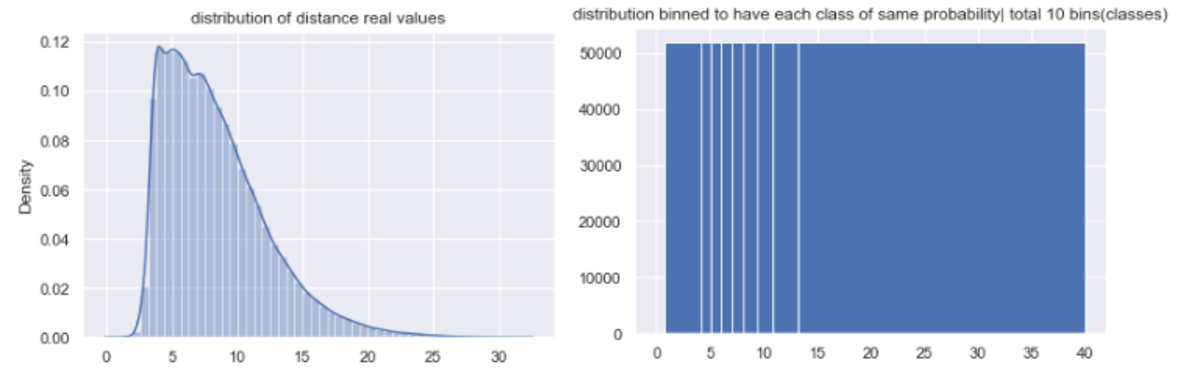}
    \caption{Histogram equalization results: the left panel shows the original distribution of distance real values; to formalize a multi-class classification where each class has equal probability, histogram equalization transforms the distribution to the right panel of 10 bins, each as a class.}
    \label{fig:hist-equal-10class}
\end{figure}

\begin{figure}[ht]
    \centering
    \includegraphics[width=\textwidth]{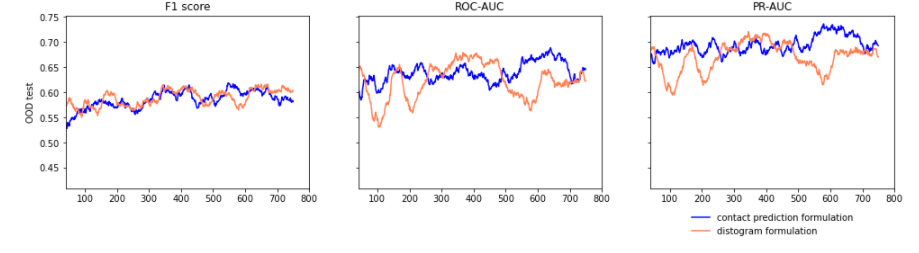}
    \caption{Ablation study to compare the impact of two formulation of protein structure distance prediction, contact prediction v.s. distogram prediction. The two variants have similar OOD-test performance.}
    \label{fig:ablation-contact-distogram}
\end{figure}

\begin{figure}[ht]
    \centering
    \includegraphics[width=\textwidth]{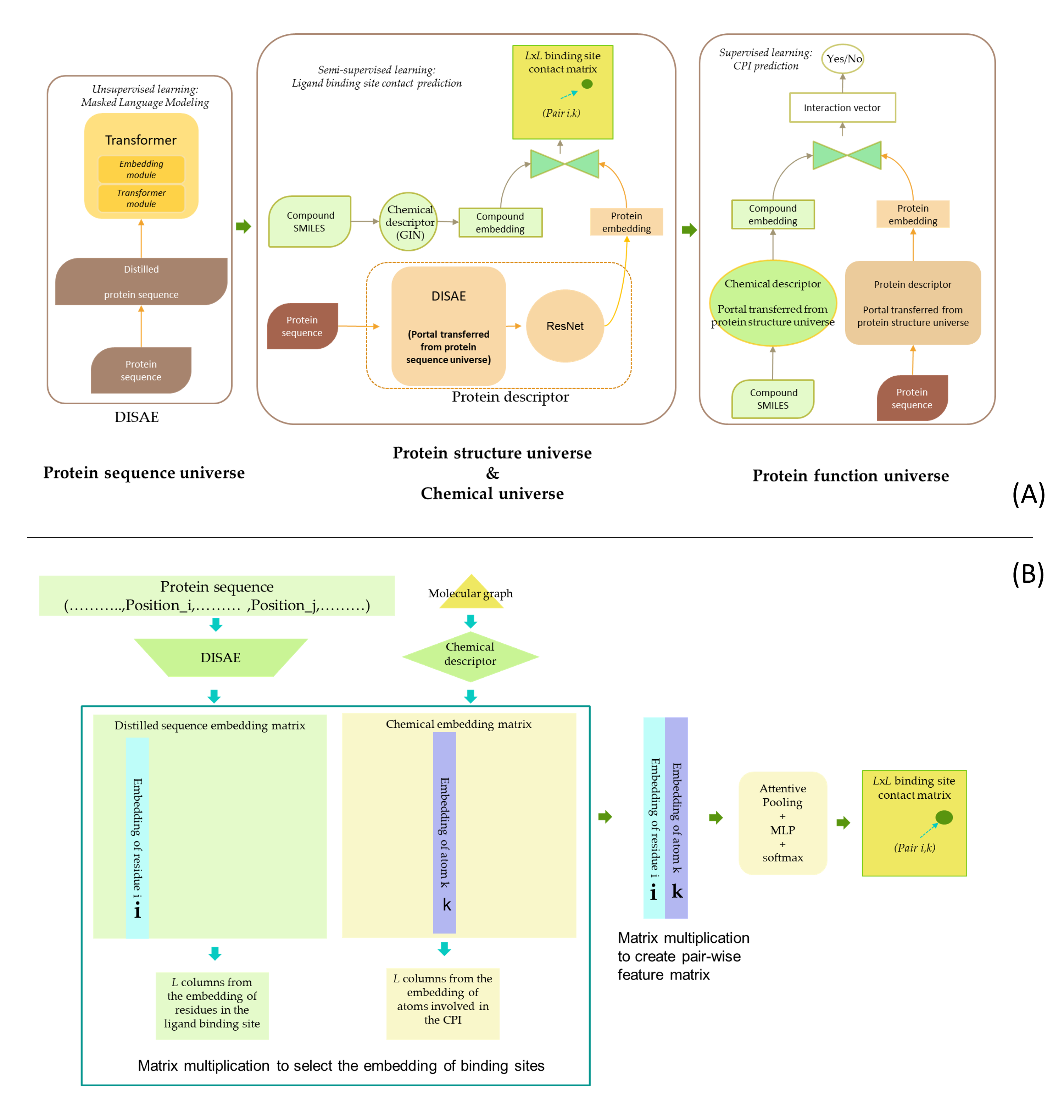}
    \caption{Illustration of PortalCG architecture: (A) the pipeline of STL (B) The model architecture for predicting binding site distance matrix. Note that Portal Learning is a general framework at the training scheme level instead of at the model architecture level.  OOC-ML as an optimization algorithm is only used in the protein function universe which is not a model architecture component (See Figure 1).}
    \label{fig:architecture-details}
\end{figure}

\addcontentsline{toc}{subsection}{Figure  \ref{fig:hist-equal-10class}:Histogram equalization results}
\addcontentsline{toc}{subsection}{Figure  \ref{fig:ablation-contact-distogram}: Ablation study}
\addcontentsline{toc}{subsection}{Figure  \ref{fig:architecture-details}: Model architecture illustration}

\end{document}